\documentclass[pra,twocolumn,preprintnumbers,showpacs,nofootinbib,eqsecnum]{revtex4-1}
\usepackage{amsmath,amssymb,bm,graphicx,hyperref}

\renewcommand\d{\partial}
\newcommand\grad{\bm{\nabla}}
\newcommand\+{\dagger}
\newcommand\<{\langle}
\renewcommand\>{\rangle}
\newcommand\up{\uparrow}
\newcommand\down{\downarrow}
\newcommand\x{{\bm{x}}}
\newcommand\y{{\bm{y}}}
\newcommand\z{{\bm{z}}}
\renewcommand\k{{\bm{k}}}
\newcommand\p{{\bm{p}}}
\newcommand\q{{\bm{q}}}
\newcommand\ek{\epsilon_\k}
\newcommand\eK{\epsilon_{\bm{K}}}
\newcommand\ep{\epsilon_\p}
\newcommand\eq{\epsilon_\q}
\newcommand\eps{\epsilon}
\renewcommand\j{\bm{j}}

\renewcommand\Re{\mathrm{Re}}
\renewcommand\Im{\mathrm{Im}}
\newcommand\kB{k_\mathrm{B}}
\newcommand\kF{k_\mathrm{F}}
\newcommand\eF{\epsilon_\mathrm{F}}

\newcommand\reg{\mathrm{reg}}
\newcommand\Born{\mathrm{Born}}
\newcommand\BCS{\mathrm{BCS}}
\newcommand\free{\mathrm{free}}
\newcommand\A{\mathcal{A}}
\newcommand\C{\mathcal{C}}
\newcommand\E{\mathcal{E}}
\newcommand\G{\mathcal{G}}
\newcommand\I{\mathcal{I}}
\newcommand\J{\mathcal{J}}
\newcommand\K{\mathcal{K}}
\renewcommand\L{\mathcal{L}}
\renewcommand\O{\mathcal{O}}
\renewcommand\S{\mathcal{S}}

\begin{document}
\preprint{MIT-CTP 4311, LA-UR-11-11809}

\title{``Hard probes'' of strongly interacting atomic gases}
\title{Probing strongly interacting atomic gases with energetic atoms}

\author{Yusuke~Nishida}
\affiliation{Center for Theoretical Physics,
Massachusetts Institute of Technology, Cambridge, Massachusetts 02139, USA}
\affiliation{Theoretical Division, Los Alamos National Laboratory,
Los Alamos, New Mexico 87545, USA}

\begin{abstract}
 We investigate properties of an energetic atom propagating through
 strongly interacting atomic gases.  The operator product expansion is
 used to systematically compute a quasiparticle energy and its
 scattering rate both in a spin-1/2 Fermi gas and in a spinless Bose
 gas.  Reasonable agreement with recent quantum Monte Carlo simulations
 even at a relatively small momentum $k/\kF\gtrsim1.5$ indicates that
 our large-momentum expansions are valid in a wide range of momentum.
 We also study a differential scattering rate when a probe atom is shot
 into atomic gases.  Because the number density and current density of
 the target atomic gas contribute to the forward scattering only, its
 contact density (measure of short-range pair correlation) gives the
 leading contribution to the backward scattering.  Therefore, such an
 experiment can be used to measure the contact density and thus provides
 a new local probe of strongly interacting atomic gases.
\end{abstract}

\date{October 2011}


\maketitle

\tableofcontents

\section{Introduction}
Strongly interacting many-body systems appear in various subfields of
physics ranging from atomic physics to condensed matter physics to
nuclear and particle physics.  An understanding of their properties is
always important and challenging.  Among others, ultracold atoms offer
ideal grounds to develop our understanding of many-body physics because
we can control their interaction strength, dimensionality of space, and
quantum statistics at
will~\cite{Bloch:2008,Giorgini:2008,Ketterle:2008}.  Many-body
properties of strongly interacting atomic gases have been probed by a
number of experimental methods including hydrodynamic
expansions~\cite{OHara:2002,Clancy:2007,Cao:2010,Riedl:2011,Trenkwalder:2011},
Bragg
spectroscopies~\cite{Veeravalli:2008,Kuhnle:2010,Kuhnle:2011,Hoinka:2012},
radio-frequency
spectroscopies~\cite{Schirotzek:2009,Pieri:2011,Sommer:2012}, precise
thermodynamic
measurements~\cite{Horikoshi:2010,Nascimbene:2010,Navon:2010,Nascimbene:2011,Navon:2011,Ku:2011,Houcke:2011},
and collisions of two atomic
clouds~\cite{Joseph:2011,Sommer:2011a,Sommer:2011b}.  In particular, the
photoemission spectroscopy employed in
Refs.~\cite{Stewart:2008,Gaebler:2010} is a direct analog of that known
to be powerful in condensed matter physics~\cite{Damascelli:2004}.

\begin{figure*}[t]
 \includegraphics[width=0.95\textwidth,clip]{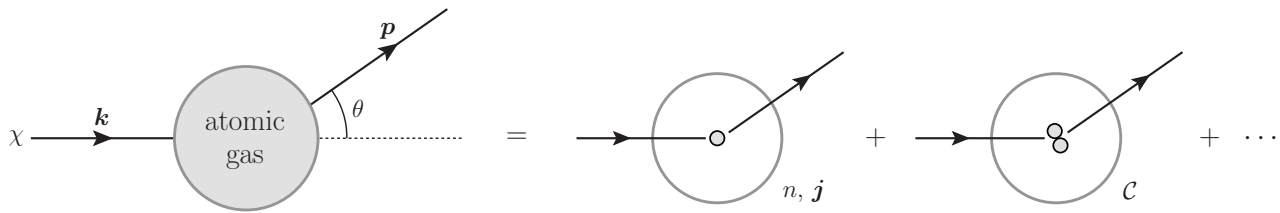}
 \caption{Schematic of a proposed scattering experiment in which we
 shoot a probe atom $\chi$ into an atomic gas with a large momentum $\k$
 and measure its differential scattering rate.  Three leading
 contributions come from two-body scatterings proportional to the number
 density $n$ and current density $\j$ of the target atomic gas and from
 a three-body scattering proportional to its contact density $\C$
 (measure of short-range pair correlation).  We will find that the
 number and current densities contribute to the forward scattering
 ($\theta<90^\circ$) only and therefore the contact density gives the
 dominant contribution to the backward scattering ($\theta>90^\circ$).
 \label{fig:scattering}}
\end{figure*}

On the other hand, often in nuclear and particle physics, high-energy
particles play important roles to reveal the nature of target
systems.  For example, neutron-deuteron or proton-deuteron scatterings
at intermediate or higher energies are important to reveal the existence
of three-nucleon forces in
nuclei~\cite{Witala:1998ey,Nemoto:1998,KalantarNayestanaki:2011wz}.
Also, two-nucleon knockout reactions by high-energy protons or electrons
have been employed to reveal short-range pair correlations in
nuclei~\cite{Subedi:2008,Arrington:2012}.  Furthermore, the discovery of
``jet quenching'' (i.e., significant energy loss of high-energy quarks
and gluons) at Relativistic Heavy Ion Collider
(RHIC)~\cite{PHENIX:2001,STAR:2003} and Large Hadron Collider
(LHC)~\cite{ATLAS:2010,CMS:2011} is one of the most striking pieces of
evidence that the matter created there is a strongly interacting
quark-gluon plasma~\cite{CERN:2010}.  Such high-energy degrees of
freedom to probe the nature of a quark-gluon plasma are generally
referred to as ``hard probes''~\cite{HardProbes2010}.  In condensed
matter physics, the use of high-energy neutrons to probe the momentum
distribution of helium atoms in liquid helium was first suggested in
Refs.~\cite{Miller:1962,Hohenberg:1966} and has been widely employed in
experiments~\cite{Woods:1973,Griffin:1993,Snow:1995}.

Now in ultracold-atom experiments, analogs of the hard probe naturally
exist because a recombination of three atoms into a two-body bound state
(dimer) produces an atom-dimer ``dijet'' propagating through the medium.
Such three-body recombinations rarely occur in spin-1/2 Fermi gases
because of the Pauli exclusion principle~\cite{Petrov:2004} but
frequently occur in spinless Bose gases and have been used as a probe of
the Efimov
effect~\cite{Kraemer:2006,Zaccanti:2009,Pollack:2009,Ferlaino:2010,Ferlaino:2011}.
While atom-dimer ``dijets'' are normally considered to simply escape
from the system, multiple collisions of the produced energetic dimer
with atoms in the atomic gas were argued in Ref.~\cite{Zaccanti:2009} to
account for the observed enhancement in atom loss.  A well-founded
understanding of collisional properties of an energetic atom or dimer in
the medium is clearly desired here (see early
works~\cite{Beijerinck:2000,Schuster:2001} and also recent
one~\cite{Machtey:2012}).  In addition to these naturally produced
energetic atoms, it is also possible to externally shoot energetic atoms
into an atomic gas in a controlled way and measure the momentum
distribution of scattered atoms.  Indeed, closely related experiments to
collide two atomic clouds have been performed successfully for Bose
gases~\cite{Thomas:2004,Buggle:2004,Kjaergaard:2004,Mellish:2007,Perrin:2007,Krachmalnicoff:2010,Williams:2012,Rakonjac:2012}
and strongly interacting Fermi
gases~\cite{Joseph:2011,Sommer:2011a,Sommer:2011b}, which have been
analyzed
theoretically~\cite{Wulin:2011,Taylor:2011,Goulko:2011,Bruun:2008}.

In this paper, we investigate various properties of an energetic atom
propagating through strongly interacting atomic gases.  Such properties
include a quasiparticle energy and a rate at which the atom is scattered
in the medium.  Both the quantities reflect many-body properties of the
atomic gas and, in particular, the scattering rate may be useful to
better understand multiple-atom loss mechanisms due to atom-dimer
``dijets'' produced by three-body recombination
events~\cite{Zaccanti:2009,Machtey:2012}.  Also we propose a scattering
experiment in which we shoot a probe atom into the atomic gas with a
large momentum and measure its differential scattering rate (see
Fig.~\ref{fig:scattering}).  Resulting scattering data must bring out
some information about the target atomic gas.  What can we learn about
the strongly interacting atomic gas from these scattering data?  This
question will be addressed in this paper.

Seemingly, these problems are difficult to tackle because of the nature
of strong interactions.  Quite remarkably, however, these problems can
be addressed in a systematic way.  This is because the atom with a large
momentum probes a short distance at which it finds only a few atoms.
Therefore, apart from probabilities of finding such few atoms in the
medium, our problem reduces to few-body scattering problems.  At the
end, we will find that such an energetic atom can be useful to locally
probe many-body aspects of strongly interacting atomic gases.

Atomic gases created in laboratories at ultralow temperatures and
quark-gluon plasmas created at RHIC and LHC at ultrahigh temperatures
are both strongly interacting many-body systems.  In spite of the fact
that they are at two extremes, various analogies have been discussed in
the literature such as hydrodynamic behaviors and small shear viscosity
to entropy density ratios~\cite{Schafer:2009dj}.  This work intends to
build a new bridge between them from the perspective of ``hard probes.''

Since this paper turns out to be long, we first summarize our main
results in Sec.~\ref{sec:summary}, discuss their consequences, and
compare them with recent quantum Monte Carlo simulations.  Our main
results consist of the quasiparticle energy and scattering rate of an
energetic atom in a spin-1/2 Fermi gas (Sec.~\ref{sec:fermi_gas}), those
in a spinless Bose gas (Sec.~\ref{sec:bose_gas}), and the differential
scattering rate of a different spin state of atoms shot into a spin-1/2
Fermi gas or a spinless Bose gas (Sec.~\ref{sec:differential}).
Furthermore, a connection of our hard-probe formula derived in
Sec.~\ref{sec:differential} with dynamic structure factors in the
weak-probe limit is elucidated in Sec.~\ref{sec:weak-probe}.  Finally,
Sec.~\ref{sec:conclusion} is devoted to conclusions of this paper and
some details of calculations are presented in
Appendices~\ref{app:wilson}--\ref{app:optical}.

Throughout this paper, we set $\hbar=1$, $\kB=1$, and use shorthand
notations $(k)\equiv(k_0,\k)$, $(x)\equiv(t,\x)$,
$kx\equiv k_0t-\k\cdot\x$, and
$\psi^\+\tensor\d\psi\equiv[\psi^\+(\d\psi)-(\d\psi^\+)\psi]/2$.
Also, note that implicit sums over repeated spin indices
$\sigma=\,\up,\down$ are {\em not\/} assumed in this paper.

\section{Summary of results and discussions \label{sec:summary}}
We suppose that atoms interact with each other by a short-range
potential and its potential range $r_0$ is much smaller than other
length scales in the atomic gas such as an $s$-wave scattering length
$a$, a mean interparticle distance $n^{-1/3}$, and a thermal de Broglie
wavelength $\lambda_T\sim1/\sqrt{mT}$.  Furthermore, we suppose that a
wavelength of the energetic atom $|\k|^{-1}$ is much smaller than the
latter length scales but still much larger than the potential range.
Therefore, the following hierarchy is assumed in the length scales:
\begin{equation}\label{eq:hierarchy}
 r_0 \ll |\k|^{-1} \ll |a|,\,n^{-1/3},\,\lambda_T.
\end{equation}
Since the potential range is much smaller than all other length scales,
we can take the zero-range limit $r_0\to0$.  Then physical observables
of our interest are expanded in terms of small quantities $1/(a|\k|)$,
$n^{1/3}/|\k|$, $1/(\lambda_T|\k|)\ll1$, which are collectively denoted
by $O(\k^{-1})$, and various contributions are organized systematically 
according to their inverse powers of $|\k|$.

\subsection{Quasiparticle energy and scattering rate}
In the case of a spin-1/2 Fermi gas with equal masses $m=m_\up=m_\down$,
the quasiparticle energy and scattering rate of spin-up fermions have
the following systematic expansions in the large-momentum limit (see
Sec.~\ref{sec:fermi_gas} for details):%
\footnote{The energy is often measured with respect to a chemical
potential.  In this case, the quasiparticle energies in
Eq.~(\ref{eq:energy_fermi}) and (\ref{eq:energy_bose}) should be
replaced with $E_\up(\k)-\mu_\up$ and $E_b(\k)-\mu_b$, respectively.}
\begin{equation}\label{eq:energy_fermi}
 E_\up(\k) = \left[1 + 32\pi\frac{n_\down}{a_f|\k|^4}
	      - 7.54\frac{\C_f}{|\k|^4} + O(\k^{-6})\right]\frac{\k^2}{2m}
\end{equation}
and
\begin{equation}\label{eq:rate_fermi}
 \begin{split}
  \Gamma_\up(\k)
  &= \biggl[32\pi\biggl(1-\frac4{a_f^2|\k|^2}\biggr)\frac{n_\down}{|\k|^3} \\
  &\quad + \left.44.2\frac{\C_f}{a_f|\k|^5} + O(\k^{-6})\right]\frac{\k^2}{2m}.
 \end{split}
\end{equation}
Here, $a_f$ is an $s$-wave scattering length between spin-up and -down
fermions, $n_\down$ is a number density of spin-down fermions, and $\C_f$
is a contact density which measures the probability of finding spin-up
and -down fermions close to each
other~\cite{Tan:2005,Braaten:2008uh,Braaten:2010if}.  The results for
spin-down fermions are obtained simply by exchanging spin indices
$\up\,\leftrightarrow\,\down$.

On the other hand, in the case of a spinless Bose gas, the quasiparticle
energy and scattering rate of bosons have the following systematic
expansions in the large-momentum limit (see Sec.~\ref{sec:bose_gas} for
details):\footnotemark[1]
\begin{equation}\label{eq:energy_bose}
 E_b(\k) = \biggl[1 + 64\pi\frac{n_b}{a_b|\k|^4}
  - 2\,\Re f\!\left(\frac{|\k|}{\kappa_*}\right)
  \frac{\C_b}{|\k|^4} + O(\k^{-5})\biggr]\frac{\k^2}{2m}
\end{equation}
and
\begin{equation}\label{eq:rate_bose}
 \Gamma_b(\k) = \left[64\pi\frac{n_b}{|\k|^3}
		 + 4\,\Im f\!\left(\frac{|\k|}{\kappa_*}\right)
		 \frac{\C_b}{|\k|^4} + O(\k^{-5})\right]\frac{\k^2}{2m}.
\end{equation}
Here, $a_b$ is an $s$-wave scattering length between two identical
bosons, $n_b$ is a number density of bosons, and $\C_b$ is a contact
density which measures the probability of finding two bosons close to
each other~\cite{Braaten:2011sz}.  $f(|\k|/\kappa_*)$ with $\kappa_*$
being the Efimov parameter is a universal log-periodic function
determined in Sec.~\ref{sec:bose_gas} [see Fig.~\ref{fig:efimov} and
Eq.~(\ref{eq:approx_bose})].  Note that the coefficient of the contact
density in the scattering rate is always negative because $\Im f$ ranges
from $-13.3$ to $-11.2$.  This rather counterintuitively means that the
energetic boson can escape from the medium easier than we naively
estimate from a binary collision.

Each term has a simple physical meaning.  Besides the free particle
kinetic energy in Eq.~(\ref{eq:energy_fermi}) or (\ref{eq:energy_bose}),
the leading term represents a contribution from a two-body scattering in
which the energetic atom collides with one atom coming from the atomic
gas.  The probability of finding such an atom in the atomic gas is
quantified by the number density $n_{\sigma,b}$.  Similarly, the
subleading term represents a contribution from a three-body scattering
in which the energetic atom collides with a small pair of two atoms
coming from the atomic gas.  The probability of finding such a small
pair in the atomic gas is quantified by the contact density $\C_{f,b}$.

These results are valid for an arbitrary many-body state with
translational and rotational symmetries (i.e., for any scattering
length, density, and temperature) as long as Eq.~(\ref{eq:hierarchy}) is
satisfied.  All nontrivial information about the many-body state is
encoded into the various densities $n_{\sigma,b}$ and $\C_{f,b}$.

\subsection{Differential scattering rate}
We then consider the proposed scattering experiment in which we shoot a
probe atom into the atomic gas and measure its differential scattering
rate (see Fig.~\ref{fig:scattering}).  We assume that the probe atom
denoted by $\chi$ is distinguishable from the rest of the atoms
constituting the target atomic gas but has the same mass $m$, which is
possible by using a different atomic spin state.  When the $\chi$ atom
is shot into a spin-1/2 Fermi gas, its differential scattering rate has
the following systematic expansion in the large-momentum limit (see
Sec.~\ref{sec:differential} for details):
\begin{widetext}
\begin{align}\label{eq:diff_fermi}
 \frac{d\Gamma_\chi(\k)}{d\Omega}
 &= \left[32\cos\theta\,\Theta(\cos\theta)\,
 \frac{n_\up(\x)+n_\down(\x)}{|\k|^3}
 + 32\left\{2\cos\theta\,\Theta(\cos\theta)\,\hat\k
 - \delta(\cos\theta)\,\hat\k + \Theta(\cos\theta)\,\hat\p\right\}
 \cdot\frac{\j_\up(\x)+\j_\down(\x)}{|\k|^4}\right. \notag\\[4pt]
 &\quad + \left.2\mspace{1mu}g\!\left(\cos\theta,\frac{|\k|}{\kappa'_*}\right)
 \frac{\C_f(\x)}{|\k|^4} + O(\k^{-5})\right]\frac{\k^2}{2m}.
\end{align}
\end{widetext}
Here, $\Theta(\,\cdot\,)$ is the Heaviside step function and $\theta$ is
a polar angle of the measured momentum $\p$ with respect to the incident
momentum chosen to be $\k=|\k|\hat\z$.  Accordingly, when a bunch of
independent $\chi$ atoms with a total number $N_\chi$ is shot into the
atomic gas, the number of scattered $\chi$ atoms measured at an angle
$(\theta,\varphi)$ is predicted to be
\begin{equation}\label{eq:diff_integrated}
 N_\mathrm{sc}(\theta,\varphi)
  = N_\chi\,\frac{m}{|\k|}\int\!dl\,\frac{d\Gamma_\chi(\k)}{d\Omega},
\end{equation}
where the line integral is taken along a classical trajectory of the
$\chi$ atom.

The differential scattering rate of the $\chi$ atom shot into a spinless
Bose gas is obtained from Eq.~(\ref{eq:diff_fermi}) by replacing the
number density $n_\up+n_\down$, the current density $\j_\up+\j_\down$,
and the contact density $\C_f$ of a spin-1/2 Fermi gas with those of a
spinless Bose gas, $n_b$, $\j_b$, and $\C_b/2$, respectively [see
Eq.~(\ref{eq:diff_bose})].  These parameters are the same as those in
Eqs.~(\ref{eq:energy_fermi})--(\ref{eq:rate_bose}), while translational
or rotational symmetries are not assumed here and thus the current
density $\j_{\sigma,b}$ can be nonzero.  On the other hand, $\kappa'_*$
in Eq.~(\ref{eq:diff_fermi}) is the Efimov parameter associated with a
three-body system of the $\chi$ atom with spin-up and -down fermions
(the $\chi$ atom with two identical bosons) and thus different from
$\kappa_*$ in Eqs.~(\ref{eq:energy_bose}) and (\ref{eq:rate_bose})
associated with three identical bosons.  We note that the dependence on
scattering lengths between the $\chi$ atom and an atom constituting the
target atomic gas appears only from $O(\k^{-5})$ in the brackets.  The
corresponding formula in the weak-probe limit can be found in
Eq.~(\ref{eq:angle_weak}).

The first two terms in Eq.~(\ref{eq:diff_fermi}) come from two-body
scatterings and are proportional to the number density and current
density of the target atomic gas.  An important observation is that,
because of kinematic constraints in the two-body scattering, they
contribute to the forward scattering ($\cos\theta>0$) only.  On the
other hand, the last term comes from a three-body scattering and is
proportional to the contact density.  Its angle distribution is
determined by a universal function $g(\cos\theta,|\k|/\kappa'_*)$, which
is mostly negative on the forward-scattering side (see
Fig.~\ref{fig:differential} in Sec.~\ref{sec:differential}).  This is no
cause for alarm, of course, because it is the subleading correction
suppressed by a power of $1/|\k|$ to the leading positive contribution
of the number density.

In contrast, $g(\cos\theta,|\k|/\kappa'_*)$ is positive everywhere
on the backward-scattering side ($\cos\theta<0$) because it is now
kinematically allowed in the three-body scattering.  Therefore, the
backward scattering is dominated by the contact density of the target
atomic gas and its measurement can be used to extract the contact
density integrated along a classical trajectory of the probe atom [see
Eq.~(\ref{eq:diff_integrated})].  Since the contact density is an
important quantity to characterize strongly interacting atomic gases, a
number of ultracold-atom experiments have been performed so far to
measure its value but integrated over the whole
volume~\cite{Partridge:2005,Kuhnle:2010,Kuhnle:2011,Stewart:2010,Navon:2010}.
Our proposed experiment provides a new way to {\em locally\/} probe the
many-body aspect of strongly interacting atomic gases.

Also we find from Eq.~(\ref{eq:diff_fermi}) that the differential
scattering rate can depend on the azimuthal angle $\varphi$ only by the
current density of the target atomic gas. Therefore, the azimuthal
anisotropy in the differential scattering rate may be useful to reveal
many-body phases accompanied by currents.

\subsection{Comparison with Monte Carlo simulations \label{sec:comparison}}
All the above results are valid for an arbitrary many-body state at a
sufficiently large momentum $|\k|$ satisfying Eq.~(\ref{eq:hierarchy}).
But how large should it be?  One can gain insight into this question by
comparing our results with other reliable results; for example, from
Monte Carlo simulations.  Currently the only available Monte Carlo
result comparable with ours is about the quasiparticle energy in a
spin-1/2 Fermi gas~\cite{Magierski:2008wa,Magierski:2011wp}.  Quite
surprisingly, we will find reasonable agreement of our result
(\ref{eq:energy_fermi}) with the recent quantum Monte Carlo simulation
even at a relatively small momentum $|\k|/\kF\gtrsim1.5$.  This
indicates that our large-momentum expansions in
Eqs.~(\ref{eq:energy_fermi})--(\ref{eq:diff_fermi}) are valid in a
momentum range wider than we naively expect.

\begin{figure*}[t]\hfill
 \includegraphics[width=0.98\columnwidth,clip]{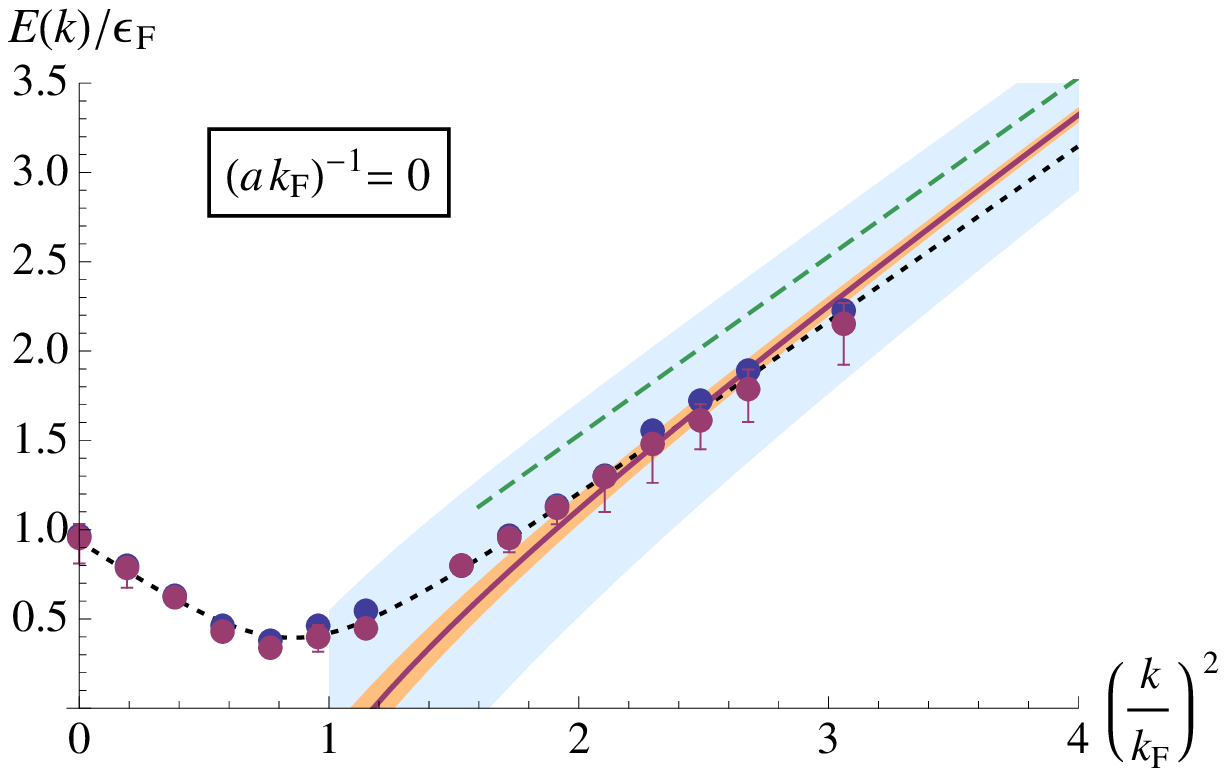}\hfill\hfill
 \includegraphics[width=0.98\columnwidth,clip]{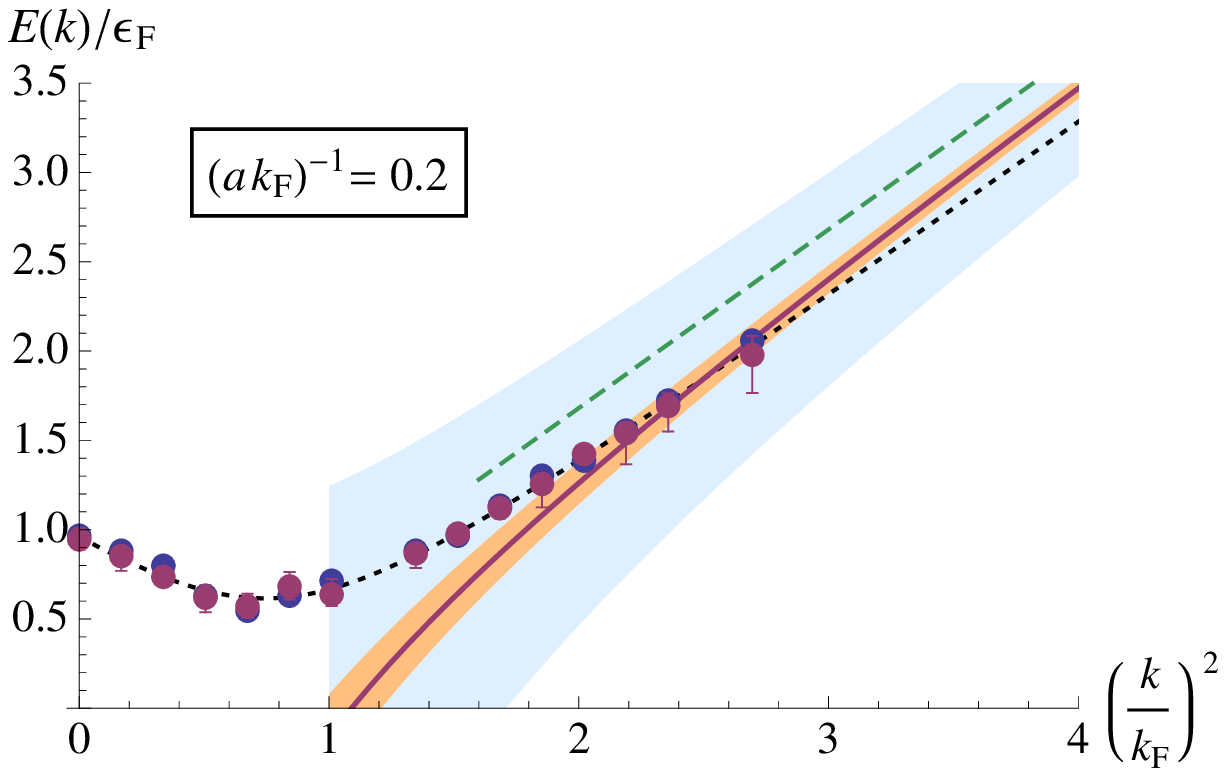}\hfill\hfill
 \caption{(Color online) Quasiparticle energies $E(\k)/\eF$ as functions
 of $(|\k|/\kF)^2$ for $(a\kF)^{-1}=0$ at $T/\eF=0.15$ (left panel) and
 for $(a\kF)^{-1}=0.2$ at $T/\eF=0.19$ (right panel).  Points are
 results extracted from the quantum Monte Carlo
 simulation~\cite{Magierski:2011wp} and dotted curves behind them are
 fits by a BCS-type formula (\ref{eq:bcs_formula}).  Solid curves are
 our results from the large-momentum expansion
 (\ref{eq:expansion_fermi}) with the use of the contact densities
 obtained in Refs.~\cite{Goulko:2010,Gandolfi:2011}.  Narrow shaded
 regions behind them correspond to the contact densities varied by
 $\pm20\%$ and broad ones indicate quasiparticle widths
 $\Gamma(\k)/\eF$ from the large-momentum expansion
 (\ref{eq:width_fermi}) with the same inputs.  For comparison, free
 particle dispersion relations, $E_\free(\k)=\k^2/(2m)-\mu$, are shown
 by dashed lines.  \label{fig:comparison}}
\end{figure*}

In Ref.~\cite{Magierski:2011wp}, P.~Magierski {\em et al.}\ extracted
the quasiparticle energy $E(\k)$ from quantum Monte Carlo data in a
balanced Fermi gas $n_\up=n_\down$ at finite temperature.  Their results
are shown by points in Fig.~\ref{fig:comparison} for $(a\kF)^{-1}=0$ at
$T/\eF=0.15$ (left panel) and for $(a\kF)^{-1}=0.2$ at $T/\eF=0.19$
(right panel) in units of the Fermi energy $\eF=\kF^2/(2m)$ as functions
of $(|\k|/\kF)^2$.

Our large-momentum expansion of the quasiparticle energy
(\ref{eq:energy_fermi}) in the same units becomes
\begin{align}\label{eq:expansion_fermi}
 \frac{E(\k)}\eF &= \left(\frac{|\k|}\kF\right)^2 - \frac\mu\eF
 + \left(\frac{16}{3\pi}\frac1{a\kF} - 7.54\,\frac\C{\kF^4}\right)
 \left(\frac\kF{|\k|}\right)^2 \notag\\
 & + O(\k^{-4}).
\end{align}
Here we used the definition of the Fermi momentum
$n_\sigma=\kF^3/(6\pi^2)$ and introduced the chemical potential $\mu$
because the quasiparticle energies in Ref.~\cite{Magierski:2011wp} are
measured with respect to $\mu$.  The values of chemical potential
obtained in Ref.~\cite{Magierski:2011wp} are
\begin{equation}
 \frac\mu\eF \approx 0.471 
  \quad\text{and}\quad 0.319 
\end{equation}
for $(a\kF)^{-1}=0$ at $T/\eF=0.15$ and for $(a\kF)^{-1}=0.2$ at
$T/\eF=0.19$, respectively, with estimated errors of about $10\%$.  Note
that the self-energy correction at $O[(\kF/|\k|)^2]$ receives two
contributions: One is from the two-body scattering which can be
attractive or repulsive depending on the sign of $(a\kF)^{-1}$.  The
other is from the three-body scattering which is proportional to
$\C/\kF^4>0$ and thus always attractive.

\begin{table}[b]
 \caption{Dimensionless contact density $\C/\kF^4$ at infinite
 scattering length $(a\kF)^{-1}=0$ from Monte Carlo simulations and
 ultracold-atom experiments at low and finite temperatures.
 \label{tab:contact}}
 \begin{ruledtabular}
  \begin{tabular}{lllclll}
   \multicolumn{3}{c}{Simulations} && \multicolumn{3}{c}{Experiments}
   \\\cline{1-3}\cline{4-7}
   \multicolumn{1}{c}{Ref.} & \multicolumn{1}{c}{$\C/\kF^4$} &
   \multicolumn{1}{c}{$T/\eF$} && \multicolumn{1}{c}{Ref.} &
   \multicolumn{1}{c}{$\C/\kF^4$} & \multicolumn{1}{c}{$T/\eF$} \\\hline
   \cite{Combescot:2006,Lobo:2006} & 0.115 & 0 &&
   \cite{Navon:2010} & 0.118(6) & 0.03(3) \\
   \cite{Gandolfi:2011} & 0.1147(3) & 0 &&
   \cite{Kuhnle:2010} & 0.101(4) & 0.10(2) \\
   \cite{Drut:2010yn} & 0.0996(34) & 0 &&
   \cite{Kuhnle:2011} & 0.105(8) & 0.09(3) \\
   \cite{Goulko:2010} & 0.1102(11) & 0.173(6) &&&& \\
   \cite{Drut:2010yn} & 0.1040(17) & 0.178 &&&&
  \end{tabular}
 \end{ruledtabular}
\end{table}

In order to make a comparison between our result and the Monte Carlo
simulation, an input into the dimensionless contact density $\C/\kF^4$
is needed ideally at the same scattering length and temperature as in
Ref.~\cite{Magierski:2011wp}.  The contact density at infinite
scattering length $(a\kF)^{-1}=0$ has been measured by a number of Monte
Carlo
simulations~\cite{Combescot:2006,Lobo:2006,Goulko:2010,Gandolfi:2011,Drut:2010yn}
and ultracold-atom
experiments~\cite{Partridge:2005,Kuhnle:2010,Kuhnle:2011,Stewart:2010,Navon:2010},
which are summarized in Table~\ref{tab:contact}.  At zero temperature,
they fall within the range of $\C/\kF^4=0.10\sim0.12$.  The
temperature dependence of the contact density was reported in
Refs.~\cite{Drut:2010yn,Kuhnle:2011}, although the situation is somewhat
controversial: The simulation observed that the contact density
increases with $T/\eF$ up to $T/\eF\approx0.4$~\cite{Drut:2010yn}, while
the experiment observed that the contact density monotonically decreases
over the temperature range $T/\eF=0.1\sim1$~\cite{Kuhnle:2011}.  Since
the precise value of the contact density at $T/\eF=0.15$ is not yet
available, we choose to use the value of Ref.~\cite{Goulko:2010} at
$T/\eF=0.173(6)$; $\C/\kF^4=0.1102(11)$.  This input fixes the
self-energy correction to be
\begin{equation}
 \left(\frac{16}{3\pi}\frac1{a\kF} - 7.54\,\frac\C{\kF^4}\right)
  \left(\frac\kF{|\k|}\right)^2 = -0.831\left(\frac\kF{|\k|}\right)^2,
\end{equation}
and our result from the large-momentum expansion
(\ref{eq:expansion_fermi}) is shown by the solid curve in
Fig.~\ref{fig:comparison} (left panel).  In order to incorporate
uncertainties of the contact density, its value is varied by $\pm20\%$
which is represented by the narrow shaded region in the same plot.  This
variation of $\pm20\%$ is a very conservative estimate of the
uncertainties because the contact density increases only by $15\%$ even
from $T/\eF\approx0$ to $0.4$ according to Ref.~\cite{Drut:2010yn}.

On the other hand, the contact density away from the infinite scattering
length is less understood, in particular, at finite temperature.
Therefore, in order to facilitate a comparison between our result and
the Monte Carlo result for $(a\kF)^{-1}=0.2$ at $T/\eF=0.19$, we choose
to use the contact density of Ref.~\cite{Gandolfi:2011} for the same
scattering length but at zero temperature; $\C/\kF^4=0.156(2)$.  This
input fixes the self-energy correction to be
\begin{equation}
 \left(\frac{16}{3\pi}\frac1{a\kF} - 7.54\,\frac\C{\kF^4}\right)
  \left(\frac\kF{|\k|}\right)^2 = -0.839\left(\frac\kF{|\k|}\right)^2,
\end{equation}
and our result from the large-momentum expansion
(\ref{eq:expansion_fermi}) is shown by the solid curve in
Fig.~\ref{fig:comparison} (right panel).  Note that the self-energy
correction at $(a\kF)^{-1}=0.2$ is close to that at $(a\kF)^{-1}=0$
because opposite changes in the contributions from two-body and
three-body scatterings happen to cancel each other.  Again uncertainties
of the contact density are incorporated by varying its value by
$\pm20\%$ which is represented by the narrow shaded region in the same
plot.

In both cases of $(a\kF)^{-1}=0$ and $0.2$, one can see from
Fig.~\ref{fig:comparison} that our results are not very sensitive to the
variations of the contact densities and, furthermore, they are in
reasonable agreement with the quantum Monte Carlo simulation even at a
relatively small momentum; $(|\k|/\kF)^2\gtrsim2$.  This indicates that
our large-momentum expansions in
Eqs.~(\ref{eq:energy_fermi})--(\ref{eq:diff_fermi}) are valid in a wide
range of momentum.

Having our large-momentum expansions tested on the quasiparticle energy,
we now present the quasiparticle width in the balanced spin-1/2 Fermi
gas.  Its large-momentum expansion (\ref{eq:rate_fermi}) in units of the
Fermi energy becomes
\begin{align}\label{eq:width_fermi}
 \frac{\Gamma(\k)}\eF &= \frac{16}{3\pi}\frac{\kF}{|\k|}
 - \frac1{a\kF}\left(\frac{64}{3\pi}\frac1{a\kF}
 - 44.2\frac{\C}{\kF^4}\right)\left(\frac{\kF}{|\k|}\right)^3 \notag\\
 & + O(\k^{-4}),
\end{align}
which is shown by the broad shaded region in Fig.~\ref{fig:comparison}
by using the same input into the contact density.  Note that the
correction at $O[(\kF/|\k|)^3]$ vanishes for $(a\kF)^{-1}=0$ (left
panel), while it is given by $+1.11\,(\kF/|\k|)^3$ for $(a\kF)^{-1}=0.2$
(right panel).  In both cases, the quasiparticle widths gradually
increase with decreasing momentum and eventually become comparable to
the quasiparticle energies.  This takes place at
$(|\k|/\kF)^2\approx2.1$ and $2.2$, respectively, which roughly
correspond to the point where our large-momentum expansions break down.

P.~Magierski {\em et al.}\ also extracted the pairing gap or pseudogap
$\Delta$, self-energy $U$, and effective mass $m^*$ parameters by
fitting a BCS-type formula
\begin{equation}\label{eq:bcs_formula}
 E_\BCS(\k) = \sqrt{\left(\frac{\k^2}{2m^*}-\mu+U\right)^2+\Delta^2}
\end{equation}
to their quasiparticle energies~\cite{Magierski:2011wp}.  However, the
fitted results (dotted curves in Fig.~\ref{fig:comparison}) do not
capture the correct asymptotic behaviors at $(|\k|/\kF)^2\gtrsim3$.
This is because $E_\BCS(\k)$ has the asymptotic expansion
\begin{equation}
 \frac{E_\BCS(\k)}\eF = \frac{m}{m^*}\left(\frac{|\k|}\kF\right)^2
  - \frac\mu\eF + \frac{U}\eF + O(\k^{-4}),
\end{equation}
in which the self-energy $U<0$ is taken to be a constant, while
according to Eq.~(\ref{eq:expansion_fermi}), $U$ should be
momentum dependent and decay as
\begin{equation}
 \frac{U}\eF \to \left(\frac{16}{3\pi}\frac1{a\kF}
  - 7.54\,\frac\C{\kF^4}\right)\left(\frac\kF{|\k|}\right)^2
\end{equation}
at $|\k|/\kF\gg1$.  Further analysis of their quantum Monte Carlo data
incorporating our exact large-momentum expansions may allow us better
access to the intriguing pseudogap physics.

\section{Spin-1/2 Fermi gas \label{sec:fermi_gas}}
Here we study properties of an energetic atom in a spin-1/2 Fermi gas
and derive its quasiparticle energy and scattering rate presented in
Eqs.~(\ref{eq:energy_fermi}) and (\ref{eq:rate_fermi}).

\subsection{Formulation}
\begin{figure}[t]
 \includegraphics[width=0.92\columnwidth,clip]{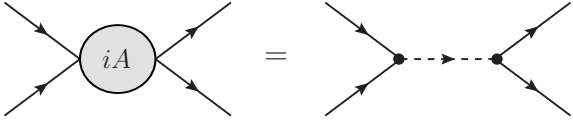}
 \caption{Two-body scattering amplitude $iA(k)$ between spin-up and
 -down fermions.  Solid (dashed) lines represent the fermion (dimer)
 propagator $iG(k)$ [$iD(k)$].  Each fermion-dimer vertex (dot) carries
 $i$, and thus, $iA(k)=-iD(k)$.  \label{fig:2-body}}
\end{figure}

The Lagrangian density describing spin-1/2 fermions with a zero-range
interaction is
\begin{equation}\label{eq:L_fermi}
 \L_F = \sum_{\sigma=\up,\down}\psi_\sigma^\+
  \left(i\d_t+\frac{\grad^2}{2m_\sigma}\right)\psi_\sigma
  + c\,\psi_\up^\+\psi_\down^\+\psi_\down\psi_\up.
\end{equation}
It is more convenient to introduce an auxiliary dimer field
$\phi=c\,\psi_\down\psi_\up$ to decouple the interaction term:
\begin{equation}
 \L_F = \sum_{\sigma=\up,\down}\psi_\sigma^\+
  \left(i\d_t+\frac{\grad^2}{2m_\sigma}\right)\psi_\sigma
  - \frac1c\phi^\+\phi + \phi^\+\psi_\down\psi_\up
  + \psi_\up^\+\psi_\down^\+\phi.
\end{equation}
For simplicity, we shall mainly consider the case of equal masses
$m=m_\up=m_\down$.  Some results in the case of unequal masses are
presented in Appendices~\ref{app:wilson} and \ref{app:integral_eq}.
The propagator of fermion field $\psi_\sigma$ in the vacuum is given by
\begin{equation}
 G(k) = \frac1{k_0-\ek+i0^+}
  \qquad \left(\ek\equiv\frac{\k^2}{2m}\right).
\end{equation}
Also by using the standard regularization procedure to relate the bare
coupling $c$ to the scattering length $a$,
\begin{equation}\label{eq:coupling_fermi}
 \frac1c = \int_{|\k|<\Lambda}\!\frac{d\k}{(2\pi)^3}\frac{m}{\k^2}
  - \frac{m}{4\pi a},
\end{equation}
the propagator of dimer field $\phi$ in the vacuum is found to be
\begin{equation}\label{eq:dimer_fermi}
 D(k) = -\frac{4\pi}m\frac1{\sqrt{\frac{\k^2}4-mk_0-i0^+}-\frac1a}.
\end{equation}
$D(k)$ coincides with the two-body scattering amplitude $A(k)$ between
spin-up and -down fermions up to a minus sign; $A(k)=-D(k)$ (see
Fig.~\ref{fig:2-body}).

Our task here is to understand the behavior of the single-particle
Green's function of spin-$\sigma$ fermions
\begin{equation}\label{eq:single-particle_fermi}
 \int\!dy\,e^{iky}\<T[\psi_\sigma(x+\tfrac{y}2)\psi_\sigma^\+(x-\tfrac{y}2)]\>
\end{equation}
in the large--energy-momentum limit $k\to\infty$ for an arbitrary
few-body or many-body state.  Without losing generality, we can consider
that of spin-up ($\sigma=\,\up$) fermions.  The result for spin-down
($\sigma=\,\down$) fermions is obtained simply by exchanging spin
indices $\up\,\leftrightarrow\,\down$.

\subsection{Operator product expansion \label{sec:ope_fermi}}
According to the operator product
expansion~\cite{Braaten:2008uh,Braaten:2008bi,Braaten:2010dv,Son:2010kq,Goldberger:2010fr,Braaten:2011sz,Barth:2011,Hofmann:2011qs,Goldberger:2011hh,Braaten:2010if},
the product of operators in Eq.~(\ref{eq:single-particle_fermi}) can be
expressed in terms of a series of local operators $\O$:
\begin{equation}\label{eq:ope_fermi}
 \int\!dy\,e^{iky}\,T[\psi_\up(x+\tfrac{y}2)\psi_\up^\+(x-\tfrac{y}2)]
  = \sum_iW_{\O_i}(k)\O_i(x).
\end{equation}
Wilson coefficients $W_\O$ depend on $k=(k_0,\k)$ and the scattering
length $a$.  When the scaling dimension of a local operator is
$\Delta_\O$, dimensional analysis implies that its Wilson coefficient
should have a form
\begin{equation}
 W_\O(k) = \frac1{|\k|^{\Delta_\O+2}}
  w_\O\!\left(\frac{\sqrt{2mk_0}}{|\k|},\frac1{a|\k|}\right).
\end{equation}
Therefore, the large--energy-momentum limit of the single-particle
Green's function is determined by Wilson coefficients of local operators
with low scaling dimensions and their expectation values with respect to
the given state.

The local operators appearing in the right-hand side of
Eq.~(\ref{eq:ope_fermi}) must have a particle number $N_\O=0$.  By
recalling $\Delta_{\psi_\sigma}=3/2$ and
$\Delta_\phi=2$~\cite{Nishida:2007pj}, we can find twelve types of local
operators with $N_\O=0$ up to scaling dimensions $\Delta_\O=5$:
\begin{equation}\label{eq:O_0_fermi}
 \openone \quad \text{(identity)}
\end{equation}
for $\Delta_\O=0$,
\begin{equation}
 \psi_\sigma^\+\psi_\sigma
\end{equation}
for $\Delta_\O=3$,
\begin{equation}
 -i\psi_\sigma^\+\tensor\d_i\psi_\sigma,\quad
  -i\d_i(\psi_\sigma^\+\psi_\sigma),\quad
  \phi^\+\phi
\end{equation}
for $\Delta_\O=4$,
\begin{subequations}\label{eq:O_5_fermi}
 \begin{equation}
  -\psi_\sigma^\+\tensor\d_i\tensor\d_j\psi_\sigma,\quad
   -\d_i(\psi_\sigma^\+\tensor\d_j\psi_\sigma),\quad
   -\d_i\d_j(\psi_\sigma^\+\psi_\sigma),
 \end{equation}
 \begin{equation}
  i\psi_\sigma^\+\tensor\d_t\psi_\sigma,\quad
   i\d_t(\psi_\sigma^\+\psi_\sigma),\quad
   -i\phi^\+\tensor\d_i\phi,\quad
   -i\d_i(\phi^\+\phi)
 \end{equation}
\end{subequations}
for $\Delta_\O=5$.  Time-space arguments of operators $(x)=(t,\x)$ are
suppressed here and below.  Operators accompanied by more spatial or
temporal derivatives have higher scaling dimensions.

Here we comment on scaling dimensions of operators involving more $\psi$
or $\phi\sim\psi_\down\psi_\up$ fields.  Scaling dimensions of operators
with three $\psi$ fields can be computed exactly by solving three-body
problems~\cite{Nishida:2007pj,Nishida:2010tm}.  For example, the lowest
two operators are $\O=2\phi(\d_i\psi_\sigma)-(\d_i\phi)\psi_\sigma$ and
$\phi\psi_\sigma$ and the products $\O^\+\O$ have scaling dimensions
$\Delta=8.54545$ and $9.33244$, respectively.  If more $\psi$ fields are
involved, it is in general difficult to compute their scaling
dimensions.  However, with the help of the operator-state
correspondence~\cite{Nishida:2007pj,Tan:2004,Werner:2006zz}, they can be
inferred from numerical calculations of energies of particles in a
harmonic potential at infinite scattering
length~\cite{Chang:2007,Stecher:2007,Blume:2007,Stecher:2008,Blume:2008,Daily:2010dz,Blume:2011,Nicholson:2011,Endres:2011,Blume:2012}.
For example, the ground-state energy of four fermions,
$E=5.01\,\hbar\omega$~\cite{Daily:2010dz}, implies that the operator
$(\phi\phi)^\+(\phi\phi)$, which involves the lowest four-body operator
$\phi\phi$, has the scaling dimension $\Delta=10.02$.  Because adding
more derivatives or fields generally increases scaling dimensions, we
conclude that the operators in
Eqs.~(\ref{eq:O_0_fermi})--(\ref{eq:O_5_fermi}) are the complete set of
local operators with $N_\O=0$ and $\Delta_\O\leq5$ in the case of equal
masses.

\begin{figure}[t]
 \includegraphics[width=0.85\columnwidth,clip]{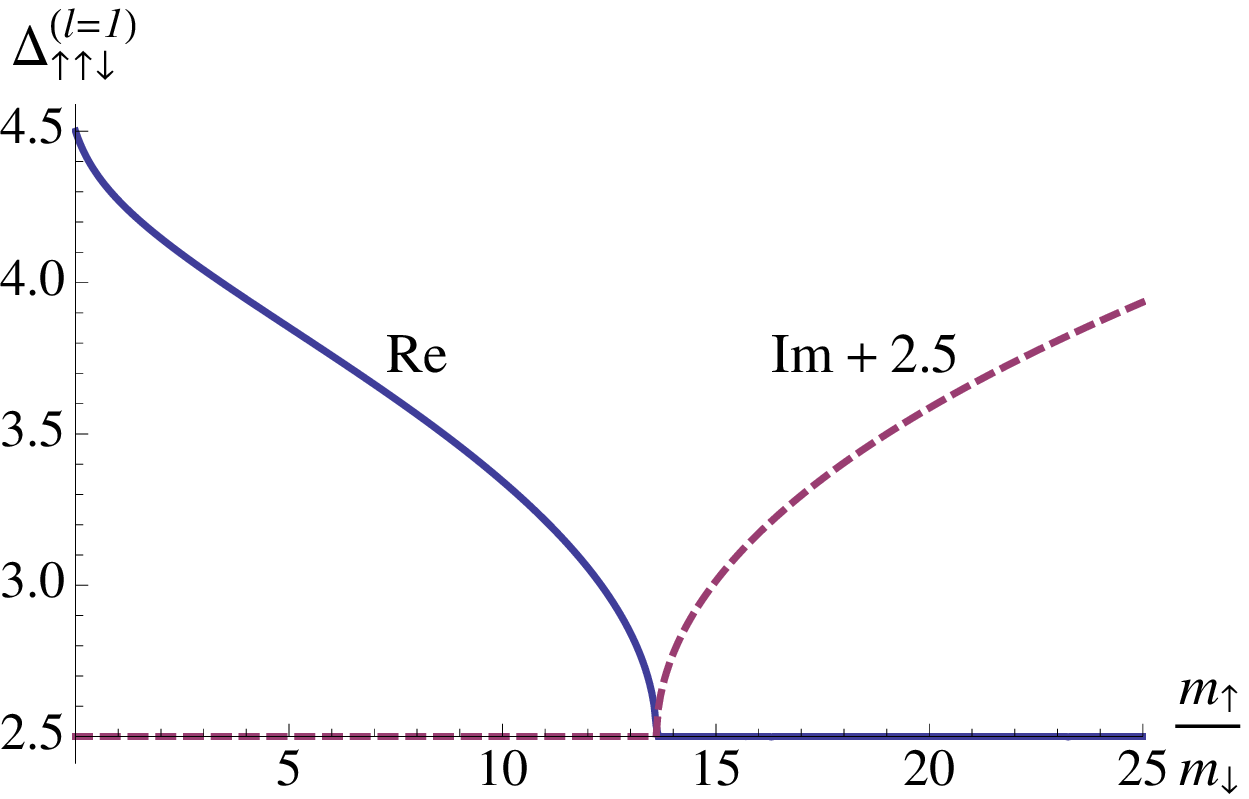}
 \caption{(Color online) Scaling dimension of the lowest three-body
 operator
 $\O_3=(m_\up+m_\down)\phi(\d_i\psi_\up)-m_\up(\d_i\phi)\psi_\up$ as a
 function of the mass ratio $m_\up/m_\down$ taken from
 Ref.~\cite{Nishida:2010tm}.  The solid curve is its real part and the
 dashed curve is its imaginary part shifted by ${+}2.5$.  The scaling
 dimension of $\O_3^\+\O_3$ is given by
 $2\mspace{1mu}\Re[\Delta_{\O_3}]$.  \label{fig:unequal_mass}}
\end{figure}

In the case of unequal masses, however, this is not always the case due
to the Efimov effect~\cite{Petrov:2003,Castin:2010}.  The scaling
dimension of the lowest three-body operator
$\O_3=(m_\up+m_\down)\phi(\d_i\psi_\up)-m_\up(\d_i\phi)\psi_\up$
decreases with increasing mass ratio $m_\up/m_\down$ and eventually
reaches $\Delta_{\O_3}=5/2$ at $m_\up/m_\down=13.607$ so that
$\Delta_{\O_3^\+\O_3}=5$  (see Fig.~\ref{fig:unequal_mass}).
Furthermore, $\Delta_{\O_3}$ develops an imaginary part for
$m_\up/m_\down>13.607$, which indicates the Efimov
effect~\cite{Nishida:2010tm}.  In general, the Efimov effect for $N$
particles implies that the corresponding $N$-body operator $\O_N$ has
the scaling dimension $\Delta_{\O_N}=5/2+i\mspace{1mu}s_\ell$ and thus
$\Delta_{\O_N^\+\O_N}=5$ with $s_\ell$ being a real number.  The recent
finding of the four-body Efimov effect for
$m_\up/m_\down>13.384$~\cite{Castin:2010} indicates the existence of a
four-body operator $\O_4$ whose scaling dimension becomes
$\Delta_{\O_4^\+\O_4}=5$ for $m_\up/m_\down>13.384$.  Therefore, only
when the mass ratio is below the lowest critical value for the Efimov
effect, the operators in
Eqs.~(\ref{eq:O_0_fermi})--(\ref{eq:O_5_fermi}) are supposed to be the
complete set of local operators with $N_\O=0$ and $\Delta_\O\leq5$.

\subsection{Wilson coefficients}
The Wilson coefficients of local operators can be obtained by matching
the matrix elements of both sides of Eq.~(\ref{eq:ope_fermi}) with
respect to appropriate few-body
states~\cite{Braaten:2008uh,Braaten:2008bi,Braaten:2010dv,Son:2010kq,Goldberger:2010fr,Braaten:2011sz,Barth:2011,Hofmann:2011qs,Goldberger:2011hh,Braaten:2010if}.
Details of such calculations are presented in
Appendix~\ref{app:wilson}.  In short, we use states
$\<\psi_\sigma(p')|$ and $|\psi_\sigma(p)\>$ to determine the Wilson
coefficients of operators of type $\psi_\sigma^\+\psi_\sigma$.  The
results are
\begin{equation}\label{eq:W_identity_fermi}
 W_{\openone}(k) = iG(k),
\end{equation}
\begin{equation}
 W_{\psi_\down^\+\psi_\down}(k) = -iG(k)^2A(k),
\end{equation}
\begin{equation}
 W_{-i\psi_\down^\+\tensor\d_i\psi_\down}(k)
 = -iG(k)^2\frac\d{\d k_i}A(k),
\end{equation}
\begin{equation}
 W_{-\psi_\down^\+\tensor\d_i\tensor\d_j\psi_\down}(k)
 = -iG(k)^2\frac12\frac{\d^2}{\d k_i\d k_j}A(k),
\end{equation}
\begin{equation}
 W_{i\psi_\down^\+\tensor\d_t\psi_\down}(k)
 = -iG(k)^2\frac\d{\d k_0}A(k),
\end{equation}
\begin{equation}
 W_{-\d_i\d_j(\psi_\down^\+\psi_\down)}(k)
 = -iA(k)\frac{G(k)^3}{4m}\!\left[\delta_{ij} + k_ik_j\frac{G(k)}m\right],
\end{equation}
\begin{equation}
 W_{-i\d_i(\psi_\down^\+\psi_\down)}(k)
 = W_{-\d_i(\psi_\down^\+\tensor\d_j\psi_\down)}(k)
 = W_{i\d_t(\psi_\down^\+\psi_\down)}(k) = 0,
\end{equation}
and all $W_\O(k)=0$ for $\sigma=\,\up$, where $A(k)$ is the two-body
scattering amplitude between spin-up and -down fermions [see
Eq.~(\ref{eq:dimer_fermi})].

\begin{figure}[t]
 \includegraphics[width=0.46\columnwidth,clip]{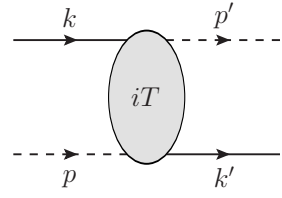}
 \caption{Three-body scattering amplitude $iT_\up(k,p;k',p')$ between
 a spin-up fermion (solid line) and a dimer (dashed line).
 \label{fig:3-body}}
\end{figure}

On the other hand, states $\<\phi(p')|$ and $|\phi(p)\>$ are used to
determine the Wilson coefficients of operators of type $\phi^\+\phi$.
The results are
\begin{equation}\label{eq:W_phi0_fermi}
 \begin{split}
  W_{\phi^\+\phi}(k) &= -iG(k)^2T_\up(k,0;k,0) \\
  & - W_{\psi_\down^\+\psi_\down}(k)
  \int\!\frac{d\q}{(2\pi)^3}\left(\frac{m}{\q^2}\right)^2 \\
  & - W_{-\psi_\down^\+\tensor\d_i\tensor\d_j\psi_\down}(k)
  \,\frac{\delta_{ij}}3\int\!\frac{d\q}{(2\pi)^3}\left(\frac{m}{\q}\right)^2 \\
  & - W_{i\psi_\down^\+\tensor\d_t\psi_\down}(k)
  \,\frac{-1}{2m}\int\!\frac{d\q}{(2\pi)^3}\left(\frac{m}{\q}\right)^2,
 \end{split}
\end{equation}
\begin{equation}\label{eq:W_phi1_fermi}
 \begin{split}
  W_{-i\phi^\+\tensor\d_i\phi}(k)
  &= -iG(k)^2\frac\d{\d p_i}T_\up(k,p;k,p)\big|_{p\to0} \\
  & - W_{-i\psi_\down^\+\tensor\d_i\psi_\down}(k)\,\frac12
  \int\!\frac{d\q}{(2\pi)^3}\left(\frac{m}{\q^2}\right)^2,
 \end{split}
\end{equation}
\begin{equation}\label{eq:W_phi2_fermi}
 W_{-i\d_i(\phi^\+\phi)}(k)
  = -iG(k)^2\frac\d{\d p_i}T_\up(k-\tfrac{p}2,\tfrac{p}2;
  k+\tfrac{p}2,-\tfrac{p}2)\big|_{p\to0}.
\end{equation}
Here, $T_\up(k,p;k',p')$ is the three-body scattering amplitude between a
spin-up fermion and a dimer with $(k,p)$ [$(k',p')$] being their initial
(final) energy-momentum (see Fig.~\ref{fig:3-body}).  As we will show in
Sec.~\ref{sec:3-body_fermi}, $T_\up(k,0;k,0)$ and
$\d T_\up(k,p;k,p)/\d p_i|_{p\to0}$ contain infrared divergences that
are canceled exactly by the second terms in Eqs.~(\ref{eq:W_phi0_fermi})
and (\ref{eq:W_phi1_fermi}), respectively.  Therefore, it is convenient
to combine them and define finite quantities by
\begin{equation}\label{eq:T_regularized_fermi}
 T_\up^\reg(k,0;k,0) \equiv T_\up(k,0;k,0)
  - A(k)\int\!\frac{d\q}{(2\pi)^3}\left(\frac{m}{\q^2}\right)^2
\end{equation}
and
\begin{align}
 &\,\frac\d{\d p_i}T_\up^\reg(k,p;k,p)\big|_{p\to0} \\
 &\equiv  \frac\d{\d p_i}T_\up(k,p;k,p)\big|_{p\to0}
 - \frac12\frac\d{\d k_i}A(k)
 \int\!\frac{d\q}{(2\pi)^3}\left(\frac{m}{\q^2}\right)^2. \notag
 \end{align}
The regularized three-body scattering amplitude $T_\up^\reg(k,0;k,0)$
will be computed in Sec.~\ref{sec:3-body_fermi}.  On the other hand, the
ultraviolet divergences in the last two terms of
Eq.~(\ref{eq:W_phi0_fermi}) are canceled by those from expectation
values of local operators as we will see below.

\subsection{Expectation values of local operators}
Now the single-particle Green's function of spin-up fermions for an
arbitrary few-body or many-body state is obtained by taking the
expectation value of Eq.~(\ref{eq:ope_fermi}):
\begin{equation}\label{eq:propagator_fermi}
 \int\!dy\,e^{iky}\<T[\psi_\up(x+\tfrac{y}2)\psi_\up^\+(x-\tfrac{y}2)]\>
  = \sum_iW_{\O_i}(k)\<\O_i(x)\>.
\end{equation}
The expectation values of local operators in
Eqs.~(\ref{eq:O_0_fermi})--(\ref{eq:O_5_fermi}) have simple physical
meanings.  For example,
\begin{equation}
 \<\psi_\sigma^\+\psi_\sigma\> = n_\sigma(x)
\end{equation}
and
\begin{equation}
 \<-i\psi_\sigma^\+\tensor\grad\psi_\sigma\> = \j_\sigma(x)
\end{equation}
are the number density and current density of spin-$\sigma$ fermions and
$\<\d_i(\psi_\sigma^\+\psi_\sigma)\>=\d_in_\sigma(x)$,
$\<\d_i\d_j(\psi_\sigma^\+\psi_\sigma)\>=\d_i\d_jn_\sigma(x)$,
$\<\d_t(\psi_\sigma^\+\psi_\sigma)\>=\d_tn_\sigma(x)$,
$\<\d_i(-i\psi_\sigma^\+\tensor\grad\psi_\sigma)\>=\d_i\j_\sigma(x)$
are their spatial or temporal derivatives.  Furthermore, it is well
known~\cite{Braaten:2008uh,Braaten:2008bi,Braaten:2010dv,Son:2010kq,Goldberger:2010fr,Braaten:2011sz,Barth:2011,Hofmann:2011qs,Goldberger:2011hh,Braaten:2010if}
that the expectation value of $\phi^\+\phi$ is related to the contact
density $\C(x)$ by
\begin{equation}\label{eq:contact_fermi}
 \<\phi^\+\phi\> = \frac{\C(x)}{m^2},
\end{equation}
and $\<\d_i(\phi^\+\phi)\>=\d_i\C(x)/m^2$ is its spatial derivative.
The contact density measures the probability of finding spin-up and
-down fermions close to each
other~\cite{Tan:2005,Braaten:2008uh,Braaten:2010if}.  
$\j_\phi(x)\equiv m^2\<-i\phi^\+\tensor\grad\phi\>$ is an analog of the
current density for dimer field $\phi$ and shall be called a contact
current density.

If the given state is translationally invariant, the expectation values
of $-\psi_\sigma^\+\tensor\d_i\tensor\d_j\psi_\sigma$ and
$i\psi_\sigma^\+\tensor\d_t\psi_\sigma$ can be expressed in terms of the
momentum distribution function of spin-$\sigma$ fermions:
\begin{equation}
 \rho_\sigma(\q) = \int\!d\y\,e^{-i\q\cdot\y}\,
  \<\psi_\sigma^\+(t,\x-\tfrac\y2)\psi_\sigma(t,\x+\tfrac\y2)\>.
\end{equation}
By using this definition, the expectation value of
$-\psi_\sigma^\+\tensor\d_i\tensor\d_j\psi_\sigma$ is found to be
\begin{equation}\label{eq:vev_1_fermi}
 \<-\psi_\sigma^\+\tensor\d_i\tensor\d_j\psi_\sigma\>
  = \int\!\frac{d\q}{(2\pi)^3}q_iq_j\rho_\sigma(\q).
\end{equation}
Similarly, the expectation value of
$i\psi_\sigma^\+\tensor\d_t\psi_\sigma$ can be evaluated by using the
equation of motion for fermion field $\psi_\sigma$:
\begin{equation}\label{eq:vev_2_fermi}
 \begin{split}
  \<i\psi_\sigma^\+\tensor\d_t\psi_\sigma\>
  &= \frac12\left\<-\psi_\sigma^\+\left(\frac{\grad^2}{2m}
  \psi_\sigma\right) - \psi_\up^\+\psi_\down^\+\phi\right\> \\
  & + \frac12\left\<-\left(\frac{\grad^2}{2m}\psi_\sigma^\+\right)
  \psi_\sigma - \phi^\+\psi_\down\psi_\up\right\> \\
  &= \int\!\frac{d\q}{(2\pi)^3}\frac{\q^2}{2m}\rho_\sigma(\q)
  - \frac1c\<\phi^\+\phi\>.
 \end{split}
\end{equation}
Both $\<-\psi_\sigma^\+\tensor\d_i\tensor\d_j\psi_\sigma\>$ and
$\<i\psi_\sigma^\+\tensor\d_t\psi_\sigma\>$ contain ultraviolet
divergences, which cancel those that already appeared in the last two
terms of Eq.~(\ref{eq:W_phi0_fermi}).

\subsection{Single-particle Green's function \label{sec:self-energy_fermi}}
Since the derivatives of $n_\sigma$, $\j_\sigma$, and $\C$ vanish for
the translationally invariant state, the single-particle Green's
function of spin-up fermions (\ref{eq:propagator_fermi}) is now written
as
\begin{align}
 i\G_\up(k) &\equiv \int\!dy\,e^{iky}
 \<T[\psi_\up(x+\tfrac{y}2)\psi_\up^\+(x-\tfrac{y}2)]\> \notag\\
 &= W_{\openone}(k) + W_{\psi_\down^\+\psi_\down}(k)\,n_\down
 + W_{-i\psi_\down^\+\tensor\grad\psi_\down}(k)\cdot\j_\down \notag\\
 & + W_{-\psi_\down^\+\tensor\d_i\tensor\d_j\psi_\down}(k)
 \int\!\frac{d\q}{(2\pi)^3}q_iq_j\rho_\down(\q) \notag\\
 & +  W_{i\psi_\down^\+\tensor\d_t\psi_\down}(k)
 \left[\int\!\frac{d\q}{(2\pi)^3}\frac{\q^2}{2m}\rho_\down(\q)
 - \frac1c\frac\C{m^2}\right] \notag\\
 & + W_{\phi^\+\phi}(k)\frac\C{m^2}
 + W_{-i\phi^\+\tensor\grad\phi}(k)\cdot\frac{\j_\phi}{m^2}
 + \cdots.
\end{align}
By using the expressions of $W_\O(k)$ obtained in
Eqs.~(\ref{eq:W_identity_fermi})--(\ref{eq:W_phi2_fermi}), we find that
$\G_\up(k)$ can be brought into the usual form
\begin{equation}\label{eq:G(k)_fermi}
 \G_\up(k) = \frac1{k_0-\ek-\Sigma_\up(k)+i0^+},
\end{equation}
where $\Sigma_\up(k)$ is the self-energy of spin-up fermions given by
\begin{align}\label{eq:Sigma_fermi}
 \Sigma_\up(k) &= -A(k)\,n_\down
 - \frac\d{\d\k}A(k)\cdot\j_\down \notag\\
 & - \frac12\frac{\d^2}{\d k_i\d k_j}A(k)\int\!\frac{d\q}{(2\pi)^3}
 \left(q_iq_j-\frac{\delta_{ij}}3\q^2\right)\rho_\down(\q) \notag\\
 & - \frac12\frac{\d^2}{\d k_i\d k_j}A(k)
 \,\frac{\delta_{ij}}3\int\!\frac{d\q}{(2\pi)^3}\q^2
 \left(\rho_\down(\q)-\frac\C{\q^4}\right) \notag\\
 & - \frac\d{\d k_0}A(k)
 \left[\int\!\frac{d\q}{(2\pi)^3}\frac{\q^2}{2m}
 \left(\rho_\down(\q)-\frac\C{\q^4}\right)
 + \frac\C{4\pi ma}\right] \notag\\
 & - T_\up^\reg(k,0;k,0)\frac\C{m^2}
 - \frac\d{\d\p}T_\up^\reg(k,p;k,p)\big|_{p\to0}\cdot
 \frac{\j_\phi}{m^2} \notag\\
 & - \cdots.
\end{align}
Here we eliminated the bare coupling $c$ by using its relationship with
the scattering length $a$ [see Eq.~(\ref{eq:coupling_fermi})].  By
recalling the large-momentum tail of the momentum distribution
function~\cite{Tan:2005,Braaten:2008uh,Braaten:2010if}
\begin{equation}
 \lim_{|\q|\to\infty}\rho_\sigma(\q) = \frac\C{\q^4} + O(\q^{-6}),
\end{equation}
one can see that the ultraviolet divergences in
Eq.~(\ref{eq:W_phi0_fermi}) and Eqs.~(\ref{eq:vev_1_fermi}) and
(\ref{eq:vev_2_fermi}) canceled out so that $\Sigma_\up(k)$ is now
manifestly finite.  Corrections to the above expression of
$\Sigma_\up(k)$ denoted by ``$\cdots$'' start with
$\sim\<\O\>/k^{\Delta_\O-2}$, where $\O$ are all possible operators with
the lowest scaling dimension at $\Delta_\O>5$.  In the case of equal
masses, $\Delta_\O=6$.

So far we only assumed that the given state is translationally
invariant.  In addition, if the given state is rotationally invariant,
$\j_\sigma$, $\j_\phi$, and
$\int d\q/(2\pi)^3(q_iq_j-\delta_{ij}\q^2/3)\rho_\sigma(\q)$ vanish.
Therefore, in this case, the self-energy of spin-up fermions is
simplified to
\begin{align}\label{eq:self-energy_fermi}
 \Sigma_\up(k) &= -A(k)\,n_\down
 - \frac\d{\d k_0}A(k)\frac\C{4\pi ma} 
 - T_\up^\reg(k,0;k,0)\frac\C{m^2} \notag\\
 & - \left[\frac{m}3\sum_{i=1}^3\frac{\d^2}{\d k_i^2}A(k)
 + \frac\d{\d k_0}A(k)\right] \notag\\
 &\quad \times \int\!\frac{d\q}{(2\pi)^3}\frac{\q^2}{2m}
 \left(\rho_\down(\q)-\frac\C{\q^4}\right) - \cdots.
\end{align}
Note that if the given state has the spin symmetry
$\rho_\up=\rho_\down$, the integral of the momentum distribution
function in the last line of Eq.~(\ref{eq:self-energy_fermi}) can be
obtained from the energy density $\E$ by using the energy
relationship~\cite{Tan:2005,Braaten:2008uh,Braaten:2010if}:
\begin{equation}
 \sum_{\sigma=\up,\down}\int\!\frac{d\q}{(2\pi)^3}\frac{\q^2}{2m}
  \left(\rho_\sigma(\q)-\frac\C{\q^4}\right)
  = \E - \frac\C{4\pi ma}.
\end{equation}

The pole of the single-particle Green's function (\ref{eq:G(k)_fermi})
determines the quasiparticle energy and scattering rate of spin-up
fermions in a many-body system~\cite{AGD,FW}:
\begin{equation}\label{eq:pole_fermi}
 k_0-\ek-\Sigma_\up(k_0,\k) = 0.
\end{equation}
Because $\Sigma_\up$ is as small as $\sim1/k$, we can set $k_0=\ek$
in $\Sigma_\up(k)$ within the accuracy of $O(\k^{-4})$.
Then the real part of the solution to Eq.~(\ref{eq:pole_fermi}) gives
the quasiparticle energy
\begin{equation}\label{eq:real_part_fermi}
 E_\up(\k) = \ek + \Re[\Sigma_\up(\ek,\k)] + O(\k^{-4}),
\end{equation}
while its imaginary part gives the scattering rate%
\footnote{The quasiparticle residue $Z_\up(\k)$ is given by
\begin{align*}
 Z_\up^{-1}(\k) &= 1 - \frac\d{\d k_0}
 \Re\bigl[\Sigma_\up(k)\bigr]_{k_0\to E_\up(\k)} \\
 &= 1 + \frac{4\pi}{\bigl[\frac14+\frac1{(a|\k|)^2}\bigr]^2}\frac{n_\down}{a|\k|^4}
 + O(\C/\k^4).
\end{align*}
Since $Z_\up(\k)=1+O(\k^{-4})$, the single-particle spectral density
function of spin-up fermions becomes
\begin{equation*}
 \A_\up(k) = -\frac1\pi\Im[\G_\up(k)]
  = \frac1{2\pi}\frac{\Gamma_\up(\k)}{[k_0-E_\up(\k)]^2+[\frac12\Gamma_\up(\k)]^2}
\end{equation*}
within the accuracy we are currently working.  Note that $\A_\up(k)$ at
a large momentum $|\k|\gg\kF$ but below Fermi sea $k_0\simeq-\ek$ (as
opposed to $k_0\simeq\ek$ in this paper) was studied in
Ref.~\cite{Schneider:2010}.  The single-particle spectral density
function was also computed in a self-consistent $T$-matrix
approximation~\cite{Haussmann:2009} and in a quantum cluster expansion
at high temperature~\cite{Hu:2010}.}
\begin{equation}\label{eq:imaginary_part_fermi}
 \Gamma_\up(\k) = -2\,\Im[\Sigma_\up(\ek,\k)] + O(\k^{-4}).
\end{equation}
Because
\begin{equation}
 \left[\frac{m}3\sum_{i=1}^3\frac{\d^2}{\d k_i^2}A(k)
  + \frac\d{\d k_0}A(k)\right]_{k_0\to\ek} = O(\k^{-4})
\end{equation}
can be found by using the expression of $A(k)$, we arrive at the
following form of the on-shell self-energy of spin-up fermions for an
arbitrary many-body state with translational and rotational symmetries:%
\footnote{In the case of unequal masses $m_\up\neq m_\down$, this
result is modified to
\begin{align*}
 \Sigma_\up(\eps_{\k\up},\k)
 &= \frac{4\pi}{i\frac{m_\down}{M}+\frac1{a|\k|}}\frac{n_\down}{2\mu|\k|}
 - \frac{i}{\Bigl(i\frac{m_\down}{M}+\frac1{a|\k|}\Bigr)^2}
 \frac{m_\up\C}{(2\mu)^2a|\k|^3} \\
 & - T_\up^\reg(k,0;k,0)\big|_{k_0=\eps_{\k\up}}\frac\C{(2\mu)^2}
 + O(\k^{-4}),
\end{align*}
where $\eps_{\k\sigma}=\k^2/(2m_\sigma)$, a total mass
$M=m_\up+m_\down$, and a reduced mass
$\mu=m_\up m_\down/(m_\up+m_\down)$.}
\begin{equation}\label{eq:on-shell_fermi}
 \begin{split}
  \Sigma_\up(\ek,\k)
  &= \frac{4\pi}{\frac{i}2+\frac1{a|\k|}}\frac{n_\down}{m|\k|}
  - \frac{i}{\bigl(\frac{i}2+\frac1{a|\k|}\bigr)^2}\frac\C{ma|\k|^3} \\
  & - t_\up^\reg(\k;\k)\frac\C{m^2} + O(\k^{-4}).
 \end{split}
\end{equation}
Here we denoted the regularized on-shell three-body scattering amplitude
by $t_\up^\reg(\k;\p)\equiv T_\up^\reg(k,0;p,k{-}p)|_{k_0=\ek,p_0=\ep}$.

The first term in the on-shell self-energy (\ref{eq:on-shell_fermi}) is
proportional to the two-body scattering amplitude $A(\ek,\k)$ and the
number density of spin-down fermions $n_\down$.  Its physical meaning is
obvious: It is the contribution from the two-body scattering of the
large-momentum spin-up fermion with a spin-down fermion in the medium.
Similarly, the last term originates from the three-body scattering of
the large-momentum spin-up fermion with a pair of spin-up and -down
fermions close to each other, which is described by the dimer field
$\phi$.  The probability of finding such a small pair in the medium is
given by the contact density
$\C=m^2\<\phi^\+\phi\>$~\cite{Tan:2005,Braaten:2008uh,Braaten:2010if}.
We note that the spin-down fermion and the small pair of spin-up and
-down fermions coming from the medium are treated as being at rest
because their characteristic momentum $\sim\kF,\lambda_T^{-1}$ are much
smaller than $|\k|$ in the large-momentum expansion [see
Eq.~(\ref{eq:hierarchy})].

Our remaining task is thus to determine the regularized on-shell
three-body scattering amplitude $t_\up^\reg(\k;\k)$ in
Eq.~(\ref{eq:on-shell_fermi}) up to $O(\k^{-4})$, which requires solving
a three-body problem.  Since it has a form of
$t_\up^\reg(\k;\k)=(m/\k^2)\tilde{t}_\up^{\,\reg}[(a|\k|)^{-1}]$, we
need to determine the first two terms in its expansion in terms of
$(a|\k|)^{-1}$.

\begin{figure*}[t]
 \includegraphics[width=0.9\textwidth,clip]{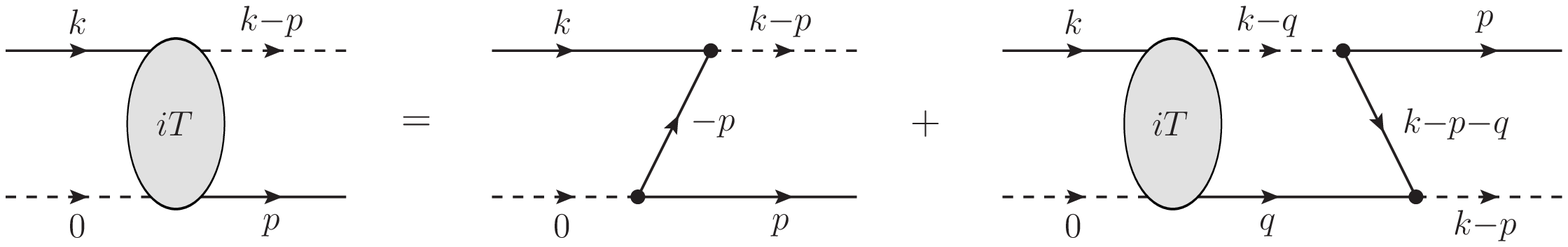}
 \caption{Integral equation for the three-body scattering amplitude
 $T_\up(k;p)\equiv T_\up(k,0;p,k{-}p)$ between a spin-up fermion and
 a dimer.  \label{fig:integral_eq}}
\end{figure*}

\subsection{Three-body problem \label{sec:3-body_fermi}}
We now compute the three-body scattering amplitude $T_\up(k,0;k,0)$.
Because $T_\up(k,0;k,0)$ does not solve a closed integral equation, 
we need to first consider $T_\up(k,0;p,k{-}p)$, which is a solution to
the integral equation depicted in Fig.~\ref{fig:integral_eq}, and then
take $p=k$.  By denoting $T_\up(k,0;p,k{-}p)$ simply by $T_\up(k;p)$,
its integral equation is written as
\begin{equation}\label{eq:bare_integral_eq_fermi}
 \begin{split}
  & T_\up(k;p) = G(-p) \\
  & - i\int\!\frac{dq_0d\q}{(2\pi)^4}\,T_\up(k;q)G(q)D(k-p)G(k-p-q).
 \end{split}
\end{equation}
Because $T_\up(k;q)$ is regular in the lower half plane of $q_0$, the
integration over $q_0$ can be easily performed to lead to
\begin{equation}
 \begin{split}
  & T_\up(k;p) = G(-p) \\
  & - \int\!\frac{d\q}{(2\pi)^3}\,
  T_\up(k;q)D(k-q)G(k-p-q)\big|_{q_0=\eq}.
 \end{split}
\end{equation}
Then by setting $k_0=\ek$, $p_0=\ep$ and defining
$t_\up(\k;\p)\equiv T_\up(\ek,\k;\ep,\p)$, we obtain an integral
equation solved by the on-shell three-body scattering amplitude
\begin{equation}\label{eq:integral_eq_fermi}
 t_\up(\k;\p) = -\frac{m}{\p^2}
  - \int\!\frac{d\q}{(2\pi)^3}\,\K_a(\k;\p,\q)\,t_\up(\k;\q),
\end{equation}
where the integral kernel $\K_a(\k;\p,\q)$ is defined by
\begin{equation}\label{eq:kernel_fermi}
 \begin{split}
  \K_a(\k;\p,\q) 
  \equiv \frac{4\pi}{\frac12\sqrt{3\q^2-\k^2-2\k\cdot\q-i0^+}-\frac1a} \\
  \times \frac1{\p^2+\q^2+\p\cdot\q-\k\cdot\p-\k\cdot\q-i0^+}.
 \end{split}
\end{equation}
This integral equation has to be solved numerically to determine
$t_\up(\k;\p)$.  We have computed $t_\up(\k;\p)$ at $(a|\k|)^{-1}=0$ and
its first derivative with respect to $(a|\k|)^{-1}$.  More details of
solving the integral equation (\ref{eq:integral_eq_fermi}) are presented
in Appendix~\ref{app:integral_eq}.

As we mentioned after Eq.~(\ref{eq:W_phi2_fermi}),
$t_\up(\k;\k)=T_\up(k,0;k,0)|_{k_0=\ek}$ contains an infrared
divergence.  This can be seen by rewriting (\ref{eq:integral_eq_fermi})
at $\p=\k$ as
\begin{equation}\label{eq:t_bare_fermi}
 \begin{split}
  t_\up(\k;\k) &= -\frac{m}{\k^2}
  + \int\!\frac{d\q}{(2\pi)^3}\,\K_a(\k;\k,\q)\frac{m}{\q^2} \\
  & - \int\!\frac{d\q}{(2\pi)^3}\,\K_a(\k;\k,\q)
  \left[t_\up(\k;\q)+\frac{m}{\q^2}\right],
 \end{split}
\end{equation}
in which the second term is infrared divergent because
\begin{equation}\label{eq:infrared_fermi}
 \K_a(\k;\k,\q) \to \frac{4\pi}{-i\frac{|\k|}2-\frac1a}\frac1{\q^2}
\end{equation}
at $|\q|\to0$.  However, this infrared divergence is canceled exactly by
the second term in Eq.~(\ref{eq:T_regularized_fermi}).  Therefore, the
regularized three-body scattering amplitude defined there is free of
divergences and its on-shell version is given by
\begin{align}\label{eq:t_regularized_fermi}
 & t_\up^\reg(\k;\k) = T_\up^\reg(k,0;k,0)|_{k_0=\ek} \notag\\
 &= -\frac{m}{\k^2} + \int\!\frac{d\q}{(2\pi)^3}\!\left[\K_a(\k;\k,\q)
 - \frac{4\pi}{-i\frac{|\k|}2-\frac1a}\frac1{\q^2}\right]\frac{m}{\q^2} \notag\\
 & - \int\!\frac{d\q}{(2\pi)^3}\,\K_a(\k;\k,\q)
 \left[t_\up(\k;\q)+\frac{m}{\q^2}\right].
\end{align}
By using the numerical solutions of $t_\up(\k;\q)$ at $(a|\k|)^{-1}=0$
and its first derivative, the expansion of $t_\up^\reg(\k;\k)$ in terms
of $(a|\k|)^{-1}$ is found to be
\begin{equation}\label{eq:solution_fermi}
 t_\up^\reg(\k;\k) = \left[3.771 + \frac{15.05}{a|\k|}i + O(\k^{-2})\right]\frac{m}{\k^2}.
\end{equation}
These numbers are universal (i.e., independent of short-range physics).
We note that the Born approximation [the first term in
Eq.~(\ref{eq:t_regularized_fermi})] gives
$t_\up^\reg(\k;\k)|_\Born=\left(-1\right)m/\k^2$, which is wrong even in
its sign.

The necessity of the above subtraction procedure is physically
understood in the following way: The ``bare'' three-body scattering
amplitude $t_\up(\k;\k)$ describes the three-body scattering of the
large-momentum spin-up fermion with a pair of spin-up and -down fermions
at rest.  $t_\up(\k;\k)$ is obtained from the integral equation depicted
in Fig.~\ref{fig:integral_eq}, which actually includes a process in
which the large-momentum spin-up fermion collides only with the
spin-down fermion coming from the pair and the other spin-up fermion
remains a spectator staying away from the scattering event.  This is
seen by recalling that the momentum $\q$ in the second term of
Eq.~(\ref{eq:t_bare_fermi}) is the relative momentum between spin-up and
-down fermions constituting the pair and thus small $|\q|$ corresponds
to the large interparticle separation.  This process is essentially the
two-body scattering which is already included in the first term of
Eq.~(\ref{eq:on-shell_fermi}).  Therefore, to avoid the double counting,
the contribution of this two-body scattering process has to be
subtracted from the bare three-body scattering amplitude.  This leads to
the regularized three-body scattering amplitude $t_\up^\reg(\k;\k)$ in
Eq.~(\ref{eq:t_regularized_fermi}), which appears in front of the
contact density in the last term of Eq.~(\ref{eq:on-shell_fermi}).

Finally, by substituting the numerical solution of the three-body
problem (\ref{eq:solution_fermi}) into Eq.~(\ref{eq:on-shell_fermi}), we
find that the on-shell self-energy of spin-up fermions has the following
systematic expansion in the large-momentum limit:
\begin{equation}
 \begin{split}
  \Sigma_\up(\ek,\k)
  &= \left[16\pi\left(-i+\frac{2}{a|\k|}+\frac4{a^2|\k|^2}i\right)
  \frac{n_\down}{|\k|^3}\right. \\
  &\quad - \left.\left(7.54+\frac{22.1}{a|\k|}i\right)
  \frac\C{|\k|^4} + O(\k^{-6})\right]\ek.
 \end{split}
\end{equation}
This result combined with Eqs.~(\ref{eq:real_part_fermi}) and
(\ref{eq:imaginary_part_fermi}) leads to the quasiparticle energy and
scattering rate of spin-up fermions presented previously in
Eqs.~(\ref{eq:energy_fermi}) and (\ref{eq:rate_fermi}).

\section{Spinless Bose gas \label{sec:bose_gas}}
Here we study properties of an energetic atom in a spinless Bose gas and
derive its quasiparticle energy and scattering rate presented in
Eqs.~(\ref{eq:energy_bose}) and (\ref{eq:rate_bose}).  The analysis is
similar to the previous case of a spin-1/2 Fermi gas.

\subsection{Formulation}
The Lagrangian density describing spinless bosons with a zero-range
interaction is
\begin{align}\label{eq:L_bose}
 \L_B &= \psi^\+\left(i\d_t+\frac{\grad^2}{2m}\right)\psi
 + \frac{c}2\,\psi^\+\psi^\+\psi\psi \\
 &= \psi^\+\left(i\d_t+\frac{\grad^2}{2m}\right)\psi
 - \frac1{2c}\phi^\+\phi + \frac12\phi^\+\psi\psi
 + \frac12\psi^\+\psi^\+\phi, \notag
\end{align}
where an auxiliary dimer field $\phi=c\,\psi\psi$ is introduced
to decouple the interaction term.  The propagator of boson field $\psi$
in the vacuum is given by
\begin{equation}
 G(k) = \frac1{k_0-\ek+i0^+}
  \qquad \left(\ek\equiv\frac{\k^2}{2m}\right).
\end{equation}
Also by using the standard regularization procedure to relate the bare
coupling $c$ to the scattering length $a$,
\begin{equation}\label{eq:coupling_bose}
 \frac1c = \int_{|\k|<\Lambda}\!\frac{d\k}{(2\pi)^3}\frac{m}{\k^2}
  - \frac{m}{4\pi a},
\end{equation}
the propagator of dimer field $\phi$ in the vacuum is found to be
\begin{equation}\label{eq:dimer_bose}
 D(k) = -\frac{8\pi}m\frac1{\sqrt{\frac{\k^2}4-mk_0-i0^+}-\frac1a}.
\end{equation}
$D(k)$ coincides with the two-body scattering amplitude $A(k)$ between
two identical bosons up to a minus sign; $A(k)=-D(k)$ (see
Fig.~\ref{fig:2-body}).  Note that $D(k)$ and $A(k)$ in the case of
bosons are twice as large as those for fermions [see
Eq.~(\ref{eq:dimer_fermi})].

Our task here is to understand the behavior of the single-particle
Green's function of bosons
\begin{equation}\label{eq:single-particle_bose}
 \int\!dy\,e^{iky}\<T[\psi(x+\tfrac{y}2)\psi^\+(x-\tfrac{y}2)]\>
\end{equation}
in the large--energy-momentum limit $k\to\infty$ for an arbitrary
few-body or many-body state.

\subsection{Operator product expansion \label{sec:ope_bose}}
According to the operator product
expansion~\cite{Braaten:2008uh,Braaten:2008bi,Braaten:2010dv,Son:2010kq,Goldberger:2010fr,Braaten:2011sz,Barth:2011,Hofmann:2011qs,Goldberger:2011hh,Braaten:2010if},
the product of operators in Eq.~(\ref{eq:single-particle_bose}) can be
expressed in terms of a series of local operators $\O$:
\begin{equation}\label{eq:ope_bose}
 \int\!dy\,e^{iky}\,T[\psi(x+\tfrac{y}2)\psi^\+(x-\tfrac{y}2)]
  = \sum_iW_{\O_i}(k)\O_i(x).
\end{equation}
The local operators $\O$ appearing in the right-hand side must have a
particle number $N_\O=0$.  By recalling $\Delta_\psi=3/2$ and
$\Delta_\phi=2$~\cite{Nishida:2007pj}, we can find thirteen types of
local operators with $N_\O=0$ up to scaling dimensions $\Delta_\O=5$:
\begin{equation}\label{eq:O_0_bose}
 \openone \quad \text{(identity)}
\end{equation}
for $\Delta_\O=0$,
\begin{equation}
 \psi^\+\psi
\end{equation}
for $\Delta_\O=3$,
\begin{equation}\label{eq:O_4_bose}
 -i\psi^\+\tensor\d_i\psi,\quad
  -i\d_i(\psi^\+\psi),\quad
  \phi^\+\phi
\end{equation}
for $\Delta_\O=4$,
\begin{subequations}\label{eq:O_5_bose}
 \begin{equation}
  -\psi^\+\tensor\d_i\tensor\d_j\psi,\quad
   -\d_i(\psi^\+\tensor\d_j\psi),\quad
   -\d_i\d_j(\psi^\+\psi),
 \end{equation}
 \begin{equation}
  i\psi^\+\tensor\d_t\psi,\quad
   i\d_t(\psi^\+\psi),\quad
   -i\phi^\+\tensor\d_i\phi,\quad
   -i\d_i(\phi^\+\phi),
 \end{equation}
 \begin{equation}\label{eq:O_3-body_bose}
  (\phi\psi)^\+(\phi\psi)
 \end{equation}
\end{subequations}
for $\Delta_\O=5$.

The striking difference between the cases of fermions and bosons is the
presence of the Efimov effect in a system of three identical
bosons~\cite{Efimov:1970}.  As we discussed at the end of
Sec.~\ref{sec:ope_fermi}, the Efimov effect implies that the
corresponding three-body operator $\phi\psi$ has the scaling dimension
$\Delta=5/2+i\mspace{1mu}s_0$ so that $(\phi\psi)^\+(\phi\psi)$ in
Eq.~(\ref{eq:O_3-body_bose}) has
$\Delta=5$~\cite{Nishida:2010tm,Braaten:2011sz}.  The determination of
its Wilson coefficient requires solving a four-body problem, which is
beyond the scope of this paper.  Therefore, in this section, we only
consider the local operators with $N_\O=0$ and $\Delta_\O\leq4$ in
Eqs.~(\ref{eq:O_0_bose})--(\ref{eq:O_4_bose}).  As before, their
expectation values have simple physical meanings such as the number
density of bosons,
\begin{equation}
 \<\psi^\+\psi\> = n(x),
\end{equation}
and its spatial derivative $\<\d_i(\psi^\+\psi)\>=\d_in(x)$, the current
density of bosons,
\begin{equation}
 \<-i\psi^\+\tensor\grad\psi\> = \j(x),
\end{equation}
and the contact density
\begin{equation}\label{eq:contact_bose}
 \<\phi^\+\phi\> = \frac{\C(x)}{m^2},
\end{equation}
which measures the probability of finding two bosons close to each
other.  This definition of the contact density coincides with that used
in Ref.~\cite{Braaten:2011sz} and thus the large-momentum tail of the
momentum distribution function of bosons is given by
$\lim_{|\q|\to\infty}\rho(\q)=\C/\q^4+O(\q^{-5})$.

\subsection{Wilson coefficients}
The Wilson coefficients of local operators can be obtained by matching
the matrix elements of both sides of Eq.~(\ref{eq:ope_bose}) with
respect to appropriate few-body
states~\cite{Braaten:2008uh,Braaten:2008bi,Braaten:2010dv,Son:2010kq,Goldberger:2010fr,Braaten:2011sz,Barth:2011,Hofmann:2011qs,Goldberger:2011hh,Braaten:2010if}.
Details of such calculations are presented in
Appendix~\ref{app:wilson}.  The results are formally equivalent to those
in the case of fermions as long as $A(k)$ in
Eqs.~(\ref{eq:W_identity_fermi})--(\ref{eq:W_phi2_fermi}) is understood
as the two-body scattering amplitude between two identical bosons [see
Eq.~(\ref{eq:dimer_bose})]:
\begin{equation}\label{eq:W_identity_bose}
 W_{\openone}(k) = iG(k),
\end{equation}
\begin{equation}
 W_{\psi^\+\psi}(k) = -iG(k)^2A(k),
\end{equation}
\begin{equation}
 W_{-i\psi^\+\tensor\d_i\psi}(k) = -iG(k)^2\frac\d{\d k_i}A(k),
\end{equation}
\begin{equation}\label{eq:W_dn_bose}
 W_{-i\d_i(\psi^\+\psi)}(k) = 0,
\end{equation}
\begin{equation}\label{eq:W_phi0_bose}
 \begin{split}
  W_{\phi^\+\phi}(k) &= -iG(k)^2T(k,0;k,0) \\
  & - W_{\psi^\+\psi}(k)\int\!\frac{d\q}{(2\pi)^3}\left(\frac{m}{\q^2}\right)^2.
 \end{split}
\end{equation}
Here, $T(k,p;k',p')$ is the three-body scattering amplitude between a
boson and a dimer with $(k,p)$ [$(k',p')$] being their initial (final)
energy-momentum (see Fig.~\ref{fig:3-body}).  Because $T(k,0;k,0)$
contains an infrared divergence that is canceled exactly by the second
term in Eq.~(\ref{eq:W_phi0_bose}), it is convenient to combine them and
define a finite quantity by
\begin{equation}\label{eq:T_regularized_bose}
 T^\reg(k,0;k,0) \equiv T(k,0;k,0)
  - A(k)\int\!\frac{d\q}{(2\pi)^3}\left(\frac{m}{\q^2}\right)^2.
\end{equation}
This regularized three-body scattering amplitude will be computed in
Sec.~\ref{sec:3-body_bose}.

\subsection{Single-particle Green's function}
Now the single-particle Green's function of bosons for an arbitrary
few-body or many-body state is obtained by taking the expectation value
of Eq.~(\ref{eq:ope_bose}).  By using the expressions of $W_\O(k)$
obtained in Eqs.~(\ref{eq:W_identity_bose})--(\ref{eq:W_phi0_bose}), we
find that it can be brought into the usual form
\begin{equation}\label{eq:G(k)_bose}
 \begin{split}
  i\G(k) &\equiv \int\!dy\,e^{iky}
  \<T[\psi(x+\tfrac{y}2)\psi^\+(x-\tfrac{y}2)]\> \\
  &= \frac{i}{k_0-\ek-\Sigma(k)+i0^+},
 \end{split}
\end{equation}
where $\Sigma(k)$ is the self-energy of bosons given by
\begin{equation}\label{eq:Sigma_bose}
 \Sigma(k) = -A(k)\,n - \frac\d{\d\k}A(k)\cdot\j
 - T^\reg(k,0;k,0)\frac\C{m^2} - \cdots.
\end{equation}
Corrections to this expression denoted by ``$\cdots$'' start with
$\sim\<\O\>/k^{\Delta_\O-2}$, where $\O$ are all operators in
Eq.~(\ref{eq:O_5_bose}) with the scaling dimension $\Delta_\O=5$.

As in Eq.~(\ref{eq:pole_fermi}), the pole of the single-particle Green's
function (\ref{eq:G(k)_bose}) determines the quasiparticle energy and
scattering rate of bosons in a many-body system.  Within the accuracy of
$O(\k^{-4})$, the real part of $\Sigma(\ek,\k)$ gives the quasiparticle
energy
\begin{equation}\label{eq:real_part_bose}
 E(\k) = \ek + \Re[\Sigma(\ek,\k)] + O(\k^{-4}),
\end{equation}
while its imaginary part gives the scattering rate
\begin{equation}\label{eq:imaginary_part_bose}
 \Gamma(\k) = -2\,\Im[\Sigma(\ek,\k)] + O(\k^{-4}).
\end{equation}
By setting $k_0=\ek$ in Eq.~(\ref{eq:Sigma_bose}), the on-shell
self-energy of bosons for an arbitrary state is found to be
\begin{equation}\label{eq:on-shell_bose}
 \begin{split}
  \Sigma(\ek,\k) &= \frac{8\pi}{\frac{i}2+\frac1{a|\k|}}\frac{n}{m|\k|}
  + \frac{4\pi i}{\bigl(\frac{i}2+\frac1{a|\k|}\bigr)^2}
  \frac{\hat\k\cdot\j}{m|\k|^2} \\
  & - t^\reg(\k;\k)\frac\C{m^2} + O(\k^{-3}),
 \end{split}
\end{equation}
where we denoted the regularized on-shell three-body scattering
amplitude by
$t^\reg(\k;\p)\equiv T^\reg(k,0;p,k{-}p)|_{k_0=\ek,p_0=\ep}$.  The
second term in Eq.~(\ref{eq:on-shell_bose}), which is proportional to
$\d A(k_0,\k)/\d\k|_{k_0\to\ek}$ and the current density $\j$,
represents the contribution from the two-body scattering in which the
large-momentum boson collides with a boson moving with a small momentum.
The physical meanings of the other two terms were discussed at the end
of Sec.~\ref{sec:self-energy_fermi}.

Our remaining task is thus to determine the regularized on-shell
three-body scattering amplitude $t^\reg(\k;\k)$ in
Eq.~(\ref{eq:on-shell_bose}) up to $O(\k^{-3})$, which requires solving
a three-body problem.  Because three identical bosons suffer from the
Efimov effect,
$(\k^2/m)\,t^\reg(\k;\k)=\tilde{t}^\reg[(a|\k|)^{-1},|\k|/\kappa_*]$
depends not only on $(a|\k|)^{-1}$ but also on $|\k|/\kappa_*$, where
$\kappa_*$ is the Efimov parameter.   As long as we are interested in
$t^\reg(\k;\k)\sim m/\k^2$ within the accuracy of $O(\k^{-3})$, we 
can set the scattering length infinite $(a|\k|)^{-1}=0$, because the
dependence on it appears only from $O(\k^{-3})$.  Then the resulting
quantity $\tilde{t}^\reg[0,|\k|/\kappa_*]$ needs to be determined, which
is a log-periodic function of $|\k|/\kappa_*$ as we will see below.

\subsection{Three-body problem \label{sec:3-body_bose}}
We now compute the three-body scattering amplitude $T(k,0;k,0)$.
Because $T(k,0;k,0)$ does not solve a closed integral equation, we need
to first consider $T(k,0;p,k{-}p)$, which is a solution to the integral
equation depicted in Fig.~\ref{fig:integral_eq}, and then take $p=k$.
By denoting $T(k,0;p,k{-}p)$ simply by $T(k;p)$, its integral equation
is written as
\begin{equation}
 \begin{split}
  & T(k;p) = -G(-p) \\
  & + i\int\!\frac{dq_0d\q}{(2\pi)^4}\,T(k;q)G(q)D(k-p)G(k-p-q).
 \end{split}
\end{equation}
Then by performing the integration over $q_0$ and defining
$t(\k;\p)\equiv T(\ek,\k;\ep,\p)$, we obtain an integral equation solved
by the on-shell three-body scattering amplitude
\begin{equation}\label{eq:integral_eq_bose}
 t(\k;\p) = \frac{m}{\p^2}
  + 2\int\!\frac{d\q}{(2\pi)^3}\,\K_a(\k;\p,\q)\,t(\k;\q),
\end{equation}
where the integral kernel $\K_a(\k;\p,\q)$ is defined in
Eq.~(\ref{eq:kernel_fermi}).  This integral equation has to be solved
numerically to determine $t(\k;\p)$ at $(a|\k|)^{-1}=0$.  

Compared to the case of fermions in Eq.~(\ref{eq:integral_eq_fermi}),
both signs in the right-hand side of Eq.~(\ref{eq:integral_eq_bose}) are
opposite due to different statistics and the second term has the factor
$2$ originating from the fact that the two-body scattering amplitude
between two identical bosons is twice larger.  The former difference
leads to the striking consequence: As is
known~\cite{Bedaque:1998kg,Braaten:2004rn}, the three-boson problem
described by the integral equation (\ref{eq:integral_eq_bose}) is ill
defined without introducing an ultraviolet momentum cutoff $\Lambda$ in
the zero orbital angular momentum channel.  $\Lambda$ can be related to
the Efimov parameter $\kappa_*$ which is defined so that three identical
bosons at infinite scattering length have the following infinite tower
of binding energies:
\begin{equation}\label{eq:efimov_bose}
 E_n \to - e^{-2\pi n/s_0}\,\frac{\kappa_*^2}m \qquad (n\to\infty),
\end{equation}
with $s_0=1.00624$~\cite{Nielsen:2001,Braaten:2004rn}.  Because
$\kappa_*$ is defined up to multiplicative factors of
$\lambda\equiv e^{\pi/s_0}=22.6944$, the solution to the integral
equation (\ref{eq:integral_eq_bose}) at $(a|\k|)^{-1}=0$ has to be a
log-periodic function of $|\k|/\kappa_*$.  We have computed such
$t(\k;\p)$ in a range $1\leq|\k|/\kappa_*\leq\lambda^2$ corresponding to
two periods.  More details of solving the integral equation
(\ref{eq:integral_eq_bose}) are presented in
Appendix~\ref{app:integral_eq}.

As we mentioned after Eq.~(\ref{eq:W_phi0_bose}),
$t(\k;\k)=T(k,0;k,0)|_{k_0=\ek}$ contains an infrared divergence.  This
can be seen by rewriting (\ref{eq:integral_eq_bose}) at $\p=\k$ as
\begin{equation}\label{eq:t_bare_bose}
 \begin{split}
  t(\k;\k) &= \frac{m}{\k^2}
  + 2\int\!\frac{d\q}{(2\pi)^3}\,\K_a(\k;\k,\q)\frac{m}{\q^2} \\
  & + 2\int\!\frac{d\q}{(2\pi)^3}\,\K_a(\k;\k,\q)
  \left[t(\k;\q)-\frac{m}{\q^2}\right],
 \end{split}
\end{equation}
in which the second term is infrared divergent at $|\q|\to0$ [see
Eq.~(\ref{eq:infrared_fermi})].  However, this infrared divergence is
canceled exactly by the second term in
Eq.~(\ref{eq:T_regularized_bose}).  Therefore, the regularized
three-body scattering amplitude defined there is free of divergences and
its on-shell version is given by
\begin{align}\label{eq:t_regularized_bose}
 & t^\reg(\k;\k) = T^\reg(k,0;k,0)|_{k_0=\ek} \notag\\
 &= \frac{m}{\k^2} + 2\int\!\frac{d\q}{(2\pi)^3}\!\left[\K_a(\k;\k,\q)
 - \frac{4\pi}{-i\frac{|\k|}2-\frac1a}\frac1{\q^2}\right]\frac{m}{\q^2} \notag\\
 & + 2\int\!\frac{d\q}{(2\pi)^3}\,\K_a(\k;\k,\q)
 \left[t(\k;\q)-\frac{m}{\q^2}\right].
\end{align}
The physical meaning of this subtraction procedure was discussed at the
end of Sec.~\ref{sec:3-body_fermi}.

\begin{figure}[t]
 \includegraphics[width=0.92\columnwidth,clip]{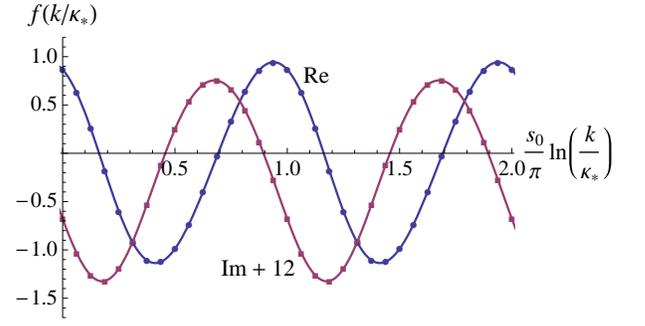}
 \caption{(Color online) Universal log-periodic function
 $f(|\k|/\kappa_*)$ defined in Eq.~(\ref{eq:solution_bose}) as a
 function of $(s_0/\pi)\ln(|\k|/\kappa_*)$.  Circles (squares) with
 steps $1/16$ correspond to its real part (imaginary part shifted by
 ${+}12$) and solid curves are fits by the approximate
 formula~(\ref{eq:approx_bose}).  Two periods in the range
 $1\leq|\k|/\kappa_*\leq e^{2\pi/s_0}$ are shown here.
 \label{fig:efimov}}
\end{figure}

By using the numerical solutions of $t(\k;\q)$ at $(a|\k|)^{-1}=0$,
$t^\reg(\k;\k)$ is computed in the range
$1\leq|\k|/\kappa_*\leq\lambda^2$ and the resulting universal function
\begin{equation}\label{eq:solution_bose}
 f\!\left(\frac{|\k|}{\kappa_*}\right)
  \equiv \frac{\k^2}m\,t^\reg(\k;\k)\big|_{(a|\k|)^{-1}=0}
\end{equation}
is shown by points in Fig.~\ref{fig:efimov}.  The logarithmic
periodicity is clearly seen and we find that our numerical results are
excellently reproduced by
\begin{equation}\label{eq:approx_bose}
 f(z) \approx X + \frac{Y_1\cos(2s_0\ln z+\delta_1)
  + iY_2\sin(2s_0\ln z+\delta_1)}{1+Z\sin(2s_0\ln z+\delta_2)},
\end{equation}
with fitting parameters $X\approx-0.09656-12.20\,i$, $Y_1\approx1.036$,
$Y_2\approx-1.032$, $Z\approx-0.08460$, and
$\delta_1\approx\delta_2\approx0.4653$.  The approximate formula
(\ref{eq:approx_bose}) is plotted by solid curves in
Fig.~\ref{fig:efimov} and differs from the numerical points only by the
amount $\lesssim4\times10^{-6}$.  Whether Eq.~(\ref{eq:approx_bose}) is
the true analytic expression of $f(|\k|/\kappa_*)$ or not needs to be
investigated further.

Finally, by substituting the numerical solution of the three-body
problem (\ref{eq:solution_bose}) into Eq.~(\ref{eq:on-shell_bose}), we
find that the on-shell self-energy of bosons has the following
systematic expansion in the large-momentum limit:
\begin{equation}
 \begin{split}
  \Sigma(\ek,\k) &= \biggl[32\pi\left(-i+\frac2{a|\k|}\right)\frac{n}{|\k|^3}
  -32\pi i\,\frac{\hat\k\cdot\j}{|\k|^4} \\
  &\quad - 2\mspace{1mu}f\!\left(\frac{|\k|}{\kappa_*}\right)
  \frac\C{|\k|^4} + O(\k^{-5})\biggr]\ek.
 \end{split}
\end{equation}
This result combined with Eqs.~(\ref{eq:real_part_bose}) and
(\ref{eq:imaginary_part_bose}) leads to the quasiparticle energy and
scattering rate of bosons presented previously in
Eqs.~(\ref{eq:energy_bose}) and (\ref{eq:rate_bose}), where the
contribution of the current density is dropped by assuming translational
and rotational symmetries.

\section{Differential scattering rate \label{sec:differential}}
So far we have studied a quasiparticle energy and a ``total'' scattering
rate of an energetic atom both in a spin-1/2 Fermi gas
(Sec.~\ref{sec:fermi_gas}) and in a spinless Bose gas
(Sec.~\ref{sec:bose_gas}).  Often in physics, differential scattering
rates or cross sections also reveal many important phenomena.  For
example, differential cross sections in neutron-deuteron or
proton-deuteron scatterings at intermediate or higher energies are
important to reveal the existence of three-nucleon forces in
nuclei~\cite{Witala:1998ey,Nemoto:1998,KalantarNayestanaki:2011wz}.
Also, momentum and angular resolutions have been essential to reveal
short-range pair correlations in nuclei from two-nucleon knockout
reactions by high-energy protons or
electrons~\cite{Subedi:2008,Arrington:2012}.  Furthermore, differential
cross sections of high-energy neutrons scattered by liquid helium have
been employed to extract the momentum distribution of helium
atoms~\cite{Miller:1962,Hohenberg:1966,Woods:1973,Griffin:1993,Snow:1995}.

Now in ultracold-atom experiments, one can in principle imagine shooting
an energetic spin-up fermion (boson) into a Fermi (Bose) gas trapped
with a finite depth and measure the angle distribution of spin-up
fermions (bosons) coming out of the trap.  Here one needs to be
cautious, however, because the incident atom cannot be distinguished
from atoms constituting the atomic gas.  For example, there is a process
in which the energetic spin-up fermion collides with a spin-down fermion
in the medium and they escape from the trap.  However, such a spin-down
fermion may be accompanied by another spin-up fermion nearby so that
they form a small pair described by the contact density.  What happens
to this spin-up fermion when its partner is kicked out by the incident
atom?  Whether it escapes from the trap to be measured or not has to be
imposed consistently on all calculations, which appears intractable in
our systematic large-momentum expansion without introducing
phenomenological procedures.

In order to avoid this problem and unambiguously determine the
differential scattering rate, it is therefore favorable to consider an
incident atom that is distinguishable from the rest of the atoms
constituting the atomic gas.  In this section, we imagine shooting a
different spin state of atoms into a spin-1/2 Fermi gas or a spinless
Bose gas with a large momentum and measure its angle distribution.  The
differential scattering rate presented in Eq.~(\ref{eq:diff_fermi}) will
be derived from the total scattering rate by using the optical theorem,
while it coincides with the one expected on physical grounds.  Here
translational or rotational symmetries are not assumed and thus the
densities depend on a time-space coordinate $(x)=(t,\x)$.

\subsection{Spin-1/2 Fermi gas}
We first consider the case of a spin-1/2 Fermi gas.  Here a probe atom
is denoted by $\chi$ and assumed to interact with spin-up and -down
fermions by scattering lengths $a_\up$ and $a_\down$, respectively.  The
Lagrangian density describing such a problem is
\begin{equation}
 \L = \chi^\+\left(i\d_t+\frac{\grad^2}{2m_\chi}\right)\chi
  + \sum_{\sigma=\up,\down}c_\sigma\,\chi^\+\psi_\sigma^\+\psi_\sigma\chi
  + \L_F,
\end{equation}
where $\L_F$ defined in Eq.~(\ref{eq:L_fermi}) describes spin-1/2
fermions interacting with each other by a scattering length $a$.  For
simplicity, we shall assume that all particles have the same mass
$m=m_\chi=m_\up=m_\down$.  In analogy with Eq.~(\ref{eq:dimer_fermi}),
the two-body scattering amplitude between the $\chi$ atom and a
spin-$\sigma$ fermion is given by
\begin{equation}\label{eq:dimer_chi-F}
 A_\sigma(k) = \frac{4\pi}m
  \frac1{\sqrt{\frac{\k^2}4-mk_0-i0^+}-\frac1{a_\sigma}}.
\end{equation}

\begin{figure*}[t]
 \includegraphics[width=0.97\textwidth,clip]{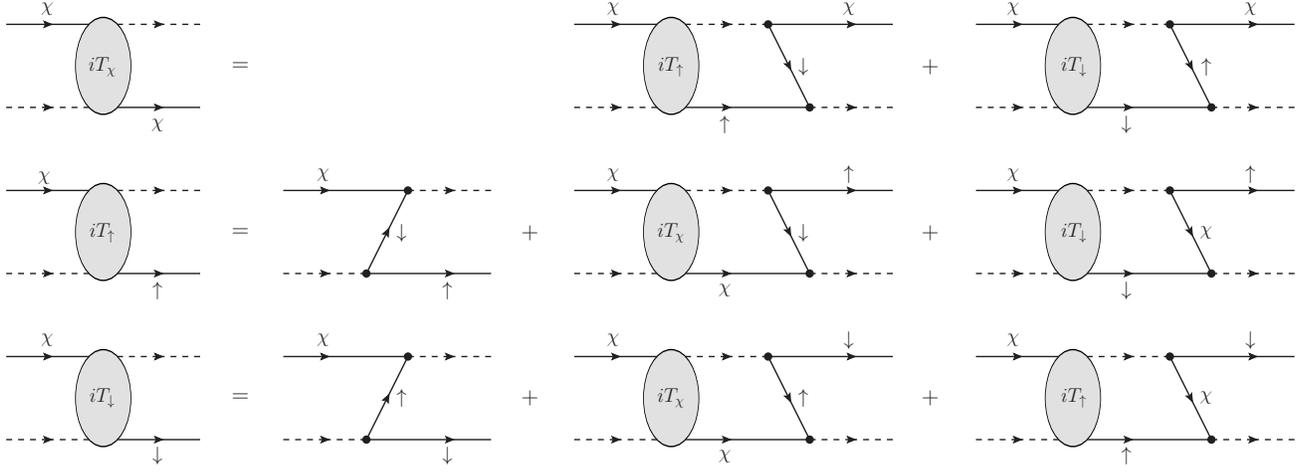}
 \caption{Set of integral equations for three-body scattering amplitudes
 $T_\chi(k;p)$, $T_\up(k;p)$, and $T_\down(k;p)$ involving three
 distinguishable particles $\chi$ and spin-$\up$ and -$\down$ fermions.
 Momentum labels are the same as those in Fig.~\ref{fig:integral_eq}.
 \label{fig:integral_eqs}}
\end{figure*}

The behavior of the single-particle Green's function of the $\chi$ atom,
\begin{equation}
 i\G_\chi(x;k) \equiv \int\!dy\,e^{iky}
  \<T[\chi(x+\tfrac{y}2)\chi^\+(x-\tfrac{y}2)]\>,
\end{equation}
in the large--energy-momentum limit $k\to\infty$ can be understood by
using the operator product expansion as in Secs.~\ref{sec:fermi_gas} and
\ref{sec:bose_gas}.  Since three distinguishable particles
($\chi,\up,\down$) with zero-range interactions suffer from the Efimov
effect~\cite{Nielsen:2001,Braaten:2004rn}, we only consider local
operators with $N_\O=0$ up to scaling dimensions $\Delta_\O=4$ (see
discussions in Sec.~\ref{sec:ope_bose}).  Then in analogy with
Eqs.~(\ref{eq:G(k)_bose}) and (\ref{eq:Sigma_bose}), $\G_\chi(x;k)$ can
be written in the usual form
\begin{equation}
 \G_\chi(x;k) = \frac1{k_0-\ek-\Sigma_\chi(x;k)+i0^+},
\end{equation}
where $\Sigma_\chi(x;k)$ is the self-energy of the $\chi$ atom given by
\begin{align}\label{eq:Sigma_chi-F}
 \Sigma_\chi(x;k) &= -\sum_{\sigma=\up,\down}
 \left[A_\sigma(k)\,n_\sigma(x)
 + \frac\d{\d\k}A_\sigma(k)\cdot\j_\sigma(x)\right] \notag\\
 & - T_\chi^\reg(k,0;k,0)\frac{\C(x)}{m^2} - \cdots.
\end{align}
Here, $n_\sigma(x)=\<\psi_\sigma^\+\psi_\sigma\>$ and
$\j_\sigma(x)=\<-i\psi_\sigma^\+\tensor\grad\psi_\sigma\>$ are the
number density and current density of spin-$\sigma$ fermions and
$\C(x)=m^2\<\phi^\+\phi\>$ is the contact density of a spin-1/2 Fermi
gas.  Note that these parameters only refer to many-body properties of
the given spin-1/2 Fermi gas and do not involve the information related
to the $\chi$ atom.  On the other hand, $T_\chi^\reg(k,0;k,0)$ is a
finite quantity defined by
\begin{align}\label{eq:T_regularized_chi-F}
 &\,T_\chi^\reg(k,0;k,0) \\
 &\equiv T_\chi(k,0;k,0) - [A_\up(k)+A_\down(k)]
 \int\!\frac{d\q}{(2\pi)^3}\left(\frac{m}{\q^2}\right)^2, \notag
\end{align}
where $T_\chi(k,p;k',p')$ is the three-body scattering amplitude between
the $\chi$ atom and a dimer composed of spin-up and -down fermions with
$(k,p)$ [$(k',p')$] being their initial (final) energy-momentum (see
Fig.~\ref{fig:3-body}).  Corrections to the above expression of
$\Sigma_\chi(x;k)$ denoted by ``$\cdots$'' start with
$\sim\<\O\>/k^{\Delta_\O-2}$, where $\O$ are all possible operators with
the scaling dimension $\Delta_\O=5$.

Then by setting $k_0=\ek$ in Eq.~(\ref{eq:Sigma_chi-F}), the on-shell
self-energy of the $\chi$ atom for an arbitrary state is found to be
\begin{align}\label{eq:on-shell_chi-F}
 & \Sigma_\chi(x;\ek,\k) \notag\\
 &= \sum_{\sigma=\up,\down}
 \left[\frac{4\pi}{\frac{i}2+\frac1{a_\sigma|\k|}}\frac{n_\sigma(x)}{m|\k|}
 + \frac{2\pi i}{\bigl(\frac{i}2+\frac1{a_\sigma|\k|}\bigr)^2}
 \frac{\hat\k\cdot\j_\sigma(x)}{m|\k|^2}\right] \notag\\
 & - t_\chi^\reg(\k;\k)\frac{\C(x)}{m^2} + O(\k^{-3}).
\end{align}
The quasiparticle energy and scattering rate of the $\chi$ atom in a
spin-1/2 Fermi gas are given by the real and imaginary parts of
$\Sigma_\chi(x;\ek,\k)$ according to
\begin{equation}\label{eq:real_part_chi-F}
 E_\chi(x;\k) = \ek + \Re[\Sigma_\chi(x;\ek,\k)] + O(\k^{-4})
\end{equation}
and
\begin{equation}\label{eq:imaginary_part_chi-F}
 \Gamma_\chi(x;\k) = -2\,\Im[\Sigma_\chi(x;\ek,\k)] + O(\k^{-4}),
\end{equation}
respectively.  Our next task is to determine the regularized on-shell
three-body scattering amplitude
$t_\chi^\reg(\k;\p)\equiv T_\chi^\reg(k,0;p,k{-}p)|_{k_0=\ek,p_0=\ep}$
in Eq.~(\ref{eq:on-shell_chi-F}), which requires solving a three-body
problem.

\subsection{Three-body problem}
We now compute the three-body scattering amplitude $T_\chi(k,0;k,0)$.
Unlike the previous cases in Secs.~\ref{sec:3-body_fermi} and
\ref{sec:3-body_bose}, $T_\chi(k;p)\equiv T_\chi(k,0;p,k{-}p)$ by itself
does not solve a closed integral equation.  To find a closed set of
integral equations, we need to introduce other three-body scattering
amplitudes $T_\sigma(k;p)$ with $\sigma=\,\up,\down$, which describe
processes where the $\chi$ atom and a dimer composed of spin-up and
-down fermions with their energy-momentum $(k,0)$ are scattered into
a spin-$\sigma$ fermion and a dimer composed of the $\chi$ atom and the
other fermion with their energy-momentum $(p,k{-}p)$.  These three
scattering amplitudes are solutions to a closed set of integral
equations depicted in Fig.~\ref{fig:integral_eqs}.  Then by following
the same procedures as in
Eqs.~(\ref{eq:bare_integral_eq_fermi})--(\ref{eq:integral_eq_fermi}),
the integral equations solved by the on-shell three-body scattering
amplitudes
$t_{\chi,\up,\down}(\k;\p)\equiv T_{\chi,\up,\down}(\ek,\k;\ep,\p)$ can
be written as
\begin{subequations}\label{eq:coupled_chi-F}
 \begin{equation}
  \begin{split}
   t_\chi(\k;\p) &= \int\!\frac{d\q}{(2\pi)^3}\,\K_{a_\down}(\k;\p,\q)\,t_\up(\k;\q) \\
   & + \int\!\frac{d\q}{(2\pi)^3}\,\K_{a_\up}(\k;\p,\q)\,t_\down(\k;\q),
  \end{split}
 \end{equation}
 \begin{equation}
  \begin{split}
   t_\up(\k;\p) &= \frac{m}{\p^2}
   + \int\!\frac{d\q}{(2\pi)^3}\,\K_a(\k;\p,\q)\,t_\chi(\k;\q) \\
   & + \int\!\frac{d\q}{(2\pi)^3}\,\K_{a_\up}(\k;\p,\q)\,t_\down(\k;\q),
  \end{split}
 \end{equation}
 \begin{equation}
  \begin{split}
   t_\down(\k;\p) &= \frac{m}{\p^2}
   + \int\!\frac{d\q}{(2\pi)^3}\,\K_a(\k;\p,\q)\,t_\chi(\k;\q) \\
   & + \int\!\frac{d\q}{(2\pi)^3}\,\K_{a_\down}(\k;\p,\q)\,t_\up(\k;\q),
  \end{split}
 \end{equation}
\end{subequations}
where the integral kernel $\K_a(\k;\p,\q)$ is defined in
Eq.~(\ref{eq:kernel_fermi}).

As long as we are interested in $t_\chi^\reg(\k;\k)\sim m/\k^2$ up to
$O(\k^{-3})$ [see Eq.~(\ref{eq:on-shell_chi-F})], we can set all three
scattering lengths infinite $(a_\up|\k|)^{-1}$, $(a_\down|\k|)^{-1}$,
$(a|\k|)^{-1}=0$, because the dependence on them appears only from
$O(\k^{-3})$.  In this case, by defining
\begin{equation}\label{eq:t_F_chi-F}
 t_F(\k;\p) \equiv \frac{2\mspace{1mu}t_\chi(\k;\p)-t_\up(\k;\p)-t_\down(\k;\p)}2
\end{equation}
and
\begin{equation}\label{eq:t_B_chi-F}
 t_B(\k;\p) \equiv \frac{t_\chi(\k;\p)+t_\up(\k;\p)+t_\down(\k;\p)}2,
\end{equation}
the three coupled integral equations (\ref{eq:coupled_chi-F}) can be
brought into two independent integral equations:
\begin{subequations}\label{eq:decoupled_chi-F}
 \begin{equation}
  t_F(\k;\p) = -\frac{m}{\p^2}
   - \int\!\frac{d\q}{(2\pi)^3}\,\K_\infty(\k;\p,\q)\,t_F(\k;\q)
 \end{equation}
 and
 \begin{equation}
  t_B(\k;\p) = \frac{m}{\p^2}
   + 2\int\!\frac{d\q}{(2\pi)^3}\,\K_\infty(\k;\p,\q)\,t_B(\k;\q).
 \end{equation}
\end{subequations}
Because these two integral equations are equivalent to
Eqs.~(\ref{eq:integral_eq_fermi}) and (\ref{eq:integral_eq_bose}) at
$(a|\k|)^{-1}=0$, their solutions are already obtained.

By using the numerical solutions obtained previously in
Eqs.~(\ref{eq:solution_fermi}) and (\ref{eq:solution_bose}),
$t_\chi^\reg(\k;\k)$ within the accuracy of $O(\k^{-3})$ is found to be
\begin{equation}\label{eq:solution_chi-F}
 \begin{split}
  t_\chi^\reg(\k;\k)
  &= \frac23\left[t_F^\reg(\k;\k) + t_B^\reg(\k;\k)\right] + O(\k^{-3}) \\
  &= \frac23\left[3.771 + f\!\left(\frac{|\k|}{\kappa'_*}\right)\right]
  \frac{m}{\k^2} + O(\k^{-3}).
 \end{split}
\end{equation}
Here, $f(|\k|/\kappa'_*)$ is the universal log-periodic function plotted
in Fig.~\ref{fig:efimov} and approximately given by
Eq.~(\ref{eq:approx_bose}) with $\kappa'_*$ being the Efimov parameter
associated with a three-body system of the $\chi$ atom with spin-up and
-down fermions.  By substituting this numerical solution of the
three-body problem into the on-shell self-energy in
Eq.~(\ref{eq:on-shell_chi-F}), the large-momentum expansions of the
quasiparticle energy and scattering rate of the $\chi$ atom in a
spin-1/2 Fermi gas are obtained from Eqs.~(\ref{eq:real_part_chi-F}) and
(\ref{eq:imaginary_part_chi-F}):
\begin{equation}
 \begin{split}
  E_\chi(x;\k) &= \Biggl[1 + 32\pi\sum_{\sigma=\up,\down}
  \frac{n_\sigma(x)}{a_\sigma|\k|^4} \\
  & - \frac43\left\{3.771 + \Re f\!\left(\frac{|\k|}{\kappa'_*}\right)\right\}
  \frac{\C(x)}{|\k|^4} + O(\k^{-5})\Biggr]\ek
 \end{split}
\end{equation}
and
\begin{equation}\label{eq:rate_chi-F}
 \begin{split}
  \Gamma_\chi(x;\k) &= \Biggl[32\pi\sum_{\sigma=\up,\down}
  \left\{\frac{n_\sigma(x)}{|\k|^3} + \frac{\hat\k\cdot\j_\sigma(x)}{|\k|^4}\right\} \\
  &\quad + \frac83\,\Im f\!\left(\frac{|\k|}{\kappa'_*}\right)
  \frac{\C(x)}{|\k|^4} + O(\k^{-5})\Biggr]\ek,
 \end{split}
\end{equation}
respectively.

\subsection{Differential scattering rate \label{sec:diff_fermi}}
Our final task is to determine the differential scattering rate
$d\Gamma_\chi(\k)/d\p$; that is, the rate at which the $\chi$ atom shot
into a spin-1/2 Fermi gas with the initial momentum $\k$ is measured at
the final momentum $\p$ (see Fig.~\ref{fig:scattering}).  We start with
the following expression for the total scattering rate obtained from
Eqs.~(\ref{eq:on-shell_chi-F}) and (\ref{eq:imaginary_part_chi-F}):
\begin{equation}\label{eq:Gamma_chi-F}
 \begin{split}
  & \Gamma_\chi(x;\k) \\
  &= \sum_{\sigma=\up,\down}\left[\frac{4\pi}{\frac14+\frac1{(a_\sigma|\k|)^2}}
  \frac{n_\sigma(x)}{m|\k|} + \frac{\pi-\frac{4\pi}{(a_\sigma|\k|)^2}}
  {\bigl[\frac14+\frac1{(a_\sigma|\k|)^2}\bigr]^2}
  \frac{\hat\k\cdot\j_\sigma(x)}{m|\k|^2}\right] \\
  & + 2\,\Im\,t_\chi^\reg(\k;\k)\frac{\C(x)}{m^2} + O(\k^{-3}) \\
  &\equiv \sum_{\sigma=\up,\down}\left[\Gamma_\chi^{(n_\sigma)}(\k)
  + \Gamma_\chi^{(\j_\sigma)}(\k)\right]
  + \Gamma_\chi^{(\C)}(\k) + O(\k^{-3}).
 \end{split}
\end{equation}
Here the contributions of the number density $n_\sigma$, current density
$\j_\sigma$, and contact density $\C$ are denoted by
$\Gamma_\chi^{(n_\sigma)}$, $\Gamma_\chi^{(\j_\sigma)}$, and
$\Gamma_\chi^{(\C)}$, respectively.

\subsubsection{Contribution of number density}
In order to extract the differential scattering rate, it is instructive
to rewrite the first term in Eq.~(\ref{eq:Gamma_chi-F}) so that its
physical meaning becomes transparent:
\begin{equation}
 \begin{split}
  \Gamma_\chi^{(n_\sigma)}(\k)
  &= n_\sigma(x)\int\!\frac{d\p\,d\q}{(2\pi)^6}|A_\sigma(\ek,\k)|^2 \\
  & \times (2\pi)^4\delta(\p+\q-\k)\delta(\ep+\eq-\ek).
 \end{split}
\end{equation}
Here, $A_\sigma(k)$ introduced in Eq.~(\ref{eq:dimer_chi-F}) is the
two-body scattering amplitude between the $\chi$ atom and a
spin-$\sigma$ fermion.  It is now obvious that
$\Gamma_\chi^{(n_\sigma)}(\k)$ represents the contribution from the
two-body scattering in which the $\chi$ atom with the initial momentum
$\k$ and a spin-$\sigma$ fermion at rest are scattered into those with
their final momenta $\p$ and $\q$, respectively.  Therefore, the
contribution of the number density of spin-$\sigma$ fermions to the
differential scattering rate of the $\chi$ atom can be read off as
\begin{equation}\label{eq:dG/dp^n_chi-F}
 \begin{split}
  \frac{d\Gamma_\chi^{(n_\sigma)}(\k)}{d\p}
  &= n_\sigma(x)\int\!\frac{d\q}{(2\pi)^6}|A_\sigma(\ek,\k)|^2 \\
  & \times (2\pi)^4\delta(\p+\q-\k)\delta(\ep+\eq-\ek).
 \end{split}
\end{equation}

Then by performing the integration over the magnitude of momentum
$|\p|^2d|\p|$, the angle distribution of the scattered $\chi$ atom is
found to be
\begin{equation}\label{eq:angle_n_chi-F}
 \frac{d\Gamma_\chi^{(n_\sigma)}(\k)}{d\Omega}
  = \frac{4\cos\theta\,\Theta(\cos\theta)}{\frac14+\frac1{(a_\sigma|\k|)^2}}
  \frac{n_\sigma(x)}{m|\k|},
\end{equation}
where $\Theta(\,\cdot\,)$ is the Heaviside step function and $\theta$ is
a polar angle of the final momentum $\p$ with respect to the initial
momentum chosen to be $\k=|\k|\hat\z$.  Because of kinematic constraints
(energy and momentum conservations) in the two-body scattering, the
number density contributes to the forward scattering ($\cos\theta>0$)
only.%
\footnote{The situation is different if the $\chi$ atom has a mass
different from that of spin-$\sigma$ fermions; $m_\chi\neq m_\sigma$.
By considering the two-body scattering in which the $\chi$ atom with the
initial momentum $\k$ collides with a spin-$\sigma$ fermion at rest, the
energy and momentum conservations constrain the final momentum $\p$ to
be on a sphere defined by
$\bigl|\p-\frac{m_\chi}{m_\chi+m_\sigma}\k\bigr|=\bigl|\frac{m_\sigma}{m_\chi+m_\sigma}\k\bigr|$.
Therefore, when $m_\chi>m_\sigma$, the $\chi$ atom can be scattered into
an angle range $\cos\theta\geq\sqrt{1-(m_\sigma/m_\chi)^2}$ only, while
when $m_\chi<m_\sigma$, it can be scattered into any angles.}

\subsubsection{Contribution of current density}
Similarly, the second term in Eq.~(\ref{eq:Gamma_chi-F}) can be
rewritten as
\begin{equation}
 \begin{split}
  & \Gamma_\chi^{(\j_\sigma)}(\k) 
  = \j_\sigma(x)\cdot\frac{\d}{\d\k}\int\!\frac{d\p\,d\q}{(2\pi)^6}
  |A_\sigma(k_0,\k)|^2 \\
  & \times (2\pi)^4\delta(\p+\q-\k)
  \delta(\ep+\eq-k_0)\Bigr|_{k_0\to\ek},
 \end{split}
\end{equation}
which represents the contribution from the two-body scattering in which
the $\chi$ atom is scattered by a spin-$\sigma$ fermion moving with a
small momentum.  Accordingly, the contribution of the current density of
spin-$\sigma$ fermions to the differential scattering rate of the $\chi$
atom can be read off as
\begin{equation}\label{eq:dG/dp^j_chi-F}
 \begin{split}
  & \frac{d\Gamma_\chi^{(\j_\sigma)}(\k)}{d\p}
  = \j_\sigma(x)\cdot\frac{\d}{\d\k}\int\!\frac{d\q}{(2\pi)^6}
  |A_\sigma(k_0,\k)|^2 \\
  & \times (2\pi)^4\delta(\p+\q-\k)
  \delta(\ep+\eq-k_0)\Bigr|_{k_0\to\ek}.
 \end{split}
\end{equation}

Then by performing the integration over the magnitude of momentum
$|\p|^2d|\p|$, the angle distribution of the scattered $\chi$ atom is
found to be
\begin{equation}\label{eq:angle_j_chi-F}
 \begin{split}
  \frac{d\Gamma_\chi^{(\j_\sigma)}(\k)}{d\Omega}
  &= \frac{2\cos\theta\,\Theta(\cos\theta)}
  {\bigl[\frac14+\frac1{(a_\sigma|\k|)^2}\bigr]^2}
  \frac{\hat\k\cdot\j_\sigma(x)}{m|\k|^2} \\
  & - \frac{\delta(\cos\theta)\,\hat\k-\Theta(\cos\theta)\,\hat\p}
  {\frac14+\frac1{(a_\sigma|\k|)^2}}\cdot\frac{4\,\j_\sigma(x)}{m|\k|^2}.
 \end{split} 
\end{equation}
Note that this differential scattering rate can be nonzero only on the
forward-scattering side again and depends on the azimuthal angle
$\varphi$ because of the $\hat\p\cdot\j_\sigma$ term.  It is easy to
check that the integration of the differential scattering rate in
Eq.~(\ref{eq:angle_n_chi-F}) or (\ref{eq:angle_j_chi-F}) over the solid
angle $d\Omega=d\cos\theta\,d\varphi$ reproduces the total scattering
rate in Eq.~(\ref{eq:Gamma_chi-F}).

\subsubsection{Contribution of contact density}
The contribution of the contact density of a spin-1/2 Fermi gas to the
differential scattering rate of the $\chi$ atom can be extracted in a
similar way.  As we discussed at the end of
Sec.~\ref{sec:self-energy_fermi}, the last term in
Eq.~(\ref{eq:Gamma_chi-F}) represents the contribution from the
three-body scattering of the $\chi$ atom with the initial momentum $\k$
with a small pair of spin-up and -down fermions at rest.  By using the
optical theorem, the imaginary part of the forward three-body scattering
amplitude $t_\chi(\k;\k)$ can be written as a form of the total
scattering rate:
\begin{align}\label{eq:optical_t}
 & 2\,\Im\,t_\chi(\k;\k) \\
 &= \int\!\frac{d\p\,d\q_\up d\q_\down}{(2\pi)^9}
 \big|t_\chi(\k;\p)A(\ek-\ep,\k-\p) \notag\\
 &\quad + t_\up(\k;\q_\up)A_\down(\ek-\eps_{\q_\up},\k-\q_\up) \notag\\
 &\quad + t_\down(\k;\q_\down)A_\up(\ek-\eps_{\q_\down},\k-\q_\down)\big|^2 \notag\\
 & \times (2\pi)^4\delta(\p+\q_\up+\q_\down-\k)
 \delta(\ep+\eps_{\q_\up}+\eps_{\q_\down}-\ek). \notag
\end{align}
Here, $\p$ and $\q_\sigma$ are momenta of the $\chi$ atom and the
spin-$\sigma$ fermion in the final state, respectively.  This equality
can be checked by a direct calculation starting with the set of integral
equations (\ref{eq:coupled_chi-F}) (see Appendix~\ref{app:optical}).

\begin{figure*}[t]\hfill
 \includegraphics[width=0.98\columnwidth,clip]{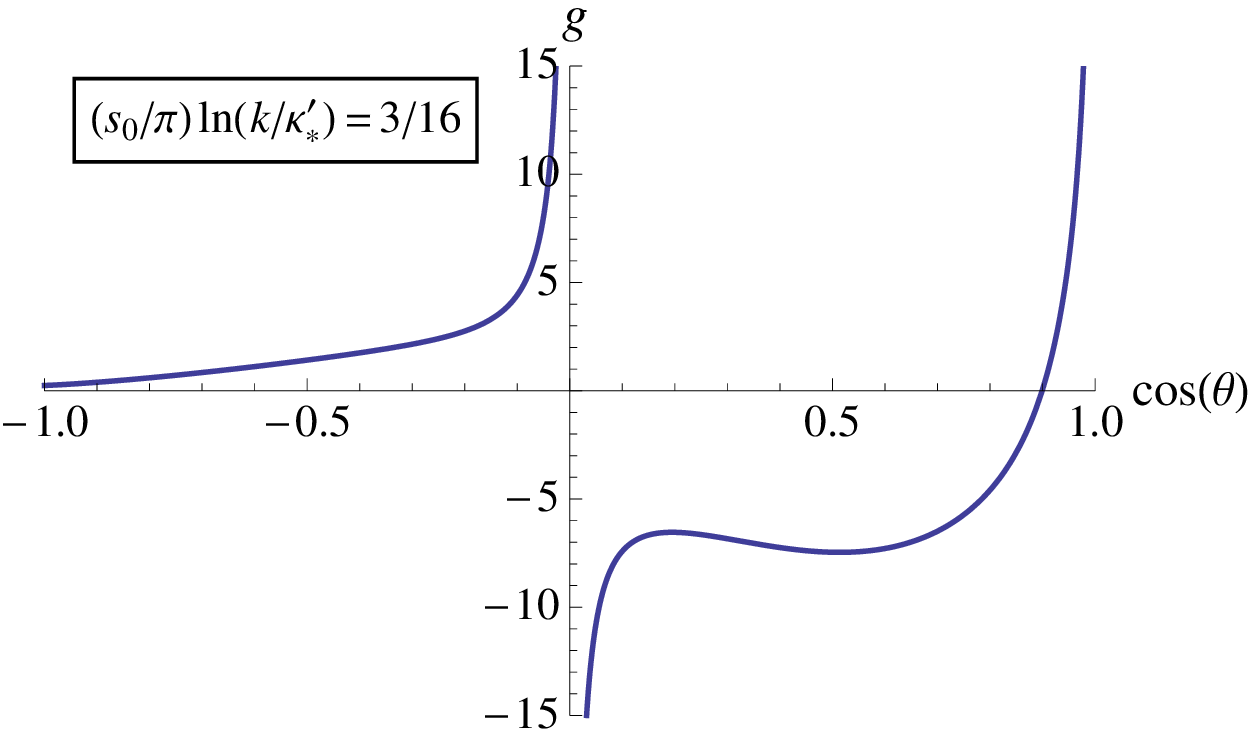}\hfill\hfill
 \includegraphics[width=0.98\columnwidth,clip]{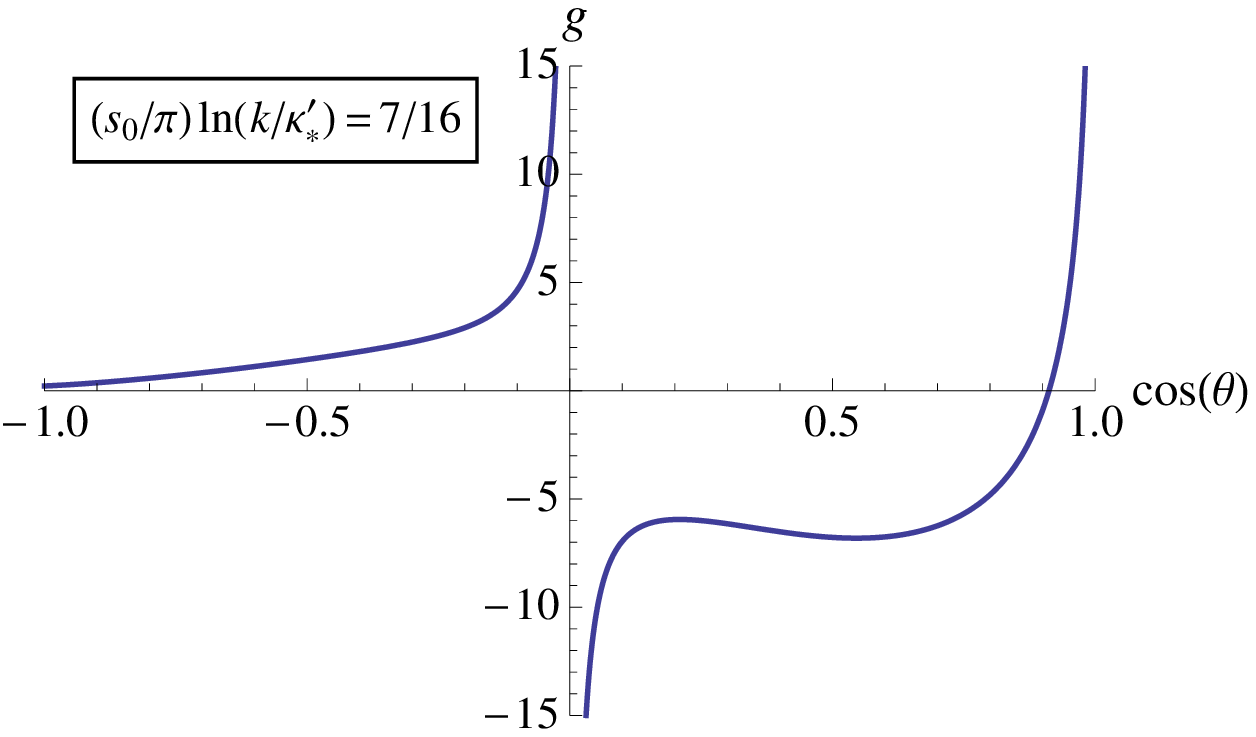}\hfill
 \vspace{2mm}\\\hfill
 \includegraphics[width=0.98\columnwidth,clip]{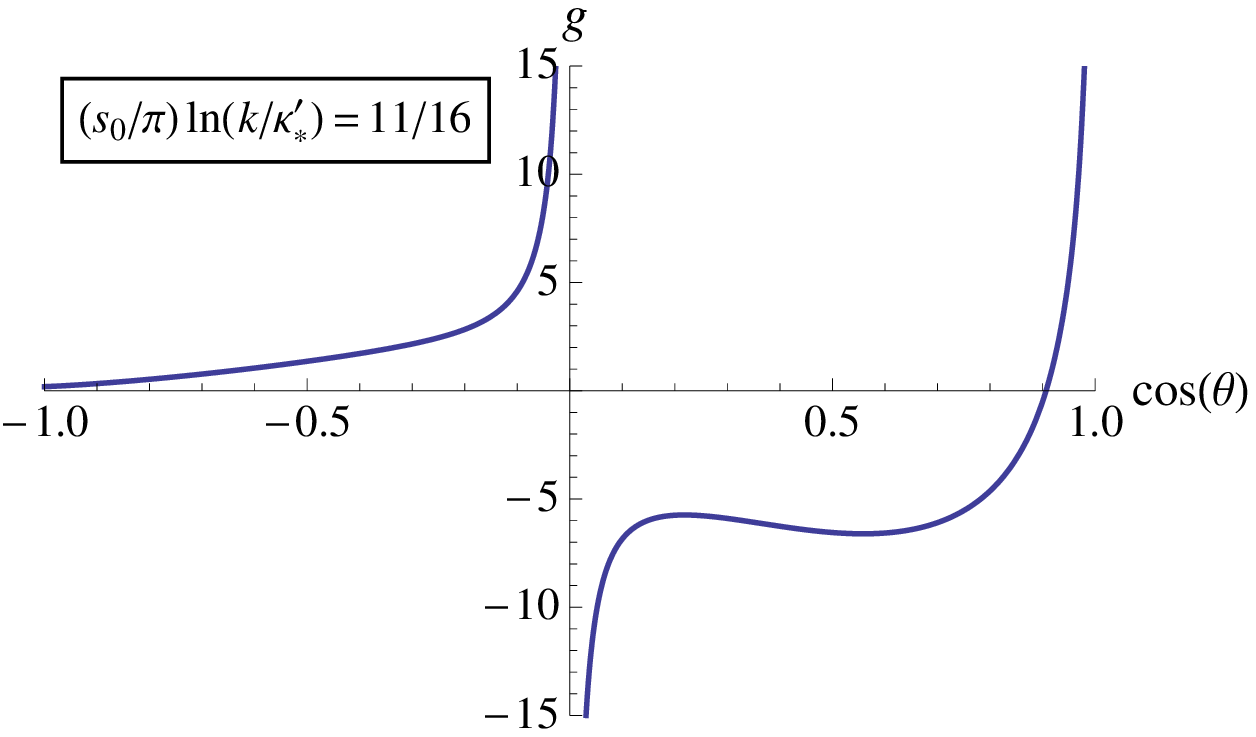}\hfill\hfill
 \includegraphics[width=0.98\columnwidth,clip]{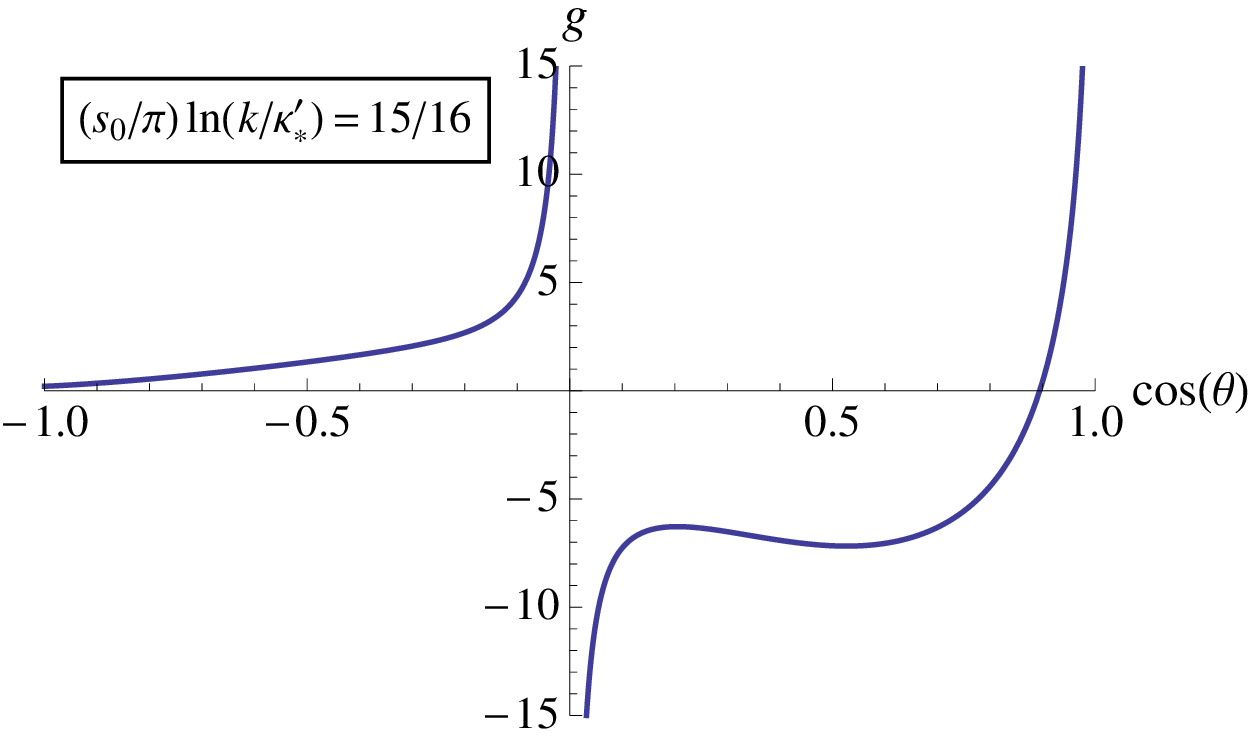}\hfill\hfill
 \caption{(Color online) Universal function
 $g(\cos\theta,|\k|/\kappa'_*)$ to determine the contribution of the
 contact density to the differential scattering rate
 (\ref{eq:angle_C_chi-F}).  $g(\cos\theta,|\k|/\kappa'_*)$ is periodic
 in terms of $0\leq(s_0/\pi)\ln(|\k|/\kappa'_*)\leq1$ and here its
 dependence on $\cos\theta$ is shown at four values of
 $(s_0/\pi)\ln(|\k|/\kappa'_*)=3/16$, $7/16$, $11/16$, $15/16$.  These
 four values roughly correspond to the minimum, inflection point,
 maximum, and inflection point of the total scattering rate,
 respectively [see Eq.~(\ref{eq:total_C_chi-F}) and
 Fig.~\ref{fig:efimov}].  \label{fig:differential}}
\end{figure*}

The right-hand side of Eq.~(\ref{eq:optical_t}) is infrared
divergent at $|\q_\sigma|\to0$ because of
$t_\sigma(\k;\q_\sigma)\to m/\q_\sigma^2$ for both $\sigma=\,\up,\down$
[see Eq.~(\ref{eq:coupled_chi-F})].  These infrared divergences are
canceled exactly by the second term in
Eq.~(\ref{eq:T_regularized_chi-F}) because its imaginary part also
contains the same form of infrared divergences which can be seen from
\begin{equation}\label{eq:optical_A}
 \begin{split}
  & 2\,\Im A_\up(\ek,\k)\int\!\frac{d\q}{(2\pi)^3}
  \biggl(\frac{m}{\q^2}\biggr)^2 \\
  &= \int\!\frac{d\p\,d\q_\up d\q_\down}{(2\pi)^9}
  \biggl(\frac{m}{\q_\down^2}\biggr)^2|A_\up(\ek,\k)|^2 \\
  & \times (2\pi)^4\delta(\p+\q_\up-\k)\delta(\ep+\eps_{\q_\up}-\ek),
 \end{split}
\end{equation}
and the same for $\up\,\leftrightarrow\,\down$.  Therefore, the
imaginary part of the regularized on-shell three-body scattering
amplitude
\begin{equation}
 \begin{split}
  &\,2\,\Im\,t_\chi^\reg(\k;\k) = 2\,\Im\,t_\chi(\k;\k) \\
  & - 2\,\Im[A_\up(\ek,\k)+A_\down(\ek,\k)]
  \int\!\frac{d\q}{(2\pi)^3}\left(\frac{m}{\q^2}\right)^2
 \end{split}
\end{equation}
appearing in $\Gamma_\chi^{(\C)}(\k)$ is free of divergences.  By
recalling that $\p$ in Eqs.~(\ref{eq:optical_t}) and
(\ref{eq:optical_A}) corresponds to the momentum of the $\chi$ atom in
the final state, the contribution of the contact density to the
differential scattering rate of the $\chi$ atom can be identified as
\newpage\begin{widetext}
\begin{equation}\label{eq:dG/dp^C_chi-F}
 \begin{split}
  \frac{d\Gamma_\chi^{(\C)}(\k)}{d\p}
  &= \int\!\frac{d\q_\up d\q_\down}{(2\pi)^9}\Biggl[
  \big|t_\chi(\k;\p)A(\ek-\ep,\k-\p)
  + t_\up(\k;\q_\up)A_\down(\ek-\eps_{\q_\up},\k-\q_\up) \\
  & \qquad + t_\down(\k;\q_\down)A_\up(\ek-\eps_{\q_\down},\k-\q_\down)\big|^2
  (2\pi)^4\delta(\p+\q_\up+\q_\down-\k)
  \delta(\ep+\eps_{\q_\up}+\eps_{\q_\down}-\ek) \\
  &\quad - \biggl(\frac{m}{\q_\up^2}\biggr)^2|A_\down(\ek,\k)|^2
  (2\pi)^4\delta(\p+\q_\down-\k)\delta(\ep+\eps_{\q_\down}-\ek) \\
  &\quad - \biggl(\frac{m}{\q_\down^2}\biggr)^2|A_\up(\ek,\k)|^2
  (2\pi)^4\delta(\p+\q_\up-\k)\delta(\ep+\eps_{\q_\up}-\ek)
  \Bigg]\frac{\C(x)}{m^2}.
 \end{split} 
\end{equation}
\end{widetext}
This differential scattering rate can be evaluated by using the on-shell
three-body scattering amplitudes $t_{\chi,\up,\down}(\k;\p)$ at infinite
scattering lengths $(a_\up|\k|)^{-1}$, $(a_\down|\k|)^{-1}$,
$(a|\k|)^{-1}=0$ obtained from the numerical solutions of
$t_{F,B}(\k;\p)$ in
Eqs.~(\ref{eq:t_F_chi-F})--(\ref{eq:decoupled_chi-F}) [note that
$t_\up(\k;\p)=t_\down(\k;\p)$ when $a_\up=a_\down$].  The physical
meaning of the subtracted terms was discussed at the end of
Sec.~\ref{sec:3-body_fermi}.

Finally, by performing the integration over the magnitude of $\p$, the
angle distribution of the scattered $\chi$ atom is given by
\begin{equation}\label{eq:angle_C_chi-F}
 \begin{split}
  \frac{d\Gamma_\chi^{(\C)}(\k)}{d\Omega}
  &= \int_0^\infty\!d|\p||\p|^2\frac{d\Gamma_\chi^{(\C)}(\k)}{d\p} \\
  &\equiv g\!\left(\cos\theta,\frac{|\k|}{\kappa'_*}\right)
  \frac{\C(x)}{m\k^2}.
 \end{split}
\end{equation}
The resulting universal function $g(\cos\theta,|\k|/\kappa'_*)$ depends
on the polar angle $\theta$ and the momentum to Efimov parameter ratio
$|\k|/\kappa'_*$ in a log-periodic way.  $g(\cos\theta,|\k|/\kappa'_*)$
as a function of $\cos\theta$ is shown in Fig.~\ref{fig:differential} at
four values of $(s_0/\pi)\ln(|\k|/\kappa'_*)=3/16$, $7/16$, $11/16$,
$15/16$.  Note that $g(\cos\theta,|\k|/\kappa'_*)$ is mostly negative on
the forward-scattering side ($\cos\theta>0$), while it is positive
everywhere on the backward-scattering side ($\cos\theta<0$).  The
divergences of $g(\cos\theta,|\k|/\kappa'_*)$ at $\cos\theta=0$ and $1$
signal poor convergences of the large-momentum expansion around these
angles.  Higher order corrections would be important as well and ideally
need to be resummed.  However, away from these two singularities, we
expect our large-momentum expansion to be valid in a wide range of
momentum based on the observation in Sec.~\ref{sec:comparison}.

The integration of the differential scattering rate
(\ref{eq:angle_C_chi-F}) over the solid angle
$d\Omega=d\cos\theta\,d\varphi$ reproduces the total scattering rate in
Eq.~(\ref{eq:rate_chi-F}):%
\footnote{Since $d\Gamma_\chi^{(\C)}(\k)/d\Omega$ has a pole at
$\cos\theta=0$ due to
$g(\cos\theta,|\k|/\kappa'_*)\sim-4/(\pi^2\cos\theta)$, the integral
over $\cos\theta$ is understood as the Cauchy principal value.}
\begin{equation}\label{eq:total_C_chi-F}
 \begin{split}
  \Gamma_\chi^{(\C)}(\k)
  &= 2\pi\int_{-1}^1\!d\cos\theta\,
  g\!\left(\cos\theta,\frac{|\k|}{\kappa'_*}\right)\frac{\C(x)}{m\k^2} \\
  &= \frac43\,\Im f\!\left(\frac{|\k|}{\kappa'_*}\right)\frac{\C(x)}{m\k^2},
 \end{split}
\end{equation}
which relates the integral of $g(\cos\theta,|\k|/\kappa'_*)$ to the
imaginary part of $f(|\k|/\kappa'_*)$.  Note that the above four values
of $(s_0/\pi)\ln(|\k|/\kappa'_*)=3/16$, $7/16$, $11/16$, $15/16$ are
chosen so that they roughly correspond to the minimum, inflection point,
maximum, and inflection point of $\Im f(|\k|/\kappa'_*)$, respectively
(see Fig.~\ref{fig:efimov}).  Therefore, although differences are
difficult to see in the differential scattering rate from
Fig.~\ref{fig:differential}, the contribution of the contact density to
the total scattering rate varies by up to $17\%$ for a different
momentum to Efimov parameter ratio $|\k|/\kappa'_*$.  By combining the
results in Eqs.~(\ref{eq:angle_n_chi-F}), (\ref{eq:angle_j_chi-F}), and
(\ref{eq:angle_C_chi-F}), we obtain the differential scattering rate of
the $\chi$ atom shot into a spin-1/2 Fermi gas up to $O(\k^{-3})$:
\begin{equation}
 \begin{split}
  \frac{d\Gamma_\chi(\k)}{d\Omega}
  &= \sum_{\sigma=\up,\down}\left[\frac{d\Gamma_\chi^{(n_\sigma)}(\k)}{d\Omega}
  + \frac{d\Gamma_\chi^{(\j_\sigma)}(\k)}{d\Omega}\right] \\
  & + \frac{d\Gamma_\chi^{(\C)}(\k)}{d\Omega} + O(\k^{-3}).
 \end{split}
\end{equation}
This result was presented previously in Eq.~(\ref{eq:diff_fermi}), where
the time dependence of the densities is suppressed by assuming a
stationary state.

\subsection{Spinless Bose gas \label{sec:diff_bose}}
The analysis in the case of a spinless Bose gas is similar.  Again a
probe atom is denoted by $\chi$ and assumed to interact with a boson by
a scattering length $a_\chi$.  The Lagrangian density describing such a
problem is
\begin{equation}
 \L = \chi^\+\left(i\d_t+\frac{\grad^2}{2m_\chi}\right)\chi
  + c_\chi\,\chi^\+\psi^\+\psi\chi + \L_B,
\end{equation}
where $\L_B$ defined in Eq.~(\ref{eq:L_bose}) describes spinless bosons
interacting with each other by a scattering length $a$.  For simplicity,
we shall assume that the $\chi$ atom and bosons have the same mass
$m=m_\chi$.  In analogy with Eq.~(\ref{eq:dimer_fermi}), the two-body
scattering amplitude between the $\chi$ atom and a boson is given by
\begin{equation}
 A_\chi(k) = \frac{4\pi}m
  \frac1{\sqrt{\frac{\k^2}4-mk_0-i0^+}-\frac1{a_\chi}}.
\end{equation}

The behavior of the single-particle Green's function of the $\chi$ atom,
\begin{equation}
 i\G_\chi(x;k) \equiv \int\!dy\,e^{iky}
  \<T[\chi(x+\tfrac{y}2)\chi^\+(x-\tfrac{y}2)]\>,
\end{equation}
in the large--energy-momentum limit $k\to\infty$ can be understood by
using the operator product expansion as in Secs.~\ref{sec:fermi_gas} and
\ref{sec:bose_gas}.  Since the $\chi$ atom and two identical bosons with
zero-range interactions suffer from the Efimov
effect~\cite{Nielsen:2001,Braaten:2004rn}, we only consider local
operators with $N_\O=0$ up to scaling dimensions $\Delta_\O=4$ (see
discussions in Sec.~\ref{sec:ope_bose}).  Then in analogy with
Eqs.~(\ref{eq:G(k)_bose}) and (\ref{eq:Sigma_bose}), $\G_\chi(x;k)$ can
be written in the usual form
\begin{equation}
 \G_\chi(x;k) = \frac1{k_0-\ek-\Sigma_\chi(x;k)+i0^+},
\end{equation}
where $\Sigma_\chi(x;k)$ is the self-energy of the $\chi$ atom given by
\begin{equation}\label{eq:Sigma_chi-B}
 \begin{split}
  \Sigma_\chi(x;k) &= -A_\chi(k)\,n(x) + \frac\d{\d\k}A_\chi(k)\cdot\j(x) \\
  & - T_\chi^\reg(k,0;k,0)\frac{\C(x)}{m^2} - \cdots.
 \end{split}
\end{equation}
Here, $n(x)=\<\psi^\+\psi\>$ and $\j(x)=\<-i\psi^\+\tensor\grad\psi\>$
are the number density and current density of bosons and
$\C(x)=m^2\<\phi^\+\phi\>$ is the contact density of a spinless Bose
gas.  Note that these parameters only refer to many-body properties of
the given spinless Bose gas and do not involve the information related
to the $\chi$ atom.  On the other hand, $T_\chi^\reg(k,0;k,0)$ is a
finite quantity defined by
\begin{equation}
 T_\chi^\reg(k,0;k,0) \equiv T_\chi(k,0;k,0) - A_\chi(k)
 \int\!\frac{d\q}{(2\pi)^3}\left(\frac{m}{\q^2}\right)^2,
\end{equation}
where $T_\chi(k,p;k',p')$ is the three-body scattering amplitude between
the $\chi$ atom and a dimer composed of two identical bosons with
$(k,p)$ [$(k',p')$] being their initial (final) energy-momentum (see
Fig.~\ref{fig:3-body}).  Corrections to the above expression of
$\Sigma_\chi(x;k)$ denoted by ``$\cdots$'' start with
$\sim\<\O\>/k^{\Delta_\O-2}$, where $\O$ are all possible operators with
the scaling dimension $\Delta_\O=5$.

Then by setting $k_0=\ek$ in Eq.~(\ref{eq:Sigma_chi-B}), the on-shell
self-energy of the $\chi$ atom for an arbitrary state is found to be
\begin{align}\label{eq:on-shell_chi-B}
 \Sigma_\chi(x;\ek,\k)
 &= \frac{4\pi}{\frac{i}2+\frac1{a_\chi|\k|}}\frac{n(x)}{m|\k|}
 + \frac{2\pi i}{\bigl(\frac{i}2+\frac1{a_\chi|\k|}\bigr)^2}
 \frac{\hat\k\cdot\j(x)}{m|\k|^2} \notag\\
 & - t_\chi^\reg(\k;\k)\frac{\C(x)}{m^2} + O(\k^{-3}).
\end{align}
The quasiparticle energy and scattering rate of the $\chi$ atom in a
spinless Bose gas are given by the real and imaginary parts of
$\Sigma_\chi(x;\ek,\k)$ according to Eqs.~(\ref{eq:real_part_chi-F}) and
(\ref{eq:imaginary_part_chi-F}), respectively.  Our next task is to
determine the regularized on-shell three-body scattering amplitude
$t_\chi^\reg(\k;\p)\equiv T_\chi^\reg(k,0;p,k{-}p)|_{k_0=\ek,p_0=\ep}$
in Eq.~(\ref{eq:on-shell_chi-B}), which requires solving a three-body
problem.

\subsection{Three-body problem}
We now compute the three-body scattering amplitude $T_\chi(k,0;k,0)$.
Unlike the previous cases in Secs.~\ref{sec:3-body_fermi} and
\ref{sec:3-body_bose}, $T_\chi(k;p)\equiv T_\chi(k,0;p,k{-}p)$ by itself
does not solve a closed integral equation.  To find a closed set of
integral equations, we need to introduce another three-body scattering
amplitude $T_\psi(k;p)$, which describes a process where the $\chi$ atom
and a dimer composed of two identical bosons with their energy-momentum
$(k,0)$ are scattered into a boson and a dimer composed of the
$\chi$ atom and the other boson with their energy-momentum $(p,k{-}p)$.
These two scattering amplitudes are solutions to a closed set of
integral equations depicted in Fig.~\ref{fig:integral_eqs} with the
identification of $T_\up$, $T_\down$ with $T_\psi$.  Then by following
the same procedures as in
Eqs.~(\ref{eq:bare_integral_eq_fermi})--(\ref{eq:integral_eq_fermi}),
the integral equations solved by the on-shell three-body scattering
amplitudes $t_{\chi,\psi}(\k;\p)\equiv T_{\chi,\psi}(\ek,\k;\ep,\p)$ can
be written as
\begin{subequations}\label{eq:coupled_chi-B}
 \begin{align}
  t_\chi(\k;\p) &= \int\!\frac{d\q}{(2\pi)^3}\,\K_{a_\chi}(\k;\p,\q)\,t_\psi(\k;\q)
  \intertext{and}
  \begin{split}
   t_\psi(\k;\p) &= \frac{m}{\p^2}
   + 2\int\!\frac{d\q}{(2\pi)^3}\,\K_a(\k;\p,\q)\,t_\chi(\k;\q) \\
   & + \int\!\frac{d\q}{(2\pi)^3}\,\K_{a_\chi}(\k;\p,\q)\,t_\psi(\k;\q),
  \end{split}
 \end{align}
\end{subequations}
where the integral kernel $\K_a(\k;\p,\q)$ is defined in
Eq.~(\ref{eq:kernel_fermi}).  Note that the factor $2$ in front of
$\K_a$ originates from the fact that the two-body scattering amplitude
between two identical bosons is twice as large as that between two
distinguishable particles [compare Eqs.~(\ref{eq:dimer_fermi}) and
(\ref{eq:dimer_bose})].

As long as we are interested in $t_\chi^\reg(\k;\k)\sim m/\k^2$ up to
$O(\k^{-3})$ [see Eq.~(\ref{eq:on-shell_chi-B})], we can set the two
scattering lengths infinite $(a_\chi|\k|)^{-1}$, $(a|\k|)^{-1}=0$,
because the dependence on them appears only from $O(\k^{-3})$.  In this
case, by defining
\begin{equation}\label{eq:t_F_chi-B}
 t_F(\k;\p) \equiv 2\mspace{1mu}t_\chi(\k;\p)-t_\psi(\k;\p)
\end{equation}
and
\begin{equation}\label{eq:t_B_chi-B}
 t_B(\k;\p) \equiv t_\chi(\k;\p)+t_\psi(\k;\p),
\end{equation}
the two coupled integral equations (\ref{eq:coupled_chi-B}) can be
brought into two independent integral equations:
\begin{subequations}\label{eq:decoupled_chi-B}
 \begin{equation}
  t_F(\k;\p) = -\frac{m}{\p^2}
   - \int\!\frac{d\q}{(2\pi)^3}\,\K_\infty(\k;\p,\q)\,t_F(\k;\q)
 \end{equation}
 and
 \begin{equation}
  t_B(\k;\p) = \frac{m}{\p^2}
   + 2\int\!\frac{d\q}{(2\pi)^3}\,\K_\infty(\k;\p,\q)\,t_B(\k;\q).
 \end{equation}
\end{subequations}
Because these two integral equations are equivalent to
Eqs.~(\ref{eq:integral_eq_fermi}) and (\ref{eq:integral_eq_bose}) at
$(a|\k|)^{-1}=0$, their solutions are already obtained.

By using the numerical solutions obtained previously in
Eqs.~(\ref{eq:solution_fermi}) and (\ref{eq:solution_bose}),
$t_\chi^\reg(\k;\k)$ within the accuracy of $O(\k^{-3})$ is found to be
\begin{equation}\label{eq:solution_chi-B}
 \begin{split}
  t_\chi^\reg(\k;\k)
  &= \frac13\left[t_F^\reg(\k;\k) + t_B^\reg(\k;\k)\right] + O(\k^{-3}) \\
  &= \frac13\left[3.771 + f\!\left(\frac{|\k|}{\kappa'_*}\right)\right]
  \frac{m}{\k^2} + O(\k^{-3}).
 \end{split}
\end{equation}
Here, $f(|\k|/\kappa'_*)$ is the universal log-periodic function plotted
in Fig.~\ref{fig:efimov} and approximately given by
Eq.~(\ref{eq:approx_bose}) with $\kappa'_*$ being the Efimov parameter
associated with a three-body system of the $\chi$ atom with two
identical bosons.  By substituting this numerical solution of the
three-body problem into the on-shell self-energy in
Eq.~(\ref{eq:on-shell_chi-B}), the large-momentum expansions of the
quasiparticle energy and scattering rate of the $\chi$ atom in a
spinless Bose gas are obtained from Eqs.~(\ref{eq:real_part_chi-F}) and
(\ref{eq:imaginary_part_chi-F}):
\begin{equation}
 \begin{split}
  E_\chi(x;\k) &= \Biggl[1 + 32\pi\frac{n(x)}{a_\chi|\k|^4} \\
  & - \frac23\left\{3.771 + \Re f\!\left(\frac{|\k|}{\kappa'_*}\right)\right\}
  \frac{\C(x)}{|\k|^4} + O(\k^{-5})\Biggr]\ek
 \end{split}
\end{equation}
and
\begin{equation}
 \begin{split}
  \Gamma_\chi(x;\k) &= \Biggl[32\pi\left\{\frac{n(x)}{|\k|^3}
  + \frac{\hat\k\cdot\j(x)}{|\k|^4}\right\} \\
  &\quad + \frac43\,\Im f\!\left(\frac{|\k|}{\kappa'_*}\right)
  \frac{\C(x)}{|\k|^4} + O(\k^{-5})\Biggr]\ek,
 \end{split}
\end{equation}
respectively.

Furthermore, by comparing
Eqs.~(\ref{eq:t_F_chi-B})--(\ref{eq:decoupled_chi-B}) with
Eqs.~(\ref{eq:t_F_chi-F})--(\ref{eq:decoupled_chi-F}), we find
$t_\psi(\k;\p)=t_{\up,\down}(\k;\p)$ and $t_\chi(\k;\p)$ in a spinless
Bose gas is a half of that in a spin-1/2 Fermi gas.  Therefore, the
contribution of the contact density to the differential scattering rate
of the $\chi$ atom in a spinless Bose gas is also a half of that in a
spin-1/2 Fermi gas (see also Appendix~\ref{app:optical}).  All
discussions given in Sec.~\ref{sec:diff_fermi} for a spin-1/2 Fermi gas
apply equally to a spinless Bose gas with minor modifications.  In
particular, the differential scattering rate of the $\chi$ atom shot
into a spinless Bose gas is given by
\begin{equation}\label{eq:diff_bose}
 \begin{split}
  \frac{d\Gamma_\chi(\k)}{d\Omega}
  &= \left[32\cos\theta\,\Theta(\cos\theta)\,\frac{n(x)}{|\k|^3}\right. \\[2pt]
  &\quad + 32\,\Bigl\{2\cos\theta\,\Theta(\cos\theta)\,\hat\k
  - \delta(\cos\theta)\,\hat\k \\
  &\qquad + \Theta(\cos\theta)\,\hat\p\Big\}\cdot\frac{\j(x)}{|\k|^4} \\[2pt]
  &\quad + \left.g\!\left(\cos\theta,\frac{|\k|}{\kappa'_*}\right)
  \frac{\C(x)}{|\k|^4} + O(\k^{-5})\right]\frac{\k^2}{2m}.
 \end{split}
\end{equation}
Compare this result with that for a spin-1/2 Fermi gas in
Eq.~(\ref{eq:diff_fermi}).

\section{Weak-probe limit \label{sec:weak-probe}}

\subsection{Spin-1/2 Fermi gas}
In the previous section, the differential scattering rate of a different
spin state of atoms shot into an atomic gas was derived in the systematic
large-momentum expansion.  For a spin-1/2 Fermi gas, it is given by
\begin{equation}\label{eq:dG/dp_chi-F}
 \begin{split}
  \frac{d\Gamma_\chi(\k)}{d\p}
  &= \sum_{\sigma=\up,\down}\left[\frac{d\Gamma_\chi^{(n_\sigma)}(\k)}{d\p}
  + \frac{d\Gamma_\chi^{(\j_\sigma)}(\k)}{d\p}\right] \\
  & + \frac{d\Gamma_\chi^{(\C)}(\k)}{d\p} + O(\k^{-6}),
 \end{split}
\end{equation}
where the three leading terms were obtained in
Eqs.~(\ref{eq:dG/dp^n_chi-F}), (\ref{eq:dG/dp^j_chi-F}), and
(\ref{eq:dG/dp^C_chi-F}), respectively.  On the other hand, there is
another limit in which a different systematic expansion is possible;
that is, the limit of $a_\sigma\to0$ where the probe atom interacts
weakly with atoms constituting the target atomic gas.  In this
``weak-probe'' limit, the self-energy of the $\chi$ atom in a spin-1/2
Fermi gas can be expanded perturbatively in terms of $a_\sigma$:
\begin{equation}
 \begin{split}
  & \Sigma_\chi(k) = -\sum_{\sigma=\up,\down}c_\sigma n_\sigma \\
  & + \sum_{\sigma,\sigma'}c_\sigma c_{\sigma'}\int\!\frac{dp}{(2\pi)^3}
  \frac{S_{\sigma\sigma'}(k-p)}{p_0-\ep+i0^+} + O(a_\sigma^3).
 \end{split}
\end{equation}
Here, $c_\sigma=-4\pi a_\sigma/m+O(a_\sigma^2)$ is the coupling strength
between the $\chi$ atom and a spin-$\sigma$ fermion and
\begin{equation}
 S_{\sigma\sigma'}(k) = \frac1{2\pi}\int\!dx\,e^{ikx}
  \<\hat n_\sigma(x)\hat n_{\sigma'}(0)\>
\end{equation}
is a dynamic structure factor of the spin-1/2 Fermi gas with
translational symmetries assumed again.  As before, the imaginary part
of the self-energy gives the scattering rate of the $\chi$ atom as
\begin{align}
 & \Gamma_\chi(\k) = -2\,\Im[\Sigma_\chi(\ek,\k)] \\
 &= \sum_{\sigma,\sigma'}c_\sigma c_{\sigma'}
 \int\!\frac{d\p}{(2\pi)^2}S_{\sigma\sigma'}(\ek-\ep,\k-\p)
 + O(a_\sigma^3). \notag
\end{align}
Because $\p$ corresponds to the momentum of the $\chi$ atom in the final
state, its differential scattering rate can be identified as
\begin{equation}\label{eq:dG/dp_weak}
 \frac{d\Gamma_\chi(\k)}{d\p}
  = \sum_{\sigma,\sigma'}\frac{c_\sigma c_{\sigma'}}{(2\pi)^2}
  S_{\sigma\sigma'}(\ek-\ep,\k-\p) + O(a_\sigma^3),
\end{equation}
which is well known in the context of inelastic neutron scattering in
condensed matter physics~\cite{vanHove:1954,Cohen:1957}.

\begin{figure}[t]
 \includegraphics[width=0.7\columnwidth,clip]{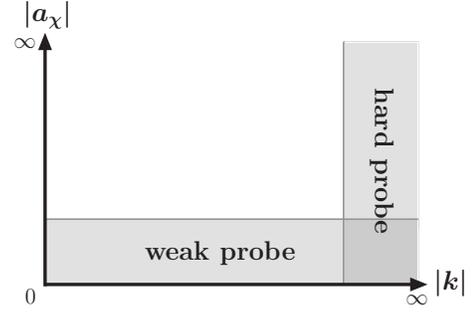}
 \caption{Schematic of valid regions of our hard-probe formula
 (\ref{eq:dG/dp_chi-F}) and the known weak-probe formula
 (\ref{eq:dG/dp_weak}) in the plane of the incident momentum $\k$ and
 the scattering length $a_\chi\sim a_\up\sim a_\down$ between the probe
 atom and an atom constituting the target atomic gas.
 \label{fig:hard_vs_weak}}
\end{figure}

This formula (\ref{eq:dG/dp_weak}) for the differential scattering rate
is valid in the weak-probe limit $a_\sigma\to0$ but for an arbitrary
incident momentum $\k$, while our formula (\ref{eq:dG/dp_chi-F}) derived
in Sec.~\ref{sec:differential} is valid in the hard-probe limit
$|\k|\to\infty$ but for an arbitrary scattering length $a_\sigma$.
Therefore, they cover different regions in the plane of $a_\sigma$ and
$|\k|$, while both the results become valid in the double limit of
$a_\sigma\to0$ and $|\k|\to\infty$ (see Fig.~\ref{fig:hard_vs_weak}).
In this section, we show that the weak-probe limit of
Eq.~(\ref{eq:dG/dp_chi-F}) is indeed equivalent to the hard-probe limit
of Eq.~(\ref{eq:dG/dp_weak}).  This serves as a nontrivial check of our
results presented in the previous section as well as a new derivation of
the large--energy-momentum expansion of the dynamic structure factor
$S_{\sigma\sigma'}$.

\subsection{Dynamic structure factor}
For simplicity, we shall consider a spin-1/2 Fermi gas at infinite
scattering length $a\to\infty$ only.  In the weak-probe limit
$a_\sigma\to0$, we can make the following expansions in each term for
the differential scattering rate in Eq.~(\ref{eq:dG/dp_chi-F}):
\begin{equation}
 A_\sigma(k) = c_\sigma + O(a_\sigma^2),
\end{equation}
\begin{equation}
 t_\sigma(\k;\p) = \frac{m}{\p^2} + O(a_\sigma),
\end{equation}
\begin{align}
 t_\chi(\k;\p) &= \int\!\frac{d\q}{(2\pi)^3}\frac{m}{\q^2}
 \frac{c_\up+c_\down}{\ep+\eq+\eps_{\k-\p-\q}-\ek-i0^+} \notag\\
 & + O(a_\sigma^2).
\end{align}
Accordingly, the contributions of the number density
(\ref{eq:dG/dp^n_chi-F}) and the current density
(\ref{eq:dG/dp^j_chi-F}) become
\begin{equation}\label{eq:dG/dp^n_weak}
 \frac{d\Gamma_\chi^{(n_\sigma)}(\k)}{d\p}
  = \frac{c_\sigma^2\,n_\sigma}{(2\pi)^2}\delta(\omega-\eK) + O(a_\sigma^3)
\end{equation}
and
\begin{equation}\label{eq:dG/dp^j_weak}
 \frac{d\Gamma_\chi^{(\j_\sigma)}(\k)}{d\p}
  = \frac{c_\sigma^2\,\j_\sigma}{(2\pi)^2}\cdot
  \frac{\d}{\d\bm{K}}\delta(\omega-\eK) + O(a_\sigma^3),
\end{equation}
respectively, where we denoted the energy-momentum transfer to the
medium by $(\omega,\bm{K})\equiv(\ek-\ep,\k-\p)$.  Similarly, after
lengthy but straightforward calculations, the contribution of the
contact density (\ref{eq:dG/dp^C_chi-F}) becomes
\begin{equation}\label{eq:dG/dp^C_weak}
 \begin{split}
  \frac{d\Gamma_\chi^{(\C)}(\k)}{d\p}
  &= \Biggl[\frac{c_\up^2+c_\down^2}{(2\pi)^2}\S^{\rm(SE)}(\omega,\bm{K})
  + \frac{(c_\up+c_\down)^2}{(2\pi)^2}\S^{\rm(AL)}(\omega,\bm{K}) \\
  &\quad + \frac{c_\up c_\down+c_\down c_\up}{(2\pi)^2}
  \S^{\rm(MT)}(\omega,\bm{K}) + O(a_\sigma^3)\Biggr]m\C,
 \end{split}
\end{equation}
where we defined%
\footnote{The second term in $\S^{\rm(SE)}(\omega,\bm{K})$ originates
from the subtracted terms in Eq.~(\ref{eq:dG/dp^C_chi-F}).  While it
does not contribute away from the single-particle peak at
$\omega=\eK$~\cite{Son:2010kq}, this term is essential to compute the
differential scattering rate in Eq.~(\ref{eq:angle_weak}) below.  This
is because its nonintegrable singularity at $q=|\bm{K}|/2$ cancels that
of the first term in $\S^{\rm(SE)}(\omega,\bm{K})$ which becomes
apparent by rewriting Eq.~(\ref{eq:S_SE}) as
\begin{align*}
 \S^{\rm(SE)}(\omega,\bm{K}) &= \frac1{(2\pi)^2}
 \int_0^\infty\!dq\frac{2q^2}{m\bigl(q^2-\frac{\bm{K}^2}4\bigr)^2} \\
 & \times \left[\delta(\omega-\tfrac\eK2-2\eq)-\delta(\omega-\eK)\right].
\end{align*}}
\begin{equation}\label{eq:S_SE}
 \begin{split}
  \S^{\rm(SE)}(\omega,\bm{K})
  &\equiv \frac{\Theta\bigl(\omega-\frac\eK2\bigr)}{(2\pi)^2}
  \left[\frac{\sqrt{m\omega-\frac{\bm{K}^2}4}}
  {\bigl(m\omega-\frac{\bm{K}^2}2\bigr)^2}\right. \\
  &\quad - \left.\int_0^\infty\!dq\frac{2q^2}{m\bigl(q^2-\frac{\bm{K}^2}4\bigr)^2}
  \delta(\omega-\eK)\right],
 \end{split}
\end{equation}
\begin{equation}
 \begin{split}
  \S^{\rm(AL)}(\omega,\bm{K})
  &\equiv \frac{\Theta\bigl(\omega-\frac\eK2\bigr)}
  {(2\pi)^2|\bm{K}|^2\sqrt{m\omega-\frac{\bm{K}^2}4}}
  \Biggl[\pi^2\Theta(\eK-\omega) \\[2pt]
  &\quad - \left.\ln^2\!
  \left(\frac{m\omega+|\bm{K}|\sqrt{m\omega-\frac{\bm{K}^2}4}}
  {\left|m\omega-\frac{\bm{K}^2}2\right|}\right)\right],
 \end{split}
\end{equation}
\begin{equation}
 \begin{split}
  \S^{\rm(MT)}(\omega,\bm{K})
  &\equiv \frac{\Theta\bigl(\omega-\frac\eK2\bigr)}{(2\pi)^2m\omega|\bm{K}|}
  \ln\!\left(\frac{m\omega+|\bm{K}|\sqrt{m\omega-\frac{\bm{K}^2}4}}
  {\left|m\omega-\frac{\bm{K}^2}2\right|}\right).
 \end{split}
\end{equation}
These three functions correspond to self-energy--, Aslamazov-Larkin--,
and Maki-Thompson--type contributions, respectively, in a direct
diagrammatic calculation~\cite{Son:2010kq,Hu:2012}.

Then by comparing Eqs.~(\ref{eq:dG/dp^n_weak}), (\ref{eq:dG/dp^j_weak}),
and (\ref{eq:dG/dp^C_weak}) with Eq.~(\ref{eq:dG/dp_weak}), we find that
each spin component of the dynamic structure factor should have the
following systematic expansion in the hard-probe limit $K\to\infty$:
\begin{equation}\label{eq:S_upup}
 \begin{split}
  & S_{\sigma\sigma}(\omega,\bm{K}) = n_\sigma\delta(\omega-\eK)
  + \j_\sigma\cdot\frac{\d}{\d\bm{K}}\delta(\omega-\eK) \\[4pt]
  & + \bigl[\S^{\rm(SE)}(\omega,\bm{K})+\S^{\rm(AL)}(\omega,\bm{K})\bigr]\,m\C
  + O(K^{-4})
 \end{split}
\end{equation}
for both $\sigma=\,\up,\down$ and
\begin{equation}\label{eq:S_updown}
 \begin{split}
  & S_{\up\down}(\omega,\bm{K}) = S_{\down\up}(\omega,\bm{K}) \\[4pt]
  &= \bigl[\S^{\rm(AL)}(\omega,\bm{K})+\S^{\rm(MT)}(\omega,\bm{K})\bigr]\,m\C
  + O(K^{-4}).
 \end{split}
\end{equation}
The sum of all four components
$S(\omega,\bm{K})\equiv\sum_{\sigma,\sigma'}S_{\sigma\sigma'}(\omega,\bm{K})$
coincides with the large--energy-momentum expansion of the total dynamic
structure factor found previously in
Refs.~\cite{Son:2010kq,Goldberger:2010fr,Hofmann:2011qs,Hu:2012,Taylor:2010}.
This establishes the equivalence between our hard-probe formula in
Eq.~(\ref{eq:dG/dp_chi-F}) and the known weak-probe formula in
Eq.~(\ref{eq:dG/dp_weak}) in the limit $a_\sigma\to0$ followed by
$|\k|\to\infty$ where both the results become valid.  However, we
emphasize again that, away from this double limit, the two results are
independent and cover different regions in the plane of $a_\sigma$ and
$|\k|$ (see Fig.~\ref{fig:hard_vs_weak}).

\subsection{Differential scattering rate}
\begin{figure}[t]
 \includegraphics[width=0.98\columnwidth,clip]{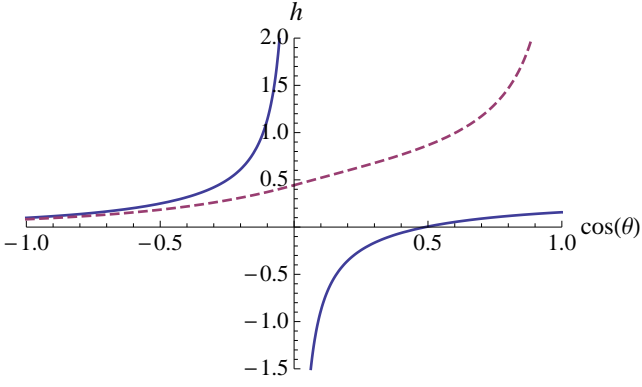}
 \caption{(Color online) Universal functions $h_\parallel(\cos\theta)$
 (solid curve) and $h_{\up\down}(\cos\theta)$ (dashed curve) to
 determine the contribution of the contact density to the differential
 scattering rate (\ref{eq:angle_weak}) in the weak-probe limit
 $a_\up,a_\down\to0$.  \label{fig:weak-probe}}
\end{figure}

Finally, we present the angle distribution of the large-momentum $\chi$
atom scattered by a spin-1/2 Fermi gas at infinite scattering length
$a\to\infty$ but in the weak-probe limit $a_\sigma\to0$.  The
integration of Eq.~(\ref{eq:dG/dp_weak}) with the dynamic structure
factor obtained in Eqs.~(\ref{eq:S_upup}) and (\ref{eq:S_updown}) over
the magnitude of momentum $|\p|^2d|\p|$ yields
\begin{equation}\label{eq:angle_weak}
 \begin{split}
  \frac{d\Gamma_\chi(\k)}{d\Omega}
  &= 4\sum_{\sigma=\up,\down}\Bigl[\cos\theta\,\Theta(\cos\theta)|\k|\,n_\sigma \\
  &\quad - \bigl\{\delta(\cos\theta)\,\hat\k-\Theta(\cos\theta)\,\hat\p\bigr\}
  \cdot\j_\sigma\Bigr]\,\frac{a_\sigma^2}{m} \\
  & + \left[h_\parallel(\cos\theta)\,\C + O(\k^{-1})\right]
  \frac{a_\up^2+a_\down^2}{2m} \\
  & + \left[h_{\up\down}(\cos\theta)\,\C + O(\k^{-1})\right]
  \frac{a_\up a_\down}{m} + O(a_\sigma^3).
 \end{split}
\end{equation}
Here, $h_\parallel(\cos\theta)$ and $h_{\up\down}(\cos\theta)$ are
universal functions plotted in Fig.~\ref{fig:weak-probe} and defined by
\begin{equation}
 \begin{split}
  h_\parallel(\cos\theta) &\equiv 8\int_0^\infty\!d|\p||\p|^2
  \bigl[\S^{\rm(SE)}(\ek-\ep,\k-\p) \\
  &\quad + \S^{\rm(AL)}(\ek-\ep,\k-\p)\bigr]
 \end{split}
\end{equation}
and
\begin{equation}
 \begin{split}
  h_{\up\down}(\cos\theta) &\equiv 8\int_0^\infty\!d|\p||\p|^2
  \bigl[\S^{\rm(AL)}(\ek-\ep,\k-\p) \\
  &\quad + \S^{\rm(MT)}(\ek-\ep,\k-\p)\bigr].
 \end{split}
\end{equation}
These two functions correspond to the weak-coupling limit of the
function $g$ introduced in Eq.~(\ref{eq:angle_C_chi-F}):
\begin{equation}
 \begin{split}
  & \lim_{a_\up,a_\down\to0}\left.
  g\!\left(\cos\theta,\frac{|\k|}{\kappa'_*}\right)\right|_{a_\sigma|\k|} \\
  &= h_\parallel(\cos\theta)\,\frac{a_\up^2+a_\down^2}2\,\k^2
  + h_{\up\down}(\cos\theta)\,a_\up a_\down\,\k^2,
 \end{split}
\end{equation}
while $g$ was evaluated in the strong-coupling limit
$a_\up,a_\down\to\infty$ in the previous section.  Note that the Efimov
effect does not appear in perturbative expansions. Again, one can see
from Eq.~(\ref{eq:angle_weak}) that the number density and current
density contribute to the forward scattering only, while
Fig.~\ref{fig:weak-probe} shows
$h_\parallel(\cos\theta)>h_{\up\down}(\cos\theta)>0$ at $\cos\theta<0$
so that the contact density gives the leading contribution to the
backward scattering.

The integration of the differential scattering rate
(\ref{eq:angle_weak}) over the solid angle
$d\Omega=d\cos\theta\,d\varphi$ yields the total scattering rate in the
weak-probe limit $a_\sigma\to0$:
\begin{equation}\label{eq:total_weak}
 \begin{split}
  \Gamma_\chi(\k)
  &= 4\pi\sum_{\sigma=\up,\down}\left[|\k|\,n_\sigma
  - \hat\k\cdot\j_\sigma\right]\frac{a_\sigma^2}{m} \\
  & + \left[1.60098\,\C + O(\k^{-1})\right]
  \frac{a_\up^2+a_\down^2}{2m} \\[4pt]
  & + \left[9.25387\,\C + O(\k^{-1})\right]
  \frac{a_\up a_\down}{m} + O(a_\sigma^3).
 \end{split}
\end{equation}
Note that, since $d\Gamma_\chi(\k)/d\Omega$ has a pole at $\cos\theta=0$
due to $h_\parallel(\cos\theta)\sim-1/(\pi^2\cos\theta)$, the integral
over $\cos\theta$ is understood as the Cauchy principal value.

\subsection{Spinless Bose gas}
Similarly, in the case of a spinless Bose gas at infinite scattering
length $a\to\infty$, the differential scattering rate of the $\chi$ atom
in the weak-probe limit $a_\chi\to0$ is given by
\begin{equation}
 \frac{d\Gamma_\chi(\k)}{d\p}
  = \frac{c_\chi^2}{(2\pi)^2}S(\ek-\ep,\k-\p) + O(a_\chi^3).
\end{equation}
Here, $S(\omega,\bm{K})$ is a dynamic structure factor of the spinless
Bose gas and has the following systematic expansion in the hard-probe
limit $K\to\infty$:
\begin{align}
 & S(\omega,\bm{K}) = n\,\delta(\omega-\eK)
 + \j\cdot\frac{\d}{\d\bm{K}}\delta(\omega-\eK) \notag\\
 & + \bigl[\S^{\rm(SE)}(\omega,\bm{K})+2\,\S^{\rm(AL)}(\omega,\bm{K})
 +\S^{\rm(MT)}(\omega,\bm{K})\bigr]\,m\C \notag\\
 & + O(K^{-4}).
\end{align}
Therefore, the contribution of the contact density to the dynamic
structure factor in a spinless Bose gas is a half of that to the total
dynamic structure factor in a spin-1/2 Fermi
gas~\cite{Goldberger:2010fr}.  Accordingly, Eqs.~(\ref{eq:angle_weak})
and (\ref{eq:total_weak}) remain valid by setting
$a_\up=a_\down\to a_\chi$, replacing $n_\up+n_\down$ ($\j_\up+\j_\down$)
with $n$ ($\j$), and dividing the coefficients of $\C$ by two.

\section{Conclusions \label{sec:conclusion}}
In this paper, we investigated various properties of an energetic atom
propagating through strongly interacting atomic gases by using
systematic large-momentum expansions.  Our main results are summarized
in Sec.~\ref{sec:summary} and consist of the quasiparticle energy and
scattering rate of an energetic atom in a spin-1/2 Fermi gas
(Sec.~\ref{sec:fermi_gas}), those in a spinless Bose gas
(Sec.~\ref{sec:bose_gas}), and the differential scattering rate of a
different spin state of atoms shot into a spin-1/2 Fermi gas or a
spinless Bose gas (Sec.~\ref{sec:differential}).  Furthermore, a
connection of our hard-probe formula derived in
Sec.~\ref{sec:differential} with dynamic structure factors in the
weak-probe limit was elucidated in Sec.~\ref{sec:weak-probe}.

Our result on the quasiparticle energy in a spin-1/2 Fermi gas
reasonably agrees with the recent quantum Monte Carlo
simulation~\cite{Magierski:2011wp} even at a relatively small momentum
$|\k|/\kF\gtrsim1.5$.  This indicates that our large-momentum expansions
are valid in a wide range of momentum.  Further analysis of quantum
Monte Carlo data incorporating our exact large-momentum expansions may
allow us better access to the intriguing pseudogap physics.  Also our
result on the rate at which the atom is scattered in the medium may be
useful to better understand multiple-atom loss mechanisms due to
atom-dimer ``dijets'' produced by three-body recombination
events~\cite{Zaccanti:2009,Machtey:2012}.  We found that the contact
density has a negative contribution to the scattering rate in a spinless
Bose gas.  This rather counterintuitively means that the energetic boson
can escape from the medium easier than we naively estimate from a binary
collision.

We also proposed a scattering experiment in which we shoot a different
spin state of atoms into an atomic gas with a large momentum and measure
its differential scattering rate (Fig.~\ref{fig:scattering}).  Here a
nice interplay between few-body physics and many-body physics can be
seen: The angle distribution of the scattered atom is determined by
few-body physics and its overall magnitude is set by many-body physics.
We elucidated that, because the number density and current density of
the target atomic gas contribute to the forward scattering only, its
contact density gives the leading contribution to the backward
scattering.  Therefore, such an experiment can be used to measure the
contact density (integrated along a classical trajectory of the probe
atom) and thus provides a new {\em local\/} probe of strongly
interacting atomic gases.  Its intriguing analogy to nuclear physics
experiments on short-range pair correlations in
nuclei~\cite{Subedi:2008,Arrington:2012} should be explored further.
Also we found that the differential scattering rate can depend on the
azimuthal angle only by the current density of the target atomic gas.
Therefore, the azimuthal anisotropy in the differential scattering rate
may be useful to reveal many-body phases accompanied by currents.  We
hope this work serves as a promising starting point for future
ultracold-atom experiments and builds a new bridge between ultracold
atoms and nuclear and particle physics from the perspective of ``hard
probes.''

\acknowledgments
This work started when the author attended the MIT nuclear and particle
physics colloquium in the fall of 2008 given by Krishna Rajagopal to
whom he is grateful.  He also thanks E.~Braaten, A.~Bulgac, J.~Carlson,
J.~E.~Drut, S.~Gandolfi, T.~Hatsuda, D.~Kang, J.~Levinsen, P.~Pieri,
D.~T.~Son, F.~Werner, G.~Wlaz{\l}owski, W.~Zwerger, M.~W.~Zwierlein,
and, in particular, Shina Tan for valuable discussions and providing
numerical data in Ref.~\cite{Magierski:2011wp}.  This work was supported
by a MIT Pappalardo Fellowship in Physics and a LANL Oppenheimer
Fellowship.  Part of numerical calculations was carried out at the YITP
computer facility in Kyoto University.

\appendix

\section{Derivation of Wilson coefficients \label{app:wilson}}
Here we show how the Wilson coefficients in
Eqs.~(\ref{eq:W_identity_fermi})--(\ref{eq:W_phi2_fermi}) and
Eqs.~(\ref{eq:W_identity_bose})--(\ref{eq:W_phi0_bose}) are derived.
For generality, we consider spin-1/2 fermions with unequal masses
$m_\up\neq m_\down$.  The propagator of fermion field $\psi_\sigma$ in
the vacuum is given by
\begin{equation}
  G_\sigma(k) = \frac1{k_0-\eps_{\k\sigma}+i0^+}
  \qquad \left(\eps_{\k\sigma}\equiv\frac{\k^2}{2m_\sigma}\right),
\end{equation}
and the two-body scattering amplitude between spin-up and -down fermions is
\begin{equation}
 A(k) = \frac{2\pi}\mu\frac1{\sqrt{\frac\mu{M}\k^2-2\mu k_0-i0^+}-\frac1a}.
\end{equation}
Here, $M=m_\up+m_\down$ is a total mass and
$\mu=m_\up m_\down/(m_\up+m_\down)$ is a reduced mass.  Results in the
case of spinless bosons are obtained by removing spin indices and
replacing the two-body scattering amplitude $A(k)$ with that between two
identical bosons [see Eq.~(\ref{eq:dimer_bose})]:
\begin{equation}
 A(k) = \frac{8\pi}m\frac1{\sqrt{\frac{\k^2}4-mk_0-i0^+}-\frac1a}.
\end{equation}

\subsection{One-body and two-body sectors}
The Wilson coefficients of local operators of type
$\psi_\sigma^\+\psi_\sigma$ in
Eqs.~(\ref{eq:O_0_fermi})--(\ref{eq:O_5_fermi}) are determined by
matching the matrix elements of both sides of
\begin{equation}\label{eq:ope_app}
 \int\!dy\,e^{iky}\,T[\psi_\up(x+\tfrac{y}2)\psi_\up^\+(x-\tfrac{y}2)]
  = \sum_iW_{\O_i}(k)\O_i(x)
\end{equation}
with respect to one-body states $\<\psi_\sigma(p')|$ and
$|\psi_\sigma(p)\>$.  The matrix element of the left-hand side is given
by
\begin{equation}\label{eq:1-body_lhs}
 \begin{split}
  & \int\!dy\,e^{iky}\<\psi_\sigma(p')|
  T[\psi_\up(x+\tfrac{y}2)\psi_\up^\+(x-\tfrac{y}2)]|\psi_\sigma(p)\> \\
  &= iG_\up(k)iG_\sigma(p)(2\pi)^4\delta(p-p') \\
  &\quad + (1-\delta_{\up\sigma})\,
  iG_\up(k-\tfrac{p-p'}2)iG_\sigma(p)iA(k+\tfrac{p+p'}2) \\
  &\quad \times iG_\sigma(p')iG_\up(k+\tfrac{p-p'}2)e^{-i(p-p')x}.
 \end{split}
\end{equation}
On the other hand, the matrix elements of local operators of type
$\psi_\sigma^\+\psi_\sigma$ are
\begin{equation}
 \<\psi_\sigma(p')|\openone|\psi_\sigma(p)\> = iG_\sigma(p)(2\pi)^4\delta(p-p'),
\end{equation}
\begin{equation}
 \<\psi_\sigma(p')|\psi_\sigma^\+\psi_\sigma(x)|\psi_\sigma(p)\>
  = iG_\sigma(p)iG_\sigma(p')e^{-i(p-p')x},
\end{equation}
\begin{equation}
 \begin{split}
  & \<\psi_\sigma(p')|-i\psi_\sigma^\+\tensor\d_i\psi_\sigma(x)|\psi_\sigma(p)\> \\
  &= \frac{p_i+p'_i}2\,iG_\sigma(p)iG_\sigma(p')e^{-i(p-p')x},
 \end{split}
\end{equation}
\begin{equation}
 \begin{split}
  & \<\psi_\sigma(p')|-i\d_i(\psi_\sigma^\+\psi_\sigma)(x)|\psi_\sigma(p)\> \\
  &= (p_i-p'_i)\,iG_\sigma(p)iG_\sigma(p')e^{-i(p-p')x},
 \end{split}
\end{equation}
\begin{equation}
 \begin{split}
  & \<\psi_\sigma(p')|-\psi_\sigma^\+\tensor\d_i\tensor\d_j\psi_\sigma(x)|\psi_\sigma(p)\> \\
  &= \frac{p_i+p'_i}2\frac{p_j+p'_j}2\,iG_\sigma(p)iG_\sigma(p')e^{-i(p-p')x},
 \end{split}
\end{equation}
\begin{equation}
 \begin{split}
  & \<\psi_\sigma(p')|-\d_i(\psi_\sigma^\+\tensor\d_j\psi_\sigma)(x)|\psi_\sigma(p)\> \\
  &= (p_i-p'_i)\,\frac{p_j+p'_j}2\,iG_\sigma(p)iG_\sigma(p')e^{-i(p-p')x},
 \end{split}
\end{equation}
\begin{equation}
 \begin{split}
  & \<\psi_\sigma(p')|-\d_i\d_j(\psi_\sigma^\+\psi_\sigma)(x)|\psi_\sigma(p)\> \\
  &= (p_i-p'_i)\,(p_j-p'_j)\,iG_\sigma(p)iG_\sigma(p')e^{-i(p-p')x},
 \end{split}
\end{equation}
\begin{equation}
 \begin{split}
  & \<\psi_\sigma(p')|i\psi_\sigma^\+\tensor\d_t\psi_\sigma(x)|\psi_\sigma(p)\> \\
  &= \frac{p_0+p'_0}2\,iG_\sigma(p)iG_\sigma(p')e^{-i(p-p')x},
 \end{split}
\end{equation}
\begin{equation}
 \begin{split}
  & \<\psi_\sigma(p')|i\d_t(\psi_\sigma^\+\psi_\sigma)(x)|\psi_\sigma(p)\> \\
  &= (p_0-p'_0)\,iG_\sigma(p)iG_\sigma(p')e^{-i(p-p')x}.
 \end{split}
\end{equation}
Therefore, the expansion of Eq.~(\ref{eq:1-body_lhs}) up to $O(p^3)$ is
reproduced by choosing their Wilson coefficients as
\begin{equation}
 W_{\openone}(k) = iG_\up(k),
\end{equation}
\begin{equation}
 W_{\psi_\down^\+\psi_\down}(k) = -iG_\up(k)^2A(k),
\end{equation}
\begin{equation}
 W_{-i\psi_\down^\+\tensor\d_i\psi_\down}(k)
  = -iG_\up(k)^2\frac\d{\d k_i}A(k),
\end{equation}
\begin{equation}
 W_{-\psi_\down^\+\tensor\d_i\tensor\d_j\psi_\down}(k)
  = -iG(k)^2\frac12\frac{\d^2}{\d k_i\d k_j}A(k),
\end{equation}
\begin{equation}
 W_{i\psi_\down^\+\tensor\d_t\psi_\down}(k)
  = -iG_\up(k)^2\frac\d{\d k_0}A(k),
\end{equation}
\begin{equation}
 W_{-\d_i\d_j(\psi_\down^\+\psi_\down)}(k)
  = -iA(k)\frac{G_\up(k)^3}{4m_\up}\!
  \left[\delta_{ij} + k_ik_j\frac{G_\up(k)}{m_\up}\right],
\end{equation}
\begin{equation}
 W_{-i\d_i(\psi_\down^\+\psi_\down)}(k)
  = W_{-\d_i(\psi_\down^\+\tensor\d_j\psi_\down)}(k)
  = W_{i\d_t(\psi_\down^\+\psi_\down)}(k) = 0,
\end{equation}
and all $W_\O(k)=0$ for $\sigma=\,\up$.

\onecolumngrid
\subsection{Three-body sector}
The Wilson coefficients of local operators of type $\phi^\+\phi$ in
Eqs.~(\ref{eq:O_0_fermi})--(\ref{eq:O_5_fermi}) are determined by
matching the matrix elements of both sides of Eq.~(\ref{eq:ope_app})
with respect to two-body states $\<\phi(p')|$ and $|\phi(p)\>$.  The
matrix element of the left-hand side is given by
\begin{equation}\label{eq:2-body_lhs}
 \begin{split}
  & \int\!dy\,e^{iky}\<\phi(p')|
  T[\psi_\up(x+\tfrac{y}2)\psi_\up^\+(x-\tfrac{y}2)]|\phi(p)\> \\
  &= iG_\up(k)iD(p)(2\pi)^4\delta(p-p') \\
  & + iG_\up(k-\tfrac{p-p'}2)iD(p)
  iT_\up(k-\tfrac{p-p'}2,\tfrac{p+p'}2+\tfrac{p-p'}2;
  k+\tfrac{p-p'}2,\tfrac{p+p'}2-\tfrac{p-p'}2)
  iD(p')iG_\up(k+\tfrac{p-p'}2)e^{-i(p-p')x}.
 \end{split}
\end{equation}
Here, $D(k)=-A(k)$ is the dimer propagator and $T_\up(k,p;k',p')$ is the
three-body scattering amplitude between a spin-up fermion and a dimer
with $(k,p)$ [$(k',p')$] being their initial (final) energy-momentum
(see Fig.~\ref{fig:3-body}).  On the other hand, the matrix elements of
local operators of type $\psi_\down^\+\psi_\down$ are
\begin{equation}
 \<\phi(p')|\openone|\phi(p)\> = iD(p)(2\pi)^4\delta(p-p'),
\end{equation}
\begin{equation}
 \begin{split}
  & \<\phi(p')|\psi_\down^\+\psi_\down(x)|\phi(p)\> \\
  &= iD(p)iD(p')e^{-i(p-p')x}\,i\int\!\frac{dq}{(2\pi)^4}
  G_\up(\tfrac{p+p'}2-q)G_\down(q+\tfrac{p+p'}2)G_\down(q-\tfrac{p+p'}2) \\
  &= iD(p)iD(p')e^{-i(p-p')x}\int\!\frac{d\q}{(2\pi)^3}
  \left(\frac{2\mu}{\q^2}\right)^2 + O(p^2),
 \end{split}
\end{equation}
\begin{equation}
 \begin{split}
  & \<\phi(p')|-i\psi_\down^\+\tensor\d_i\psi_\down(x)|\phi(p)\> \\
  &= iD(p)iD(p')e^{-i(p-p')x}\,i\int\!\frac{dq}{(2\pi)^4}\,q_i\,
  G_\up(\tfrac{p+p'}2-q)G_\down(q+\tfrac{p+p'}2)G_\down(q-\tfrac{p+p'}2) \\
  &= iD(p)iD(p')e^{-i(p-p')x}\left(\frac{m_\down}M\frac{p_i+p'_i}2\right)
  \int\!\frac{d\q}{(2\pi)^3}\left(\frac{2\mu}{\q^2}\right)^2 + O(p^2),
 \end{split}
\end{equation}
\begin{equation}
 \begin{split}
  & \<\phi(p')|-i\d_i(\psi_\down^\+\psi_\down)(x)|\phi(p)\> \\
  &= iD(p)iD(p')e^{-i(p-p')x}\,i\int\!\frac{dq}{(2\pi)^4}(p_i-p'_i)
  G_\up(\tfrac{p+p'}2-q)G_\down(q+\tfrac{p+p'}2)G_\down(q-\tfrac{p+p'}2) \\
  &= iD(p)iD(p')e^{-i(p-p')x}(p_i-p'_i)
  \int\!\frac{d\q}{(2\pi)^3}\left(\frac{2\mu}{\q^2}\right)^2 + O(p^2),
 \end{split}
\end{equation}
\begin{equation}
 \begin{split}
  & \<\phi(p')|-\psi_\down^\+\tensor\d_i\tensor\d_j\psi_\down(x)|\phi(p)\> \\
  &= iD(p)iD(p')e^{-i(p-p')x}\,i\int\!\frac{dq}{(2\pi)^4}q_iq_j
  G_\up(\tfrac{p+p'}2-q)G_\down(q+\tfrac{p+p'}2)G_\down(q-\tfrac{p+p'}2) \\
  &= iD(p)iD(p')e^{-i(p-p')x}\,\frac{\delta_{ij}}3
  \int\!\frac{d\q}{(2\pi)^3}\,\q^2\left(\frac{2\mu}{\q^2}\right)^2 + O(p^2),
 \end{split}
\end{equation}
\begin{equation}
 \begin{split}
  & \<\phi(p')|-\d_i(\psi_\down^\+\tensor\d_j\psi_\down)(x)|\phi(p)\> \\
  &= iD(p)iD(p')e^{-i(p-p')x}\,i\int\!\frac{dq}{(2\pi)^4}(p_i-p'_i)q_j
  G_\up(\tfrac{p+p'}2-q)G_\down(q+\tfrac{p+p'}2)G_\down(q-\tfrac{p+p'}2) \\
  &= O(p^2),
 \end{split}
\end{equation}
\begin{equation}
 \begin{split}
  & \<\phi(p')|-\d_i\d_j(\psi_\down^\+\psi_\down)(x)|\phi(p)\> \\
  &= iD(p)iD(p')e^{-i(p-p')x}\,i\int\!\frac{dq}{(2\pi)^4}(p_i-p'_i)(p_j-p'_j)
  G_\up(\tfrac{p+p'}2-q)G_\down(q+\tfrac{p+p'}2)G_\down(q-\tfrac{p+p'}2) \\
  &= O(p^2),
 \end{split}
\end{equation}
\begin{equation}
 \begin{split}
  & \<\phi(p')|i\psi_\down^\+\tensor\d_t\psi_\down(x)|\phi(p)\> \\
  &= iD(p)iD(p')e^{-i(p-p')x}\,i\int\!\frac{dq}{(2\pi)^4}q_0
  G_\up(\tfrac{p+p'}2-q)G_\down(q+\tfrac{p+p'}2)G_\down(q-\tfrac{p+p'}2) \\
  &= iD(p)iD(p')e^{-i(p-p')x}\int\!\frac{d\q}{(2\pi)^3}
  \left(-\frac{\q^2}{2m_\up}\right)\left(\frac{2\mu}{\q^2}\right)^2 + O(p^2).
 \end{split}
\end{equation}
\begin{equation}
 \begin{split}
  & \<\phi(p')|i\d_t(\psi_\down^\+\psi_\down)(x)|\phi(p)\> \\
  &= iD(p)iD(p')e^{-i(p-p')x}\,i\int\!\frac{dq}{(2\pi)^4}(p_0-p'_0)
  G_\up(\tfrac{p+p'}2-q)G_\down(q+\tfrac{p+p'}2)G_\down(q-\tfrac{p+p'}2) \\
  &= O(p^2).
 \end{split}
\end{equation}
Because the matrix elements of local operators of type $\phi^\+\phi$ are
given by
\begin{equation}
 \<\phi(p')|\phi^\+\phi(x)|\phi(p)\> = iD(p)iD(p')e^{-i(p-p')x},
\end{equation}
\begin{equation}
 \<\phi(p')|-i\phi^\+\tensor\d_i\phi(x)|\phi(p)\>
  = \frac{p_i+p'_i}2\,iD(p)iD(p')e^{-i(p-p')x},
\end{equation}
\begin{equation}
 \<\phi(p')|-i\d_i(\phi^\+\phi)(x)|\phi(p)\>
  = (p_i-p'_i)\,iD(p)iD(p')e^{-i(p-p')x},
\end{equation}
the expansion of Eq.~(\ref{eq:2-body_lhs}) up to $O(p^2)$ is reproduced
by choosing their Wilson coefficients as
\begin{equation}
 \begin{split}
  W_{\phi^\+\phi}(k) &= -iG(k)^2T_\up(k,0;k,0)
  - W_{\psi_\down^\+\psi_\down}(k)
  \int\!\frac{d\q}{(2\pi)^3}\left(\frac{2\mu}{\q^2}\right)^2 \\
  & - W_{-\psi_\down^\+\tensor\d_i\tensor\d_j\psi_\down}(k)
  \,\frac{\delta_{ij}}3\int\!\frac{d\q}{(2\pi)^3}\left(\frac{2\mu}{\q}\right)^2
  - W_{i\psi_\down^\+\tensor\d_t\psi_\down}(k)
  \,\frac{-1}{2m_\up}\int\!\frac{d\q}{(2\pi)^3}\left(\frac{2\mu}{\q}\right)^2,
 \end{split}
\end{equation}
\begin{equation}
 W_{-i\phi^\+\tensor\d_i\phi}(k)
  = -iG_\up(k)^2\frac\d{\d p_i}T_\up(k,p;k,p)\big|_{p\to0}
  - W_{-i\psi_\down^\+\tensor\d_i\psi_\down}(k)\,\frac{m_\down}M
  \int\!\frac{d\q}{(2\pi)^3}\left(\frac{2\mu}{\q^2}\right)^2,
\end{equation}
\begin{equation}
 W_{-i\d_i(\phi^\+\phi)}(k)
  = -iG_\up(k)^2\frac\d{\d p_i}T_\up(k-\tfrac{p}2,\tfrac{p}2;
  k+\tfrac{p}2,-\tfrac{p}2)\big|_{p\to0}.
\end{equation}
These results are presented in
Eqs.~(\ref{eq:W_identity_fermi})--(\ref{eq:W_phi2_fermi}) for a spin-1/2
Fermi gas and in Eqs.~(\ref{eq:W_identity_bose})--(\ref{eq:W_phi0_bose})
for a spinless Bose gas.

\section{Details of solving integral equations \label{app:integral_eq}}
Here we discuss details of solving the integral equations in
Eq.~(\ref{eq:integral_eq_fermi}) for a spin-1/2 Fermi gas and in
Eq.~(\ref{eq:integral_eq_bose}) for a spinless Bose gas.

\subsection{Spin-1/2 Fermi gas}
For generality, we consider spin-1/2 fermions with unequal masses
$m_\up\neq m_\down$, in which the integral equation
(\ref{eq:integral_eq_fermi}) is modified to
\begin{equation}\label{eq:integral_eq_fermi_app}
 t_\up(\k;\p) = -\frac{2\mu}{\p^2}
  - \int\!\frac{d\q}{(2\pi)^3}\,\K_a(\k;\p,\q)\,t_\up(\k;\q),
\end{equation}
where the integral kernel $\K_a(\k;\p,\q)$ is given by
\begin{equation}
 \K_a(\k;\p,\q) = \frac{2\pi}\mu\frac1{\sqrt{\frac\mu{M}(\k-\q)^2
  -2\mu\left(\eps_{\k\up}-\eps_{\q\up}\right)-i0^+}-\frac1a}
  \frac1{\eps_{\k-\p-\q\down}+\eps_{\p\up}+\eps_{\q\up}-\eps_{\k\up}-i0^+}.
\end{equation}
For unequal masses, the quantity we would like to compute
(\ref{eq:t_regularized_fermi}) becomes
\begin{equation}\label{eq:t_regularized_fermi_app}
 t_\up^\reg(\k;\k) = -\frac{2\mu}{\k^2}
  + \int\!\frac{d\q}{(2\pi)^3}\!\left[\K_a(\k;\k,\q)
  - \frac{4\pi}{-i\frac{m_\down}M|\k|-\frac1a}\frac1{\q^2}\right]
  \frac{2\mu}{\q^2} - \int\!\frac{d\q}{(2\pi)^3}\,\K_a(\k;\k,\q)
  \left[t_\up(\k;\q) + \frac{2\mu}{\q^2}\right].
\end{equation}
The second term is a simple integral and has the following analytic
expression for its expansion in terms of $(a|\k|)^{-1}$:
\begin{equation}\label{eq:analytic_app}
 \begin{split}
  & \int\!\frac{d\q}{(2\pi)^3}\!\left[\K_a(\k;\k,\q) - \frac{4\pi}
  {-i\frac{m_\down}M|\k|-\frac1a}\frac1{\q^2}\right]\frac{2\mu}{\q^2} \\
  &= \Biggl[\frac{(u+1)\sqrt{2u+1}+\frac{(u+1)^3}u
  \arcsin\bigl(\frac{u}{u+1}\bigr)}\pi + \frac{(u+1)^3}{a|\k|}i
  + \cdots\Biggr]_{u=\frac{m_\up}{m_\down}}\frac{2\mu}{\k^2}.
 \end{split}
\end{equation}
Therefore, the nontrivial task is the calculation of the last term in
Eq.~(\ref{eq:t_regularized_fermi_app}) at $(a|\k|)^{-1}=0$ and its first
derivative with respect to $(a|\k|)^{-1}$, which requires solving the
two-dimensional integral equation (\ref{eq:integral_eq_fermi_app})
numerically.

For numerical purposes, it is more convenient to shift momentum
variables as $\p\,(\q)\to\p\,(\q)+\frac{m_\up}{M+m_\up}\k$ to move to
the center-of-mass frame and introduce a dimensionless function
\begin{equation}
 s_\up(\p) \equiv \frac{\k^2}{2\mu}\,t_\up\bigl(\k;\p+\tfrac{m_\up}{M+m_\up}\k\bigr).
\end{equation}
This function solves a simpler integral equation
\begin{equation}
 s_\up(\p) = -\I(\p)
  - \int\!\frac{d\q}{(2\pi)^3}\J_a(\p,\q)\,s_\up(\q),
\end{equation}
where the new inhomogeneous term $\I(\p)$ and integral kernel
$\J_a(\p,\q)$ are defined by
\begin{equation}
 \I(\p) \equiv \frac{\k^2}{\bigl(\p+\frac{m_\up}{M+m_\up}\k\bigr)^2}
\end{equation}
and
\begin{equation}
 \begin{split}
  \J_a(\p,\q) &\equiv \K_a\bigl(\k;\p+\tfrac{m_\up}{M+m_\up}\k,
  \q+\tfrac{m_\up}{M+m_\up}\k\bigr) \\
  &= \frac{4\pi}{\sqrt{m_\down\frac{M+m_\up}{M^2}\q^2
  -\frac{m_\down}{M+m_\up}\k^2-i0^+}-\frac1a}
  \frac1{\p^2+\q^2+\frac{2m_\up}M\p\cdot\q-\frac{m_\down}{M+m_\up}\k^2-i0^+}.
 \end{split}
\end{equation}
Since $\J_a(\p,\q)$ now depends on the angle only between $\p$ and $\q$,
each component of the partial-wave expansion
\begin{equation}
 s_\up(\p) = \sum_{\ell=0}^\infty s_\up^{(\ell)}(p)P_\ell(\cos\theta)
  \qquad (\cos\theta\equiv\hat\k\cdot\hat\p)
\end{equation}
solves an independent one-dimensional integral equation
\begin{equation}\label{eq:integral_eq_s_l_fermi}
 s_\up^{(\ell)}(p) = -\I^{(\ell)}(p)
  - \int_0^\infty\!dq\,\J_a^{(\ell)}(p,q)\,s_\up^{(\ell)}(q),
\end{equation}
where the partial-wave projections of the inhomogeneous term and integral
kernel are given by
\begin{equation}\label{eq:inhomo_partial_app}
 \I^{(\ell)}(p)
  = \frac{2\ell+1}2\int_{-1}^1\!d\cos\theta\,P_\ell(\cos\theta)\,\I(\p)
  = \frac{2\ell+1}2\int_{-1}^1\!d\cos\theta\,\frac{P_\ell(\cos\theta)}
  {p^2+\bigl(\frac{u}{2u+1}\bigr)^2+\frac{2u}{2u+1}p\cos\theta}
\end{equation}
and
\begin{equation}\label{eq:kernel_partial_app}
 \begin{split}
  \J_a^{(\ell)}(p,q) &= \frac{2\pi q^2}{(2\pi)^3}
  \int_{-1}^1\!d\cos\theta\,P_\ell(\cos\theta)\,\J_a(\p,\q) \\
  &= \frac{q^2}{\pi}
  \frac1{\sqrt{\frac{2u+1}{(u+1)^2}q^2-\frac1{2u+1}-i0^+}-\frac1{ak}}
  \int_{-1}^1\!d\cos\theta\,\frac{P_\ell(\cos\theta)}
  {p^2+q^2+\frac{2u}{u+1}p\,q\cos\theta-\frac1{2u+1}-i0^+}.
 \end{split}
\end{equation}
Here, $p=|\p|/|\k|$, $q=|\q|/|\k|$ are dimensionless momenta and the mass
ratio is denoted by $u=m_\up/m_\down$.  Note that the integrations over
$\cos\theta$ can be done analytically by using Gauss's hypergeometric
function:
\begin{equation}
 \int_{-1}^1\!d\cos\theta\,\frac{P_\ell(\cos\theta)}{x+y\cos\theta}
  = \frac{2\,\ell!}{x\,(2\ell+1)!!}\left(-\frac{y}{x}\right)^\ell
  {}_{2\!}F_1\!\left[\frac{\ell+1}2,\frac{\ell+2}2;
		\ell+\frac32;\left(\frac{y}{x}\right)^2\right].
\end{equation}

In terms of $s_\up^{(\ell)}(p)$, the last term in
Eq.~(\ref{eq:t_regularized_fermi_app}) is written as
\begin{equation}
 \begin{split}
  - \int\!\frac{d\q}{(2\pi)^3}\,\K_a(\k;\k,\q)
  \left[t_\up(\k;\q) + \frac{2\mu}{\q^2}\right]
  &= - \int\!\frac{d\q}{(2\pi)^3}\J_a\bigl(\tfrac{M}{M+m_\up}\k,\q\bigr)
  \left[s_\up(\q) + \I(\q)\right]\frac{2\mu}{\k^2} \\
  &= - \sum_{\ell=0}^\infty\int_0^\infty\!dq\,\J_a^{(\ell)}\bigl(\tfrac{u+1}{2u+1},q\bigr)
  \left[s_\up^{(\ell)}(q) + \I^{(\ell)}(q)\right]\frac{2\mu}{\k^2} \\
  &= \sum_{\ell=0}^\infty\left[s_\up^{(\ell)}\bigl(\tfrac{u+1}{2u+1}\bigr)
  + \I^{(\ell)}\bigl(\tfrac{u+1}{2u+1}\bigr) - \int_0^\infty\!dq\,
  \J_a^{(\ell)}\bigl(\tfrac{u+1}{2u+1},q\bigr)\,\I^{(\ell)}(q)\right]\frac{2\mu}{\k^2}.
 \end{split}
\end{equation}
In the last line, we used the integral equation for $s_\up^{(\ell)}(p)$
in Eq.~(\ref{eq:integral_eq_s_l_fermi}).  Because $s_\up^{(\ell)}(p)$ is
singular at $p=\left|\frac{u^2-(2u+1)}{(u+1)(2u+1)}\right|$ and
$\frac{u+1}{2u+1}$, it is better to work on the following function:
\begin{equation}
 \delta s_\up^{(\ell)}(p) \equiv s_\up^{(\ell)}(p) + \I^{(\ell)}(p)
  - \int_0^\infty\!dq\,\J_a^{(\ell)}(p,q)\,\I^{(\ell)}(q),
\end{equation}
in which the singularity at
$p=\left|\frac{u^2-(2u+1)}{(u+1)(2u+1)}\right|$ is eliminated and the
singularity at $p=\frac{u+1}{2u+1}$ is weaker.  This new function solves
an integral equation
\begin{equation}\label{eq:integral_eq_delta_s_l_fermi}
 \delta s_\up^{(\ell)}(p) = -\int_0^\infty\!dq\int_0^\infty\!dq'
  \,\J_a^{(\ell)}(p,q)\,\J_a^{(\ell)}(q,q')\,\I^{(\ell)}(q')
  - \int_0^\infty\!dq\,\J_a^{(\ell)}(p,q)\,\delta s_\up^{(\ell)}(q),
\end{equation}
and the last term in Eq.~(\ref{eq:t_regularized_fermi_app}) is given by
its value at $p=\frac{u+1}{2u+1}$:
\begin{equation}
 \begin{split}
  - \int\!\frac{d\q}{(2\pi)^3}\,\K_a(\k;\k,\q)
  \left[t_\up(\k;\q) + \frac{2\mu}{\q^2}\right]
  &= \left[\sum_{\ell=0}^\infty\delta s_\up^{(\ell)}
  \bigl(\tfrac{u+1}{2u+1}\bigr)\right]\frac{2\mu}{\k^2}.
 \end{split}
\end{equation}
In the case of equal masses $u=1$, we numerically solved the integral
equation (\ref{eq:integral_eq_delta_s_l_fermi}) at $(ak)^{-1}\simeq0$
for $0\leq\ell\leq\ell_{\max}$ up to $\ell_{\max}=20$.  By extrapolating
the result to $\ell_{\max}\to\infty$, we obtain
\begin{equation}
  \sum_{\ell=0}^\infty\delta s_\up^{(\ell)}\bigl(\tfrac23\bigr)
   \simeq 2.335 + \frac{7.047}{a|\k|}i + O(\k^{-2}).
\end{equation}
Since we find that the imaginary part of $O(\k^0)$ term is as small as
$\sim2\times10^{-6}$ and the real part of $O(\k^{-1})$ term is as small
as $\sim4\times10^{-5}$, we assume that their actual values are zero.
This result combined with Eqs.~(\ref{eq:t_regularized_fermi_app}) and
(\ref{eq:analytic_app}) is presented in Eq.~(\ref{eq:solution_fermi}).

\subsection{Spinless Bose gas}
In the case of spinless bosons, the quantity we would like to compute
is Eq.~(\ref{eq:t_regularized_bose}):
\begin{equation}\label{eq:t_regularized_bose_app}
 t^\reg(\k;\k) = \frac{m}{\k^2}
  + 2\int\!\frac{d\q}{(2\pi)^3}\!\left[\K_a(\k;\k,\q)
  - \frac{4\pi}{-i\frac{|\k|}2-\frac1a}\frac1{\q^2}\right]\frac{m}{\q^2}
  + 2\int\!\frac{d\q}{(2\pi)^3}\,\K_a(\k;\k,\q)
  \left[t(\k;\q)-\frac{m}{\q^2}\right],
\end{equation}
in which $t(\k;\p)$ solves the integral equation in
Eq.~(\ref{eq:integral_eq_bose}):
\begin{equation}
 t(\k;\p) = \frac{m}{\p^2}
  + 2\int\!\frac{d\q}{(2\pi)^3}\,\K_a(\k;\p,\q)\,t(\k;\q).
\end{equation}
Since the second term in Eq.~(\ref{eq:t_regularized_bose_app}) is twice
as large as that in Eq.~(\ref{eq:t_regularized_fermi_app}), its
expansion in terms of $(a|\k|)^{-1}$ is obtained from
Eq.~(\ref{eq:analytic_app}) with $u=1$ by multiplying it by two.  In
analogy with the case of spin-1/2 fermions, we shift momentum variables
as $\p\,(\q)\to\p\,(\q)+\k/3$ to move to the center-of-mass frame and
introduce a dimensionless function
\begin{equation}
 s(\p) \equiv \frac{\k^2}{m}\,t\bigl(\k;\p+\tfrac\k3\bigr).
\end{equation}
Each component of its partial-wave expansion
\begin{equation}
 s(\p) = \sum_{\ell=0}^\infty s^{(\ell)}(p)P_\ell(\cos\theta)
  \qquad (\cos\theta\equiv\hat\k\cdot\hat\p)
\end{equation}
solves an independent one-dimensional integral equation
\begin{equation}
 s^{(\ell)}(p) = \I^{(\ell)}(p)
  + 2\int_0^\infty\!dq\,\J_a^{(\ell)}(p,q)\,s^{(\ell)}(q).
\end{equation}
Here the inhomogeneous term $\I^{(\ell)}(p)$ and integral kernel
$\J_a^{(\ell)}(p,q)$ are obtained from Eqs.~(\ref{eq:inhomo_partial_app})
and (\ref{eq:kernel_partial_app}), respectively, by setting $u=1$.

Then by defining the better behaving function
\begin{equation}
 \delta s^{(\ell)}(p) \equiv s^{(\ell)}(p) - \I^{(\ell)}(p)
  - 2\int_0^\infty\!dq\,\J_a^{(\ell)}(p,q)\,\I^{(\ell)}(q),
\end{equation}
with its integral equation
\begin{equation}\label{eq:integral_eq_delta_s_l_bose}
 \delta s^{(\ell)}(p) = 4\int_0^\infty\!dq\int_0^\infty\!dq'
  \,\J_a^{(\ell)}(p,q)\,\J_a^{(\ell)}(q,q')\,\I^{(\ell)}(q')
  + 2\int_0^\infty\!dq\,\J_a^{(\ell)}(p,q)\,\delta s^{(\ell)}(q),
\end{equation}
the last term in Eq.~(\ref{eq:t_regularized_bose_app}) is given by its
value at $p=2/3$:
\begin{equation}
 \begin{split}
  2\int\!\frac{d\q}{(2\pi)^3}\,\K_a(\k;\k,\q)
  \left[t(\k;\q) - \frac{m}{\q^2}\right]
  &= \left[\sum_{\ell=0}^\infty\delta s^{(\ell)}
  \bigl(\tfrac23\bigr)\right]\frac{m}{\k^2}.
 \end{split}
\end{equation}
We numerically solved the integral equation
(\ref{eq:integral_eq_delta_s_l_bose}) at $(ak)^{-1}=0$ for
$1\leq\ell\leq\ell_{\max}$ up to $\ell_{\max}=20$.  By extrapolating the
result to $\ell_{\max}\to\infty$, we obtain
\begin{equation}\label{eq:nonzero_l_bose_app}
  \sum_{\ell=1}^\infty\delta s^{(\ell)}\bigl(\tfrac23\bigr)
   \simeq -2.044 + O(\k^{-1}).
\end{equation}
Since we find that the imaginary part of $O(\k^0)$ term is as small as
$\sim4\times10^{-6}$, we assume that its actual value is zero.  On the
other hand, the $\ell=0$ channel has to be treated specially due to the
Efimov effect.

As is known~\cite{Bedaque:1998kg,Braaten:2004rn}, the integral equation
(\ref{eq:integral_eq_delta_s_l_bose}) in the $\ell=0$ channel is ill
defined without introducing an ultraviolet momentum cutoff $\Lambda$:
\begin{equation}\label{eq:integral_eq_l=0_bose}
 \delta s^{(0)}(p) = 4\int_0^\infty\!dq\int_0^\infty\!dq'
  \,\J_a^{(0)}(p,q)\,\J_a^{(0)}(q,q')\,\I^{(0)}(q')
  + 2\int_0^{\Lambda/|\k|}\!dq\,\J_a^{(0)}(p,q)\,\delta s^{(0)}(q).
\end{equation}
First we find that $\delta s^{(0)}(2/3)$ obtained by solving this
integral equation is a log-periodic function of $|\k|/\Lambda$ in the
limit $\Lambda/|\k|\to\infty$, which is approximated by
\begin{equation}\label{eq:l=0_bose_app}
 \delta s^{(0)}\bigl(\tfrac23\bigr) \approx -3.9246-i\,12.20
  + \frac{1.036\,\cos(2s_0\ln|\k|/\Lambda+3.9604)
  - i\,1.032\,\sin(2s_0\ln|\k|/\Lambda+3.9604)}
  {1-0.08460\,\sin(2s_0\ln|\k|/\Lambda+3.9604)},
\end{equation}
with $s_0=1.00624$ being the solution to the transcendental equation
\begin{equation}
 \frac{8}{\sqrt3\,s_0}
  \frac{\sinh(\frac\pi6s_0)}{\cosh(\frac\pi2s_0)} = 1.
\end{equation}
Then this artificial cutoff $\Lambda$ has to be related to the physical
Efimov parameter $\kappa_*$ defined in Eq.~(\ref{eq:efimov_bose}).  To
this end, we observe that Eq.~(\ref{eq:integral_eq_l=0_bose}) without
the inhomogeneous term
\begin{equation}
 \delta s^{(0)}(p) = \frac2\pi\int_0^{\Lambda}\!dq
  \frac{q^2}{\sqrt{\frac34q^2-\frac{k^2}3-i0^+}-\frac1a}
  \int_{-1}^1\!d\cos\theta\,\frac{\delta s^{(0)}(q)}
  {p^2+q^2+p\,q\cos\theta-\frac{k^2}3-i0^+}
\end{equation}
is the Skorniakov--Ter-Martirosian equation to determine the binding
energy of three identical bosons by identifying $k^2/3$ as the collision
energy $mE$~\cite{Braaten:2004rn}.  By solving this homogeneous integral
equation at infinite scattering length $a\to\infty$, we find an infinite
tower of binding energies given by
\begin{equation}
 mE_n \to - e^{-2\pi n/s_0}\left(\frac\Lambda{5.67865}\right)^2
  \qquad (n\to\infty),
\end{equation}
from which we can read off the relationship between $\Lambda$ and
$\kappa_*$ as
\begin{equation}
 \Lambda = 5.67865 \times \kappa_*.
\end{equation}
This result combined with Eqs.~(\ref{eq:analytic_app}),
(\ref{eq:t_regularized_bose_app}), (\ref{eq:nonzero_l_bose_app}), and
(\ref{eq:l=0_bose_app}) is presented in Eq.~(\ref{eq:approx_bose}),
which determines the universal log-periodic function $f(|\k|/\kappa_*)$
introduced in Eq.~(\ref{eq:solution_bose}).

\section{Derivation of optical theorem in Eq.~(\ref{eq:optical_t}) \label{app:optical}}
Here we derive the optical theorem used in Eq.~(\ref{eq:optical_t}) by a
direct calculation starting with the set of integral equations in
Eq.~(\ref{eq:coupled_chi-F}).  We only consider the case in which
$a_\up^{-1}=a_\down^{-1}=a^{-1}=0$ and thus
$t_\psi(\k;\p)\equiv t_\up(\k;\p)=t_\down(\k;\p)$ because this case is
sufficient for the analysis presented in the text.  By using the
definitions of $t_{F,B}(\k;\p)$ in Eqs.~(\ref{eq:t_F_chi-F}) and
(\ref{eq:t_B_chi-F}), the three coupled integral equations in
Eq.~(\ref{eq:coupled_chi-F}) can be brought into the two independent
integral equations in Eq.~(\ref{eq:decoupled_chi-F}):
\begin{subequations}
 \begin{equation}
  t_F(\k;\p) = -\frac{m}{\p^2} - \int\!\frac{d\q}{(2\pi)^3}\,t_F(\k;\q)
   A(\ek-\eq,\k-\q)\frac1{\ep+\eq+\eps_{\k-\p-\q}-\ek-i0^+}
 \end{equation}
 and
 \begin{equation}
  t_B(\k;\p) = \frac{m}{\p^2} + 2\int\!\frac{d\q}{(2\pi)^3}\,t_B(\k;\q)
   A(\ek-\eq,\k-\q)\frac1{\ep+\eq+\eps_{\k-\p-\q}-\ek-i0^+}.
 \end{equation}
\end{subequations}
Here we used the expression of $\K_a(\k;\p,\q)$ in
Eq.~(\ref{eq:kernel_fermi}) and $A(k)=-D(k)$ is the two-body scattering
amplitude in Eq.~(\ref{eq:dimer_fermi}) at infinite scattering length.
Their complex conjugates are
\begin{subequations}
 \begin{equation}
  t_F^*(\k;\p) = -\frac{m}{\p^2} - \int\!\frac{d\q}{(2\pi)^3}\,t_F^*(\k;\q)
   A^*(\ek-\eq,\k-\q)\frac1{\ep+\eq+\eps_{\k-\p-\q}-\ek+i0^+}
 \end{equation}
 and
 \begin{equation}
  t_B^*(\k;\p) = \frac{m}{\p^2} + 2\int\!\frac{d\q}{(2\pi)^3}\,t_B^*(\k;\q)
   A^*(\ek-\eq,\k-\q)\frac1{\ep+\eq+\eps_{\k-\p-\q}-\ek+i0^+}.
 \end{equation}
\end{subequations}
By using these complex conjugates, $t_{F,B}(\k;\p)$ at $\p=\k$ can be
written as
\begin{subequations}
 \begin{equation}
  \begin{split}
   t_F(\k;\k) &= -\frac{m}{\p^2} - \int\!\frac{d\p}{(2\pi)^3}
   \,t_F(\k;\p)A(\ek-\ep,\k-\p)\frac{m}{\p^2} \\
   &= -\frac{m}{\p^2} + \int\!\frac{d\p}{(2\pi)^3}\,|t_F(\k;\p)|^2A(\ek-\ep,\k-\p) \\
   & + \int\!\frac{d\p\,d\q}{(2\pi)^6}\,t_F(\k;\p)A(\ek-\ep,\k-\p)
   \,t_F^*(\k;\q)A^*(\ek-\eq,\k-\q)\frac1{\ep+\eq+\eps_{\k-\p-\q}-\ek+i0^+}
  \end{split} 
 \end{equation}
 and
 \begin{equation}
  \begin{split}
   t_B(\k;\k) &= \frac{m}{\p^2} + 2\int\!\frac{d\p}{(2\pi)^3}
   \,t_B(\k;\p)A(\ek-\ep,\k-\p)\frac{m}{\p^2} \\
   &= \frac{m}{\p^2} + 2\int\!\frac{d\p}{(2\pi)^3}\,|t_B(\k;\p)|^2A(\ek-\ep,\k-\p) \\
   & - 4\int\!\frac{d\p\,d\q}{(2\pi)^6}\,t_B(\k;\p)A(\ek-\ep,\k-\p)
   \,t_B^*(\k;\q)A^*(\ek-\eq,\k-\q)\frac1{\ep+\eq+\eps_{\k-\p-\q}-\ek+i0^+}.
  \end{split} 
 \end{equation}
\end{subequations}
Then by using the identity
\begin{equation}
 2\,\Im A(\ek-\ep,\k-\p) = \int\!\frac{d\q_\up d\q_\down}{(2\pi)^6}
  |A(\ek-\ep,\k-\p)|^2(2\pi)^4\delta(\p+\q_\up+\q_\down-\k)
  \delta(\ep+\eps_{\q_\up}+\eps_{\q_\down}-\ek),
\end{equation}
the imaginary parts of $t_{F,B}(\k;\k)$ become
\begin{subequations}
 \begin{equation}
  \begin{split}
   2\,\Im\,t_F(\k;\k) &= \frac12\int\!\frac{d\p\,d\q_\up d\q_\down}{(2\pi)^9}\!
   \left|t_F(\k;\p)A(\ek-\ep,\k-\p) 
   - t_F(\k;\q_\up)A(\ek-\eps_{\q_\up},\k-\q_\up)\right|^2 \\
   & \times (2\pi)^4\delta(\p+\q_\up+\q_\down-\k)
   \delta(\ep+\eps_{\q_\up}+\eps_{\q_\down}-\ek)
  \end{split} 
 \end{equation}
 and
 \begin{equation}
  \begin{split}
   2\,\Im\,t_B(\k;\k) &= \frac23\int\!\frac{d\p\,d\q_\up d\q_\down}{(2\pi)^9}\!
   \left|t_B(\k;\p)A(\ek-\ep,\k-\p)
   + t_B(\k;\q_\up)A(\ek-\eps_{\q_\up},\k-\q_\up)\right. \\
   &\quad + \left.t_B(\k;\q_\down)A(\ek-\eps_{\q_\down},\k-\q_\down)\right|^2
   (2\pi)^4\delta(\p+\q_\up+\q_\down-\k)
   \delta(\ep+\eps_{\q_\up}+\eps_{\q_\down}-\ek).
  \end{split} 
 \end{equation}
\end{subequations}
Finally, by using the definitions of $t_{F,B}(\k;\p)$ in
Eqs.~(\ref{eq:t_F_chi-F}) and (\ref{eq:t_B_chi-F}) again, the imaginary
part of the forward three-body scattering amplitude $t_\chi(\k;\k)$ is
found to be
\begin{equation}\label{eq:optical_t-F}
 \begin{split}
  2\,\Im\,t_\chi(\k;\k) &= \frac43\,\Im\left[t_F(\k;\k)+t_B(\k;\k)\right] \\
  &= \int\!\frac{d\p\,d\q_\up d\q_\down}{(2\pi)^9}\!
  \left|t_\chi(\k;\p)A(\ek-\ep,\k-\p)
  + t_\psi(\k;\q_\up)A(\ek-\eps_{\q_\up},\k-\q_\up)\right. \\
  &\quad + \left.t_\psi(\k;\q_\down)A(\ek-\eps_{\q_\down},\k-\q_\down)\right|^2
  (2\pi)^4\delta(\p+\q_\up+\q_\down-\k)
  \delta(\ep+\eps_{\q_\up}+\eps_{\q_\down}-\ek),
 \end{split}
\end{equation}
which is equivalent to Eq.~(\ref{eq:optical_t}) when
$a_\up^{-1}=a_\down^{-1}=a^{-1}=0$.

On the other hand, in the case of spinless bosons, the definitions of
$t_{F,B}(\k;\p)$ in Eqs.~(\ref{eq:t_F_chi-B}) and (\ref{eq:t_B_chi-B})
lead to
\begin{equation}\label{eq:optical_t-B}
 \begin{split}
  2\,\Im\,t_\chi(\k;\k) &= \frac23\,\Im\left[t_F(\k;\k)+t_B(\k;\k)\right] \\
  &= \int\!\frac{d\p\,d\q_\up d\q_\down}{(2\pi)^9}\,
  \frac12\left|2\mspace{1mu}t_\chi(\k;\p)A(\ek-\ep,\k-\p)
  + t_\psi(\k;\q_\up)A(\ek-\eps_{\q_\up},\k-\q_\up)\right. \\
  &\quad + \left.t_\psi(\k;\q_\down)A(\ek-\eps_{\q_\down},\k-\q_\down)\right|^2
  (2\pi)^4\delta(\p+\q_\up+\q_\down-\k)
  \delta(\ep+\eps_{\q_\up}+\eps_{\q_\down}-\ek).
 \end{split}
\end{equation}
Since $t_\chi(\k;\p)$ in Eq.~(\ref{eq:optical_t-B}) is a half of that in
Eq.~(\ref{eq:optical_t-F}), the integrand in Eq.~(\ref{eq:optical_t-B})
is also a half of that in Eq.~(\ref{eq:optical_t-F}).  This establishes
that the contribution of the contact density to the differential
scattering rate of the $\chi$ atom in a spinless Bose gas is a half of
that in a spin-1/2 Fermi gas.

\twocolumngrid

\end{document}